\documentclass[11pt]{article}

\usepackage{geometry,setspace}
\small\normalsize
\usepackage{hyperref}
\hypersetup{
    colorlinks=true,
    citecolor=teal,
    linkcolor=teal
}
\usepackage{caption}
\usepackage{enumerate}
\usepackage{graphicx}
\usepackage{rotating,multirow}
\usepackage[table,dvipsnames]{xcolor}
\definecolor{DarkGreen}{rgb}{0,0.6,0}
\definecolor{royalblue}{rgb}{0.25, 0.41, 0.88}
\definecolor{darkcandyapplered}{rgb}{0.64, 0.0, 0.0}
 \usepackage{amsmath,amsfonts,amssymb,mathrsfs,amsthm}
\DeclareMathOperator{\Var}{Var}
 \usepackage{mathtools}
\usepackage{bbm}% Doublestruck maths characters like the indicator function 
\usepackage{bm}% Proper bold characters including Greek letters
\usepackage{float}
\usepackage[
  style=authoryear,
  maxbibnames=9,
  maxcitenames=2,
  mincitenames=1,
  uniquename=false,      % prevent initials in citations
  autocite=inline,
  giveninits=false,
  backend=biber,
  natbib
]{biblatex}
\DeclareNameAlias{author}{family-given}
\DeclareFieldFormat[article,periodical]{volume}{\mkbibbold{#1}}

\usepackage{booktabs}
\usepackage{adjustbox}
\usepackage{pdflscape}
\usepackage{todonotes}
\usepackage{authblk}
\usepackage{xr}
\usepackage{tikz-cd}
\usepackage{framed}

\usepackage{xfrac}

\usepackage{caption}
\usepackage[T1]{fontenc}
\input{glyphtounicode}
\usepackage{tikz}
\usepackage{ subcaption, cleveref}
\usepackage{prodint}
\usepackage{stackengine}
\newcommand{\indep}{\perp \!\!\! \perp}
\usepackage{pifont}   
\usepackage{enumitem}

\usetikzlibrary{arrows,positioning,calc} 
\usetikzlibrary{decorations.pathreplacing}
\tikzset{
    >=stealth',
    punkt/.style={
           rectangle,
           rounded corners,
           draw=black, thick,
           text width=5.5em,
           minimum height=2em,
           text centered},
    punktl/.style={
           rectangle,
           rounded corners,
           draw=black, thick,
           text width=7em,
           minimum height=2em,
           text centered},
    pil/.style={
           ->,
           shorten <=4pt,
       shorten >=4pt
    },
    pildotted/.style={
           ->,
           shorten <=4pt,
           shorten >=4pt,
  dotted,
  },
    punktf/.style={
           rectangle,
           text width=4.0em,
           minimum height=1.5em,
           text centered},
    punktfTop/.style={
           rectangle,
           text width=4.0em,
           minimum height=1.5em,
           text centered,
           append after command={
               [thick,shorten >=0.2bp, shorten <=0.2bp]
               (\tikzlastnode.north west)edge(\tikzlastnode.north east)
}
    },
    punktfBot/.style={
           rectangle,
           text width=4.0em,
           minimum height=1.5em,
           text centered,
           append after command={
               [thick,shorten >=0.2bp, shorten <=0.2bp]
               (\tikzlastnode.south west)edge(\tikzlastnode.south east)
            }
    }
}
\newtheorem{proposition}{Proposition}

\newtheorem{corollary}{Corollary}
\theoremstyle{definition}
\newtheorem{definition}{Definition}
 \newtheorem{assumption}{Assumption}
\newtheorem{remark}{Remark}
\newcommand{\Tr}{{\mathsf T}}
\newcommand{\calF}{{\mathcal F}}
\newcommand{\Cov}{{\operatorname{Cov}}}

\usepackage{accents}

\newcommand{\E}{\mathbb{E}}

\renewcommand{\leq}{\leqslant}

\DeclareMathOperator*{\argmin}{arg\,min}

\usepackage{listings}
\begin{document}
\title{Forecasting sub-population mortality using credibility theory}
\author{Mathias Lindholm}
 \affil{Department of Mathematics, Stockholm University}
\author{Gabriele Pittarello}
 \affil{Section of Biostatistics, University of Copenhagen}
 %\author{\mbox{ }}   % uncomment for blinded version
\date{\today}
 \maketitle

\begin{abstract}
	The focus of the present paper is to forecast mortality rates for small sub-populations that are parts of a larger super-population. In this setting the assumption is that it is possible to produce reliable forecasts for the super-population, but the sub-populations may be too small or lack sufficient history to produce reliable forecasts if modelled separately. This setup is aligned with the ideas that underpin credibility theory, and in the present paper the classical credibility theory approach is extended to be able to handle the situation where future mortality rates  are driven by a latent stochastic process, as is the case for, e.g., Lee-Carter type models.
    This results in sub-population credibility predictors that are weighted averages of expected future super-population mortality rates and expected future sub-population specific mortality rates. Due to the predictor's simple structure it is possible to derive  an explicit expression for the expected quadratic forecast error. Moreover, the proposed credibility modelling approach does not depend on the specific form of the super-population model, making it broadly applicable regardless of the chosen forecasting model for the super-population.
    The performance of the suggested sub-population credibility predictor is illustrated on simulated population data. These illustrations highlight how the credibility predictor serves as a compromise between only using a super-population model, and only using a potentially unreliable sub-population specific model. 
\end{abstract}

\section{Introduction}

Human mortality data is often aggregated into age-period tables by populations with common characteristics for a variety of applications ranging from epidemiology and demography to actuarial science. Modelling the dynamics of the (central) mortality rate of sub-populations can be challenging, especially when the sub-population sizes are small {\color{black} or the history is insufficient, see e.g.~\citet{millosovich14}}. In this situation it may be difficult to find reliable estimates for the parameters of a mortality model, which implies difficulties in providing reliable forecasts in this situation.%for noisy data.

\begin{figure}[ht]  % h: Place figure here (use t=top, b=bottom, p=page)
    \centering
    \includegraphics[width=0.7\textwidth]{ 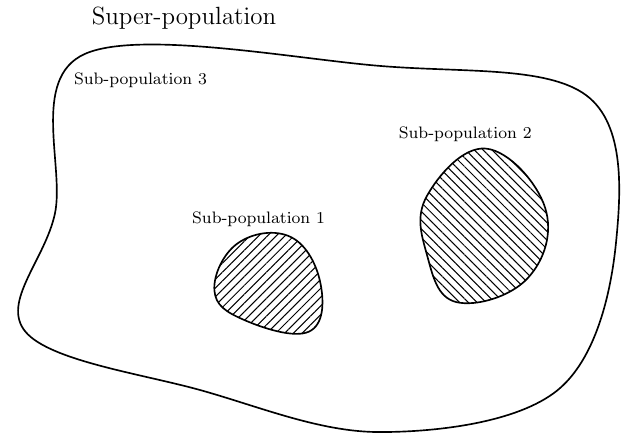}  % Adjust width as needed
    \caption{Sub-populations 1, and 2 are part of a larger reference super-population. {\color{black} Sub-population 3 contains the individuals belonging to the reference super-population and that are not part of either of sub-population 1 or 2. } While the mortality trend of the two smaller populations may be influenced by group-specific effects, the literature on this type of problem, which we discuss in our paper, usually assumes that they have some overall mortality trend in common with the super-population.}
    \label{fig:superpop-subpop}
\end{figure}

Further, while these small sub-populations, at least partly, will share global mortality dynamics with the overall super-population to which they belong, the sub-population mortality dynamics are also likely influenced by sub-population specific effects like biological markers, demographic variables or behavioural factors, see e.g.~\citet{dimai24}. 
{\color{black} \Cref{fig:superpop-subpop} provides a schematic representation of a setting involving a common super-population, within which two small sub-populations (sub-populations 1 and 2) are explicitly identified. The present paper focuses on modelling and forecasting mortality rates in such small sub-populations; specifically, groups possibly comprising relatively few individuals who are known to share homogeneous characteristics, as exemplified by sub-populations 1 and 2 in the figure. Moreover, in addition it is assumed that} 

\begin{itemize}
	
    \item[$(i)$] {\color{black} it is possible to reliably model and forecast the mortality rates of the super-population, but}
	
	\item[$(ii)$] (some of) the sub-populations are too small to {\color{black} allow for reliable direct modelling of mortality rates.}
\end{itemize}

{\color{black} Sub-population 3, by contrast, represents a residual group consisting of individuals that are not part of either sub-population 1 or 2, but that still belong to the overall super-population.}

Concerning $(i)$, we assume that we have access to reasonable mortality rate forecasts for, e.g., the entire Italian or Swedish population. Examples of commonly used models for mortality forecasting are the Poisson Lee-Carter model from \citet{brouhns02} and the generalised age-period-cohort models from \citet{hunt21}, see also the survey \citet{haberman11}. More recent data driven approaches are e.g. \citet{perla21, lindholm22, robben25}. Further, regarding $(ii)$, this suggests that detailed modelling is not guaranteed to be possible at the sub-population level. Instead of directly modelling a simpler model at the sub-population level, an alternative is to consider a credibility theory based approach. This approach is closely connected to modelling the sub-population mortality rates as a random effect type relative survival model {\color{black} in the fashion of \citet{buckley84}}, where the reference model is given by the super-population model. This approach is common to use in so-called claim-frequency modelling in non-life insurance pricing, see e.g.~\citet{buhlmann05}. Also note that by using a relative risk type model for the sub-population mortality introduces parameter sharing between the super- and sub-population models, which reduces the overall number of parameters. In the current paper the classical credibility theory approach to claim-frequency modelling is adjusted to handle that future mortality rates are modelled as a stochastic process, as in, e.g., the Poisson Lee-Carter model, see again \citet{brouhns02}.

In particular, by adjusting the assumptions that underpin the classical credibility theory approach to claim-frequency modelling it is possible to derive a simple credibility predictor for future expected sub-population mortality rates. The paper first derives the mortality predictor analytically as a function of population-level expected quantities, and then describe how these quantities can be replaced by computable plug-in estimates. The credibility predictor can be expressed as a weighted average between the expected mortality rate estimated on the super-population and the same expected mortality rate corrected by a random effect estimate that can be interpreted as an interaction between sub-population and age. The weights appearing in this averaging, i.e.~the credibility weights, provide an automatic calibration of the sub-population mortality as a compromise between a pure super-population model and a sub-population specific relative mortality model. The credibility weights will favour the super-population model if the the sub-population has a too low (effective risk) exposure, or if the super-population provides a sufficiently good fit for the sub-population; and vice versa. 

Relating to the current literature on models for mortality forecasting, the suggested credibility approach connects both to multi-population models and credibility theory. In terms of multi-population models, there is a vast literature {\color{black} that started with the seminal work of \citet{li05} and the related works on common factor models, see e.g.~\citet{booth02}}. 
These models all use more or less complex parameter sharing setups, generalising the simple relative survival {\color{black} construction} used in the present paper; {\color{black} see also the discussion in Section~\ref{sec: sub-pop mortality} below.} 
{\color{black} Credibility theory was first applied in the context of multi-population mortality models by \citet{tsai15, tsai17}.}
These papers, however, apply the credibility theory approach only to the latent stochastic process part of the mortality model, whereas in the current paper the credibility modelling also takes the Poisson death count variation into account. A related branch of modelling is the frailty approach taken in \citet{jarner11}, which also takes the Poisson variation into account. Furthermore, the credibility theory approach is closely connected to Bayesian modelling: the credibility estimator is known to be the \textit{best linear estimator} that approximates an underlying Bayesian model. In this context \textit{best} refers to its optimality with respect to the quadratic loss function, see e.g.~Theorem 2.5 in \citet{buhlmann05} and, as we will discuss in this paper, it is \textit{linear} with respect to the past observations.

The modelling approach suggested in the present paper is closely connected to the Bayesian multi-population model discussed in \citet{vanberkum17}. The main difference between the current approach and the approach taken in \citet{vanberkum17}, is that the linear credibility theory mortality predictor does not rely on explicit distributional assumptions other than with respect to the {\color{black} first two } moments and dependencies across ages, time periods and (sub-)populations {\color{black} of the mortality rates. This is particularly useful in situations with small sub-populations having small (effective) size, where a Bayesian approach may rely heavily on prior assumptions due to limited available data.}

The remainder of the paper is structured as follows: Section~\ref{sec: sub-pop mortality} starts with a general discussion of modelling of sub-population mortality and mortality forecasting, Section~\ref{sec:model} introduces the credibility theory setup. More specifically, \Cref{ss:theoretical-framework} presents the theoretical framework, while \Cref{ss:estimation} is devoted to deriving the credibility predictor and its expected quadratic forecast error. This is followed by a discussion of plug-in estimation in Section~\ref{sec: plug-in}. An empirical illustration of the proposed framework is given in \Cref{sec: numerical illustrations}. After introducing the simulation study setup in \Cref{sec: simulating sub-population data}, we discuss global mortality model selection in \Cref{ss:bic-comparison-models}. In \Cref{ss:simulation-one-data-set}, we first examine the estimated credibility model and discuss how the credibility predictor changes as more information becomes available in a single data set, in comparison with the global mortality model and the relative survival model. We then turn to a broader numerical illustration, where the proposed credibility predictor is evaluated against a number of benchmark models using proper scoring rules. Finally, the paper concludes with a number of closing remarks in \Cref{sec:conclusions}.

\section{Modelling the mortality of sub-populations}\label{sec: sub-pop mortality}

The perhaps most well-known paper on multi-population mortality models is \citet{li05}, which is an extension of the Lee-Carter model introduced in \citet{lee92}. Let us denote age with index $x \in \mathbb{X}$ and calendar time with $t \in \mathbb{T}$, with $\mathbb{X}, \mathbb{T} \subset \mathbb{N}^+$. The original formulation of the Lee-Carter model is a Gaussian model for the one-year logarithmic difference of central mortality rates, and was phrased as a Poisson death count model in \citet{brouhns02} according to
\begin{align}\label{eq: Brouhns et al}
	D_{x, t} \mid E_{x, t}, \mu_{x, t} \sim \operatorname{Poisson}(E_{x, t}  \mu_{x, t}),
\end{align}
where the number of deaths at calendar time $t$ and age $x$ is denoted with $D_{x, t}$ conditionally on the number of individuals exposed to risk at age $x$ and time $t$ ($E_{x, t}$) and on a mortality trend $\mu_{x, t}$. The model specification from \eqref{eq: Brouhns et al} relies on a Poisson likelihood equivalence based on using piecewise constant mortality rates and assuming that all individuals are i.i.d., see e.g.~\citet{andersson22}. This likelihood equivalence stresses that both the $D_{x, t}$s and $E_{x, t}$s are random and only known in retrospect, which motivates the conditioning seen in \eqref{eq: Brouhns et al}. Further, henceforth, $E_{x, t}$ will be referred to as ``exposure'', following common notation in the literature on human mortality models. Moreover, also to ease the exposition, $\mu_{x, t}$ will be referred to as the (central) ``mortality rate''.

{\color{black} Continuing, in \citet{lee92} and \citet{brouhns02} the mortality rates are described by a log-bilinear function given by}
\begin{align}\label{eq: Poisson LC}
	\log(\mu_{x, t}) := \alpha_x +  \beta_x \kappa_t,
\end{align}
where the $\kappa_t$s are modelled as a Gaussian time series, see Section~\ref{sec: time series forecasting} below.

{\color{black} Further, models focusing on an overall population as described by \eqref{eq: Brouhns et al} will be referred to as a ``global mortality model''.} Analogously, the multi-population extension of the Lee-Carter model from \citet{li05} can be phrased as a Poisson death count model as in \citet{li13} according to
\[
	D_{x, t}^i \mid E_{x, t}^i, \mu_{x, t}^i \sim \operatorname{Poisson}(E_{x, t}^i \mu_{x, t}^i),
\]
{\color{black} where $i \in \mathbb{I}$ with $\mathbb{I} \subset \mathbb{N}^+$ is the index denoting the sub-population and }
\begin{align}\label{eq: Li and Lee}
	\log(\mu_{x, t}^i) :=&~ \alpha_x^i +  \beta_x \kappa_t + \beta_x^i \kappa_t^i\nonumber\\
		=&~ \log(\mu_{x, t}) + \alpha_x^i - \alpha_x + \beta_x^i \kappa_t^i\nonumber\\
		=&~ \log(\mu_{x, t}) + \widetilde\alpha_x^i + \beta_x^i \kappa_t^i\nonumber\\
		=&~ \log(\mu_{x, t}) + \log(v_{x, t}^i).
\end{align}
% I would remove: That is, 

As discussed in \citet{vanberkum17}, the rates from \eqref{eq: Li and Lee} can be estimated using a two-step frequentist approach by
\begin{enumerate}
	\item treating the $\mu_{x, t}$s as deterministic and estimating the $\mu_{x, t}$s by  maximising the likelihood under the model assumption of the standard Poisson Lee-Carter model given by \eqref{eq: Brouhns et al} and \eqref{eq: Poisson LC},
	
	\item given the  maximum-likelihood-estimate (MLE) $\widehat\mu_{x, t}$ of $\mu_{x, t}$, estimate $v_{x, t}^i$ from \eqref{eq: Li and Lee} as the MLE of the subsequent sub-population Poisson death count model using $E_{x, t}^i \widehat\mu_{x, t}$ as log-offset.
\end{enumerate}
This two-step setup is similar to what will be pursued in the credibility setting introduced below, but where the second step will be replaced by a credibility adjustment.

As an alternative to the above two-step procedure one can view \eqref{eq: Li and Lee} directly as a generalised non-linear Poisson model with $E_{x, t}^i$ as log-offset.

\subsection{Relative survival models and modelling small sub-populations}

In the current paper, focus is on small sub-populations such that the part corresponding to $v_{x, t}^i$ in \eqref{eq: Li and Lee}  is not believed to be possible to estimate and forecast reliably, which is the motivation for instead considering models of the type given by

\begin{align}\label{eq: basic cred model}
	D_{x, t}^i \mid E_{x, t}^i, \Theta_{x}^i, \mu_{x, t} \sim \operatorname{Poisson}(E_{x, t}^i \Theta_{x}^i \mu_{x, t}),
\end{align}

where $\Theta_{x}^i$ is an age and population dependent random effect, and where $\mu_{x, t}$ corresponds to a given global mortality model.

The parametrisation of the mortality rates from \eqref{eq: basic cred model} corresponds to using
\begin{align}\label{eq: relative survival}
	\mu_{x, t}(\Theta_x^i) := \mu_{x, t} \Theta_x^i,
\end{align}
which can be thought of as a random effects version of a relative survival model, see e.g.~p.~413 in \citet{andersen93}. That is, if we replace the random effect $\Theta_x^i$ in \eqref{eq: basic cred model} by a deterministic parameter $\theta_x^i$, i.e.
\begin{align}\label{eq: relative survival model}
	\mu_{x, t}^i :=  \mu_{x, t} \theta_x^i,
\end{align}
it follows that
\[
	\frac{\mu_{x, t}^i}{\mu_{x, t}^j} = \frac{\theta_x^i}{\theta_x^j}.
\]
Further, $\theta_x^i$ corresponds to an adjustment of the intercept of $\mu_{x, t}$ on the log-link scale by $\log(\theta_x^i)$, i.e., $\log(\theta_x^i)$ corresponds to $\widetilde\alpha_x^i$ in \eqref{eq: Li and Lee}.

\begin{remark}~
\label{remark:general-mortality-models-obs}
\begin{enumerate}[label = (\roman*)]
	\item Note that the mortality rate given by \eqref{eq: relative survival} resembles that of a (stratified) proportional frailty model. This resemblance, however, is faltering, since this would require that the $\Theta_x^i$s are assigned independently at birth, see e.g. \citet{vaupel79}. {\color{black} A related well-defined frailty based approach is the so-called SAINT model, see \citet{jarner11}.} 

	\item  Consider the model
	\[
	     D^i_{x,t}\mid E^i_{x,t}, \mu_{x, t} \sim \text{Poisson}\left(E^i_{x,t}\mu_{x,t}\right),
    	\]
	and let
	\[
		D_{x, t} = \sum_i D^i_{x,t} \quad \text{and}\quad E_{x, t} = \sum_i E^i_{x,t}.
	\]
 	Assuming conditional independence across all $(x, t, i)$s, given piece-wise constant $\mu_{x, t}$s over integer ages and years, provides us with a log-likelihood given by

\begin{align}
\label{eq: likelihood GM}
	\ell((\mu_{x, t})_{x, t}) &\propto \sum_{x, t, i} D^i_{x,t} \log( E^i_{x,t} \mu_{x,t}) - E^i_{x,t}\mu_{x,t} \nonumber \\
		&\propto \sum_{x, t} D_{x,t} \log( \mu_{x,t}) - E_{x,t}\mu_{x,t}.
\end{align}

This allows us to estimate $\mu_{x, t}$ using specific parametrisations such as, e.g., the Lee-Carter parametrisation from \eqref{eq: Poisson LC}. Further, note that \eqref{eq: likelihood GM} is likelihood equivalent to the Poisson model given by
\[    
     D_{x,t} \mid E_{x,t}, \mu_{x, t} \sim \text{Poisson}\left(E_{x,t} \mu_{x,t}\right)
\]
as it should.
	
	\item Note that if we replace the random effect $\Theta_x^i$ in \eqref{eq: basic cred model} by a deterministic parameter $\theta_x^i$, i.e.~use \eqref{eq: relative survival model} and assume

\begin{align}\label{eq: Poisson relative survival model}
	D^i_{x,t} \mid E^i_{x,t}, \mu_{x, t}  \sim \text{Poisson}\left(E^i_{x,t}\theta^i_x \mu_{x,t}\right),
\end{align}
and treat $\mu_{x, t}$ as known, the MLE of $\theta_x^i$ is then given by
\begin{align}\label{eq: Theta MLE}
	\widehat \theta_x^i := \frac{\sum_v E^i_{x,v}F^i_{x,v}}{\sum_v E^i_{x,v} \mu_{x,v}},
\end{align}
where $F^i_{x,t}:=D^i_{x,t}/E^i_{x,t}$ are the observed mortality ratios for $i \in \mathbb{I}$, $x \in \mathbb{X}$ and $t \in \mathbb{T}$.
\end{enumerate}
\end{remark}

\subsection{Extrapolating a calendar period and cohort component}\label{sec: time series forecasting}

The sub-population models that are considered in the current paper are of type \eqref{eq: basic cred model} which means that the calendar period effect will only be introduced via the $t$-dimension of the global mortality rate model $\mu_{x, t}$. The models that will be used  in the numerical illustrations are all of Poisson {\color{black}Generalised Age-Period-Cohort} type, {\color{black} see Table~\ref{tab:label_model_predictor_constraints} in Section~\ref{sec: numerical illustrations} and \Cref{appendix:gapc}}, which means that all parameters can be estimated using a generalised non-linear Poisson model with log-link given a set of suitable identification constraints.

For projecting central mortality rates, it is necessary to extrapolate the calendar period effects for $t > t'$ and cohorts $t-x > t' - 1$, where $t' := \max{\mathbb{T}}$ and $\mathbb{T}$ corresponds to the set of observed calendar times. Following the original Lee-Carter approach from \citet{lee92} this corresponds to modelling the $\kappa_t$s as time series. To be precise, the fitted $\kappa_t$s, the $\widehat\kappa_t$s, are modelled as time series belonging in the ARIMA($p, q, d$)-class of models, see e.g.~{\color{black} Chapter 3 in \citet{shumway00}}. This is in line with the implementation of the procedures used in the \texttt{StMoMo} \texttt{R}-package that later will be used in the numerical illustrations in Section~\ref{sec: numerical illustrations}. By  using this package, model selection of the time series part of the modelling will be carried out using the Bayesian information criterion (BIC), see {\color{black} e.g.~Chapter 2 in \citet{shumway00}}.

\begin{remark}
One can consider {\color{black} different approaches for extrapolating the calendar time effect $\kappa_t$}, {\color{black} for example using a multivariate time-series model to jointly model the calendar time effect of the sub-populations in a similar fashion to the vector auto-regressive moving average models (VARIMA) discussed in \citet{tiao89}.} However, the focus of the present paper is rather to illustrate how credibility theory can be used to combine sub-population information. Assessing different choices for extrapolation of calendar period effects is outside of the scope of the paper.
\end{remark}

\section{Credibility theory for mortality rate forecasting}
\label{sec:model}

In this Section, we present the credibility theory model for forecasting the mortality rate of sub-populations. As discussed in the beginning of \Cref{sec: sub-pop mortality}, the two-step estimation procedure described for fitting the Poisson Li-Lee model from \citet{li13} will be generalised to the situation when starting from an arbitrary global $\mu_{x, t}$ model that will serve as a basis for sub-population mortality models using credibility theory. The  focus of this section  is to  derive such a credibility estimator generalising the setup used for non-life frequency modelling described in Chapter 4 in \citet{buhlmann05}. After presenting the theoretical framework in \Cref{ss:theoretical-framework}, we study the credibility estimator in \Cref{ss:estimation}. Under our assumptions, this first yields a theoretical linear credibility estimator expressed in terms of the past observed mortality rates. Since some of the quantities entering this estimator appear as conditional expectations and variances, the estimator is not directly computable. In \Cref{sec: plug-in} we discuss a computable plug-in version.
Note that, in practical applications, the central mortality rate $\mu_{x,t}$ must itself be estimated and forecast over the relevant ages, periods, and sub-populations. For ease of exposition, however, we treat $\mu_{x,t}$ and its moments as known, as the main focus of the paper is the development of the credibility theory model.

\subsection{Theoretical Framework}
\label{ss:theoretical-framework}

To be able to define the credibility estimator, we need to introduce a number of assumptions on the underlying data generating process:

\begin{assumption}
\label{ass:CM1}
Conditionally on a random parameter $\Theta_x^i \in \mathbb{R}$ and $\mu_{x, t}$, we have that for age $x \in \mathbb{X}$, sub-population $i \in \mathbb{I}$ and $t \in \mathbb{T}$
    $$
    D^i_{x,t} \mid E^i_{x,t}, \Theta^i_x, \mu_{x, t}  \sim \text{Poisson}\left(E^i_{x,t}\Theta^i_x \mu_{x,t}\right).
    $$
\end{assumption}    

\begin{assumption}

\label{ass:CM2}
$\Theta_x^i \indep \mu_{x, t}$, for all $x, t, i$, and ${\mathbb{E}}\left[\Theta_x^i\right]=1$

\end{assumption}

\begin{assumption}
\label{ass:CM3}
    For $i \in \mathbb{I}$, $x \in \mathbb{X}$ and $t,v \in \mathbb{T}$ with $v \neq t$, it holds that
    \[
        D^i_{x,t} \indep D^i_{x,v} \mid \Theta^i_x, E_{x, t}^i, E_{x, v}^i, \mu_{x, t}, \mu_{x, v}.
    \]
\end{assumption}    

\vspace{3mm}

\Cref{ass:CM1} is added explicitly based on the Poisson likelihood equivalence discussed in Section~\ref{sec: sub-pop mortality}. If we let $t' := \max \mathbb{T}$, and let
\[
  \calF_{t'} := \{\mu_{x, s} : x \in \mathbb{X}, s \in \mathbb{T}\},
\]
\Cref{ass:CM1} also provides us with the following relations
\[
  \mu_{x,t} \left(\Theta^i_x\right) = \mu_{x,t}\Theta^i_x = \mathbb{E}\left[F^i_{x,t} \mid \Theta^i_x, \mathcal{F}_{t^{\prime}}\right].
\]

\Cref{ass:CM2} introduces $\Theta_x^i$ as a random age-group effect per sub-population that is stable over calendar time. This aims at capturing the hetereogeneity within an age-group and sub-population; for more on this, see Section~\ref{ss:estimation}. Further, note that the assumption that the $\Theta_x^i$s are independent of the possibly stochastic baseline mortality rates $\mu_{x,t}$ is an assumption about the underlying stochastic dynamics. In practice, however, we will observe dependence between the estimated counterparts of $\mu_{x,t}$ and $\Theta_x^i$. Furthermore, this assumption ensures that
\begin{align}
  \mathbb{E}\left[\mu_{x,t} \left(\Theta^i_x\right) \mid \mathcal{F}_{t^{\prime}} \right] =&~ \mathbb{E}[F_{x, t}^i \mid \mathcal{F}_{t^{\prime}}] = \mu_{x,t},
\label{eq: expectaion of Fxti}
\end{align}
holds, which implies that the sub-population-specific mortality projections will not diverge. For more on this, see e.g.~\citet{li05, vanberkum17}. Further, Assumptions \ref{ass:CM1}-\ref{ass:CM2} also provides us with
\begin{align}
  \Var\left(\mu_{x,t} \left(\Theta^i_x\right) \mid  \mathcal{F}_{t^{\prime}} \right) =&~ (\mu_{x,t})^2 \Var(\Theta^i_x ),\label{eq:varianceoftheta}
\end{align}
and
\begin{align}\label{eq: variance of Fxti}
    \Var\left(F^i_{x,t} \mid  \mathcal{F}_{t^{\prime}} \right) &= (\mu_{x,t})^2\Var(\Theta^i_x ) + \mu_{x,t}\E\left[\frac{\Theta_x^i}{E^i_{x,t}} ~\Bigg|~ \mathcal{F}_{t^{\prime}}\right],
\end{align}
due to that the exposures are random; recall the Poisson likelihood equivalence discussed in Section~\ref{sec: sub-pop mortality}. Equations \eqref{eq:varianceoftheta} -- \eqref{eq: variance of Fxti} also illustrate why a plug-in estimation step is needed for practical implementation, since $\E[\Theta_x^i / E_{x,t}^i \mid \calF_{t'}]$ and $\Var(\Theta^i_x )$ are not directly computable and will therefore be replaced by computable counterparts. This is discussed in Section~\ref{sec: plug-in}.

Moreover, for the purpose of the present paper, from a mathematical point of view it is sufficient to assume that $\Cov(\mu_{x, t}, \Theta_x^i) = 0$. But, since this is an assumption about the underlying data generating process we believe that it is conceptually more natural to assume independence in Assumption~\ref{ass:CM2}.
\vspace{3mm}

\Cref{ass:CM3} states that observations from disjoint time intervals are conditionally independent, given exposures, intensities, and $\Theta_x^i$. In the present setting, this can be interpreted as a form of conditional independence across birth cohorts. Furthermore, since Assumptions \ref{ass:CM1} and \ref{ass:CM3} treat the exposures as random, by using Assumptions~\ref{ass:CM1} -- \ref{ass:CM3} it holds that
\begin{align}\label{eq: covariance raw mortality rates}
    \text{Cov}(F^i_{x,t}, F^i_{x,v} \mid  \mathcal{F}_{t^{\prime}}) &= \mu_{x,t}\mu_{x,v} \Var(\Theta^i_x).
\end{align}
That is, by not conditioning on exposures, $D_{x, t}^i$ and $D_{x, v}^i$ are neither uncorrelated or independent.

Again, for the mathematical derivations below it would suffice to assume conditional uncorrelatedness.
\vspace{3mm}

To conclude, Assumptions~\ref{ass:CM1} -- \ref{ass:CM3} can be seen as extensions of the assumptions used in \citet[Ch. 4]{buhlmann05}, and we will return to a more detailed comparison of the resulting credibility estimators in Section~\ref{sec: buhlmann comparison} below.

\subsection{Estimation}
\label{ss:estimation}

The objective of this paper is to derive a linear credibility estimator $\widehat{\mu_{x, t'+h} \left(\Theta^i_x\right)}$ of the future central mortality rate $\mu_{x, t'+h} \left(\Theta^i_x\right)$ under Assumptions \ref{ass:CM1} -- \ref{ass:CM3} over a forecasting horizon $h \in \mathbb{N}^+$, given the information $\mathcal{F}_{t^{\prime}}$.

In credibility theory, estimation can be performed parametrically using a pure Bayesian approach, assuming a distribution for the data and a prior distribution over the group-specific characteristics. Alternatively, the Bayesian result can be approximated non-parametrically with the so-called (inhomogeneous) linear credibility model, see e.g. Chapter 4 in \citet{buhlmann05}. The two approaches are equivalent when the past losses are assumed to follow a distribution from the exponential family and the group-specific characteristics are chosen as a conjugate prior, see e.g.~\citet{jewell74}. Under standard regularity conditions \citep[Theorem 3.15][]{buhlmann05}, the linear credibility estimator, given sub-populations $i, i \in \mathbb{I}$ and ages $x \in \mathbb{X}$, is expressed in terms of parameters $\boldsymbol\omega_x^i \in \mathbb{R}^{|\mathbb{T}| + 1}$.

\begin{definition}\label{definition: linear cred}
    The linear credibility estimator $\widehat{\mu_{x, t'+h} \left(\Theta^i_x\right)}$ of the future central mortality rate $\mu_{x, t'+h} \left(\Theta^i_x\right)$, for sub-population $i \in \mathbb{I}$, and age $x \in \mathbb{X}$, based on the observed mortality ratios $\mathbf{F}^i_x = \left(F^i_{x,t}\right)_{t \in \mathbb{T}}$, with $\mathbf{F}^i_x \in \mathbb{R}^{|\mathbb{T}|}$, is the estimator in the class
    $$
    \mathcal{H}(\mathbf{F}^i_x) = \left\{h(\mathbf{F}^i_x) : \; h(\mathbf{F}^i_x)=\omega^i_{x, 0} + \sum_{v \in \mathbb{T}} \omega^i_{x,v} F^i_{x,v} \; ; \;\boldsymbol\omega_x^i \in \mathbb{R}^{|\mathbb{T}| + 1}\right\},
    $$
    such that
    \[
      \widehat{\mu_{x, t'+h} \left(\Theta^i_x\right)}  = \argmin_{h(\mathbf{F}^i_x) \in  \mathcal{H}(\mathbf{F}^i_x) } \mathbb{E}\left[\left(\mu_{x, t'+h} \left(\Theta^i_x\right) - h(\mathbf{F}^i_x) \right)^2 ~ \Bigg|~ \mathcal{F}_{t^{\prime}}  \right].
    \]
\end{definition}

\begin{proposition}
\label{proposition:credibilityestimator}
    Under Assumptions \ref{ass:CM1} -- \ref{ass:CM3} the (inhomogeneous) linear credibility estimator of the central mortality rate from Definition~\ref{definition: linear cred} for sub-population $ i \in \mathbb{I}$, and age $x \in \mathbb{X}$ over a forecasting horizon $h \in \mathbb{N}^+$ is given by
    \begin{align}\label{eq: linear cred no plug-in}
      \widehat{\mu_{x, t'+h} \left(\Theta^i_x\right)} := (1 - z_x^i) \overline \mu_{x, t' + h} + z_x^i  \mu_{x, t' + h}^i,
    \end{align}
    where    
    \begin{align*}
      \overline \mu_{x, t' + h} :=&~ \mathbb{E}[\mu_{x, t' + h} \mid \mathcal{F}_{t'}],\\
      z_x^i :=&~ \frac{\sum_{v \in \mathbb{T}}  \mu_{x,v} / \mathbb{E}\left[ \frac{\Theta^i_x}{E^i_{x,v}} \mid  \mathcal{F}_{t^{\prime}}\right]}{ \frac{1}{\Var(\Theta^i_x)} +  \sum_{v \in \mathbb{T}}  \mu_{x,v} / \mathbb{E}\left[ \frac{\Theta^i_x}{E^i_{x,v}} \mid  \mathcal{F}_{t^{\prime}}\right] },\\
       \mu_{x, t' + h}^i :=&~ \overline \mu_{x, t' + h} R_x^i,
   \end{align*}
   together with
    \[
      R_x^i := \frac{\sum_{v \in \mathbb{T}}  F_{x,v}^i / \mathbb{E}\left[ \frac{\Theta^i_x}{E^i_{x,v}} \mid  \mathcal{F}_{t^{\prime}}\right]}{ \sum_{v \in \mathbb{T}}  \mu_{x,v} / \mathbb{E}\left[ \frac{\Theta^i_x}{E^i_{x,v}} \mid  \mathcal{F}_{t^{\prime}}\right] }.
    \]
\end{proposition}

\begin{proof}[Sketch of Proof]
    The solution to the minimisation from Definition~\ref{definition: linear cred} is equivalent to solving
\[
  \widehat \omega_x^i = \argmin_{ \boldsymbol\omega_x^i \in \mathbb{R}^{|\mathbb{T}| + 1} } \mathbb{E}\left[\left(\mu_{x, t'+h} \left(\Theta^i_x\right) - \omega^i_{0} - \sum_{v \in \mathbb{T}} \omega^i_{x,v} F^i_{x,v} \right)^2 ~ \Bigg|~ \mathcal{F}_{t^{\prime}}  \right].
\]
Under  first order conditions the solution to this minimisation problem is given by
\begin{align*}
    \widehat \omega^i_{x, 0}  &:=  \bar \mu_{x,t^\prime+h}-\sum_{v \in \mathbb{T}} \hat  \omega^i_{x,v} \mu_{x,v}, \\
    \widehat \omega^i_{x,t} &:= \frac{\bar\mu_{x,t^\prime+h}/ \mathbb{E}\left[ \frac{\Theta^i_x}{E^i_{x,t}} \mid  \mathcal{F}_{t^{\prime}}\right]}{ \frac{1}{\Var(\Theta^i_x)} +  \sum_{v \in \mathbb{T}}  \mu_{x,v} / \mathbb{E}\left[ \frac{\Theta^i_x}{E^i_{x,v}} \mid  \mathcal{F}_{t^{\prime}}\right] },
\end{align*}
and where
$$
\widehat{\mu_{x, t'+h} \left(\Theta^i_x\right)} = \widehat \omega^i_{x, 0} + \sum_{v \in \mathbb{T}} \widehat \omega_{x, v} F_{x, v}^i.
$$
The remaining part of the proof follows by algebraic manipulations identifying $z_x^i$ and $\mu_{x, t}^i$.
\end{proof}

A full proof of \Cref{proposition:credibilityestimator} is given in \Cref{appendix:proof_cm}. We note that, throughout the paper, the overline notation is used to denote future expected mortality rates, namely $\overline{\mu}_{x,t'+h} := \mathbb{E}[\mu_{x,t'+h}\mid \mathcal{F}_{t'}]$, with $h>0.$ By contrast, the in-sample quantities $\mu_{x,t}$ are treated as exogenous, in line with the interpretation discussed at the beginning of \Cref{sec:model}.
\vspace{2mm}

The quantity $R_x^i$ introduced in Proposition~\ref{proposition:credibilityestimator} can be interpreted as a sub-population-specific correction factor with unit conditional expectation, $\mathbb{E}[R_x^i \mid \mathcal{F}_{t'}] = 1$. In this sense, $R_x^i$ plays a role similar to $\Theta_x^i$, capturing the relative mortality level of sub-population $i$ compared with the global mortality model. The difference is that $\Theta_x^i$ is a latent parameter of the data-generating process, whereas $R_x^i$ is a quantity derived from the credibility formulas. Hence, $\widehat{\mu_{x,t'+h}(\Theta_x^i)}$ in \Cref{eq: linear cred no plug-in} may be interpreted as a weighted average of a global mortality model and the relative mortality model $\mu_{x, t' + h}^i := \overline \mu_{x, t' + h} R_x^i$.
\vspace{3mm}

Further, note that the credibility estimator from \Cref{proposition:credibilityestimator} can be rewritten according to
\begin{align}\label{eq: cred relative survival}
	\widehat{\mu_{x, t'+h} \left(\Theta^i_x\right)} :=  \overline \mu_{x, t' + h} \widehat \Theta_x^i,
\end{align}
where
\begin{align}\label{eq: cred scale factor}
	\widehat \Theta_x^i := 1 + z^i_{x}(R_x^i - 1).
\end{align}
Thus, \eqref{eq: cred relative survival} and \eqref{eq: cred scale factor} illustrate that one could equivalently define the credibility estimator in terms of the relativities introduced through the random effects $\Theta_x^i$ directly.

Furthermore, concerning the credibility weights $z_x^i$ from Proposition~\ref{proposition:credibilityestimator} we see that these will be close to 1 if the exposures tend to be large, which means that the credibility estimator collapses to the relative survival model for sub-population $i$. Continuing, by noting that
\[
	\operatorname{CV}(\mu_{x, t}(\Theta_x^i) \mid  \mathcal{F}_{t^{\prime}}) := \frac{\sqrt{\Var(\mu_{x, t}(\Theta_x^i) \mid  \mathcal{F}_{t^{\prime}})}}{\mathbb{E}[\mu_{x, t}(\Theta_x^i) \mid  \mathcal{F}_{t^{\prime}}]} = \frac{\sqrt{\mathbb{E}[(\mu_{x, t}(\Theta_x^i) - \mu_{x, t})^2 \mid  \mathcal{F}_{t^{\prime}}]}}{\mu_{x, t}} = \sqrt{\Var\left(\Theta^i_x\right)},
\]
it follows that
\[
	\frac{1}{\Var\left(\Theta^i_x\right)} = \frac{1}{\operatorname{CV}(\mu_{x, t}(\Theta_x^i) \mid  \mathcal{F}_{t^{\prime}})^2}.
\]
Hence, if the coefficient of variation of $\mu_{x, t}(\Theta_x^i)$ is large there is a large unexplained source of variation when trying to describe the mean of sub-population $i$ using $\mu_{x, t}$. This again motivates collapsing the credibility estimator to the relative survival model for sub-population $i$. Furthermore, note that if the true underlying process would violate the assumption that $\Theta_x^i$ does not depend on calendar time, this would result in a higher estimate of the variance of $\Theta_x^i$, placing more weight on the sub-population specific mortality model.

To conclude, when the amount of data in sub-population $i$ is sufficiently large, or if the fit when describing the mean of sub-population $i$ using the global mean is too poor, the credibility estimator from \Cref{proposition:credibilityestimator} reduces to the sub-population specific  relative survival model.
\vspace{3mm}

Concerning forecasting uncertainty, we consider the expected quadratic forecast error of the linear credibility estimator defined according to
\[
  \operatorname{Q}(\mu_{x,t^\prime+h} \left(\Theta^i_x\right), \widehat{ \mu_{x,t^\prime+h} \left(\Theta^i_x\right)}\mid \calF_{t'}) := \mathbb{E}\left[(\mu_{x,t^\prime+h} \left(\Theta^i_x\right)-\widehat{ \mu_{x,t^\prime+h} \left(\Theta^i_x\right)})^2 \mid \mathcal{F}_{t^{\prime}}\right],
\]
for which we have the following result:
\begin{proposition}\label{thm: MSEP}
    The expected quadratic forecast error of the linear credibility estimator from Proposition~\ref{proposition:credibilityestimator} is given by
    \begin{equation}
        \operatorname{Q}(\mu_{x,t^\prime+h} \left(\Theta^i_x\right), \widehat{ \mu_{x,t^\prime+h} \left(\Theta^i_x\right)}\mid \calF_{t'}) = \Var\left(\mu_{x,t^\prime+h} \left(\Theta^i_x\right) \mid \mathcal{F}_{t^{\prime}}\right)+(z^i_x)^2(\overline{\mu}_{x,t^\prime+h})^2 \Var\left(R^i_x \mid \mathcal{F}_{t^{\prime}}\right)
    \end{equation}
where 
$$
\Var\left(\mu_{x,t^\prime+h} \left(\Theta^i_x\right) \mid \mathcal{F}_{t^{\prime}}\right)=\overline{\sigma}^2_{x,t'  + h}(\Var\left(\Theta^i_x\right)+1)+(\overline{\mu}_{x,t^\prime+h})^2 \Var\left(\Theta^i_x\right)
$$
with $\overline\sigma_{x, t' + h}^2 := \Var(\mu_{x, t' + h} \mid \mathcal{F}_{t^{\prime}})$, and where
$$
\Var\left(R^i_x \mid \mathcal{F}_{t^{\prime}}\right) = \Var\left(\Theta^i_x\right)+\frac{1}{\sum_v \mu_{x,v} / \mathbb{E}\left[ \frac{\Theta^i_x}{E^i_{x,v}} \mid  \mathcal{F}_{t^{\prime}}\right]}.
$$
\end{proposition}
The proof of Proposition~\ref{thm: MSEP} is provided in Appendix~\ref{app: proof MSEP}.

Concerning the expression for the quadratic forecast error from Proposition~\ref{thm: MSEP}, this illustrates that by using the suggested credibility theory approach introduces an additional source of variation. In particular, even in the boundary cases $z_x^i = 0$ and $z_x^i = 1$, corresponding to full weight on the super-population model and the sub-population model, respectively, the expected quadratic forecast error still depends on $\Var(\Theta_x^i)$. This results in an over-dispersion effect relative to a model without sub-population heterogeneity.

\subsubsection{Plug-in estimation}\label{sec: plug-in}

To make Propositions~\ref{proposition:credibilityestimator}--\ref{thm: MSEP} practically implementable, we need computable counterparts for $\Var(\Theta_x^i)$ and $\E[\Theta_x^i / E_{x,t}^i \mid \calF_{t'}]$, while treating the underlying mortality model $\mu_{x,t}$ and its moments as given exogenously. For notational simplicity, we avoid hats on, e.g., $\mu_{x,t}$, although in applications they are understood as plug-in estimates of the corresponding population quantities. The global mortality model will be based on the plug-in estimates obtained from applying the methods discussed in Section~\ref{sec: time series forecasting}; see Table~\ref{tab:label_model_predictor_constraints} in Section~\ref{sec: numerical illustrations} for specific models that will be used for illustration purposes. Here, it is worth noting that although Poisson Lee--Carter type models are used in the present paper to model the global mortality rates, it is also possible to use crude (observed) death rates $F^i_{x,t}$ for the in-sample estimation of $\mu_{x,t'+h}$.

For plug-in estimation of $\E[\Theta_x^i / E_{x, t}^i \mid \calF_{t'}]$ we will use its sample size one counterpart, i.e.
\begin{align}\label{eq: sample size on exp}
  \widehat \E\left[\frac{\Theta_x^i}{E_{x, t}^i}~ \Big|~ \calF_{t'} \right] := \frac{1}{E_{x, t}^i}.
\end{align}

For plug-in estimation of $\Var(\Theta^i_x)$ we will use a moment based estimator, given estimates of the global mortality $\mu_{x, t}$. For other approaches to estimate $\Var(\Theta^i_x)$, see e.g.~the discussion in \citet[Ch.~4.8]{buhlmann05} and \citet{norberg82}.

The estimator that will be used is obtained by combining

\begin{align*}
\widehat\Var\left(\sum^{t^\prime}_{t=1} F_{x, t}^i ~\Bigg|~ \mathcal{F}_{t^{\prime}}\right) &:= \mathbb{E} \left[\left(\sum^{t^\prime}_{t=1} F_{x, t}^i- \mathbb{E}\left[\sum^{t^\prime}_{t=1} F_{x, t}^i ~\Bigg|~ \mathcal{F}_{t^{\prime}}\right]\right)^2 ~\Bigg|~ \mathcal{F}_{t^{\prime}}\right]\\
&\approx \left(\sum^{t^\prime}_{t=1} F_{x, t}^i- \sum^{t^\prime}_{t=1}\mu_{x, t} \right)^2,
\end{align*}
with
\begin{align*}
\Var\left(\sum^{t^\prime}_{t=1} F_{x, t}^i ~\Bigg|~ \mathcal{F}_{t^{\prime}}\right) &= \sum^{t^\prime}_{t=1} \Var(F_{x, t}^i \mid \mathcal{F}_{t^{\prime}}) + 2 \sum_{v>t} \Cov(F_{x, v}^i, F_{x, t}^i \mid \mathcal{F}_{t^{\prime}})\\
  &= \left(\sum_{t \in \mathbb{T}} \mu_{x, t}\right)^2 \Var(\Theta_x^i) + \sum_{t \in \mathbb{T}} \frac{\mu_{x,t}}{E_{x, t}^i},
\end{align*}

and solving for $\Var(\Theta_x^i)$. This results in the following moment based plug-in variance estimator

\begin{align}\label{eq: plug-in variance Theta}
	\widehat { \Var}_{\text{m}} (\Theta^i_x) := \frac{\left((\sum_{t \in \mathbb{T}}  F^i_{x,t}- \sum_{t \in \mathbb{T}}  \widehat \mu_{x,t})^2 - \sum_{t \in \mathbb{T}} \frac{\mu_{x, t}}{E_{x, t}^i} \right)}{ (\sum_{t \in \mathbb{T}} mu_{x,t})^2}.
\end{align}

Further, note that the plug-in estimator $\widehat \Var_{\text{m}} (\Theta^i_x)$from \eqref{eq: plug-in variance Theta} is not guaranteed to be non-negative due to that the underlying death counts may contain too many zeros, which suggests the following zero-adjusted estimator:

\begin{align}\label{eq: plug-in variance Theta zero-adjusted}
	\widehat{\Var}(\Theta^i_x) := \max\left(\widehat{\Var_{\text{m}}} (\Theta^i_x), 0\right).
\end{align}

Here one can note that when \eqref{eq: plug-in variance Theta zero-adjusted} equals 0, it means that the credibility weight is set to 0 and the credibility estimator will place all weight on the global MLE, which is the natural baseline when not having a sufficient amount of data.

Using the above, we can now state a computable plug-in version of Proposition~\ref{proposition:credibilityestimator}:

\begin{corollary}
\label{corollary:credibilityestimator}
    The computable plug-in estimator of the linear credibility estimator from Proposition~\ref{proposition:credibilityestimator} using \eqref{eq: sample size on exp} and \eqref{eq: plug-in variance Theta zero-adjusted} is given by
    $$
    \widehat{\widehat{\mu_{x, t'+h} \left(\Theta^i_x\right)}}  = (1- \widehat z^i_x) \overline \mu_{x, t' + h} + \widehat z^i_x \widehat{\mu}^i_{x, t' + h},
    $$
    where
\begin{align*}
	\widehat z^i_{x} :=&~ \frac{\sum_{v \in \mathbb{T}}E^i_{x,v}\mu_{x,v}}{\frac{1}{\widehat \Var\left(\Theta^i_x \right)} +	\sum_{v \in \mathbb{T}} E^i_{x,v} \mu_{x,v}},\\
  \widehat{\mu}^i_{x, t' + h} :=&~ \overline \mu_{x, t' + h} \widehat R_x^i,\\
      \widehat R_x^i :=&~ \frac{\sum_v E^i_{x,v}F^i_{x,v}}{\sum_v E^i_{x,v} \mu_{x,v}} =: \widehat \theta_x^i,
    \end{align*}
  where $\widehat \theta_x^i$ is the MLE from \eqref{eq: Theta MLE}.
  \end{corollary}

Continuing, since the estimator is based on a Poisson assumption and the minimisation of an $L^2$-distance, it is sufficient to only specify the first two moments of $\Theta_x^i$ in order to obtain an explicit credibility estimator. This can be compared to the Bayesian analog of \eqref{eq: basic cred model} discussed in \citet{vanberkum17}, which relies on a number of additional distributional assumptions on $\Theta_x^i$ and on the parameters of the global mortality model.

As with Proposition~\ref{proposition:credibilityestimator}, we provide a computable version of Proposition~\ref{thm: MSEP}:

\begin{corollary}\label{cor: plug-in MSEP}
  The computable plug-in estimator of the expected quadratic forecast error from Proposition~\ref{thm: MSEP} using \eqref{eq: sample size on exp} and \eqref{eq: plug-in variance Theta zero-adjusted} is given by
  \begin{equation}
        \widehat{\operatorname{Q}}(\mu_{x,t^\prime+h} \left(\Theta^i_x\right), \widehat{ \mu_{x,t^\prime+h} \left(\Theta^i_x\right)}\mid \calF_{t'}) = \widehat{\Var}\left(\mu_{x,t^\prime+h} \left(\Theta^i_x\right) \mid \mathcal{F}_{t^{\prime}}\right)+(\widehat{z}^i_x)^2(\overline{\mu}_{x,t^\prime+h})^2 \widehat{\Var}\left(R^i_x \mid \mathcal{F}_{t^{\prime}}\right)
    \end{equation}
where 
$$
\widehat{\Var}\left(\mu_{x,t^\prime+h} \left(\Theta^i_x\right) \mid \mathcal{F}_{t^{\prime}}\right)=\overline{\sigma}^2_{x,t'  + h}(\widehat{\Var}\left(\Theta^i_x\right)+1)+(\overline{\mu}_{x,t^\prime+h})^2 \widehat{\Var}\left(\Theta^i_x\right)
$$
with $\overline\sigma_{x, t' + h}^2 := \Var(\mu_{x, t' + h} \mid  \mathcal{F}_{t^{\prime}})$, and where
$$
\widehat{\Var}\left(R^i_x \mid \mathcal{F}_{t^{\prime}}\right) = \widehat{\Var}\left(\Theta^i_x\right)+\frac{1}{\sum_v E_{x, v}^i \mu_{x,v}}.
$$

\end{corollary}

Here one can note that $\widehat\Var(R_x^i \mid \calF_{t'}) = \Var(\widehat \theta_x^i \mid \calF_{t'}, (E_{x,t}^i)_{t \in \mathbb{T}})$. Consequently, as discussed after Proposition~\ref{thm: MSEP}, it is seen that the credibility approach is a random effects model that although being estimated as a weighted average between a global model and a local relative survival MLE, will have greater variance than the corresponding marginal models even when $\widehat{z}_x^i$ is 0 or 1.

Furthermore, as just noted, $\widehat\Var(R_x^i \mid \calF_{t'})\neq \Var(\widehat R_x^i \mid \calF_{t'})$. Hence, for the plug-in version of the expected quadratic forecast error from Corollary~\ref{cor: plug-in MSEP} it follows that
\[
  \widehat{\operatorname{Q}}(\mu_{x,t^\prime+h} \left(\Theta^i_x\right), \widehat{ \mu_{x,t^\prime+h} \left(\Theta^i_x\right)}\mid \calF_{t'}) \neq \operatorname{Q}(\mu_{x,t^\prime+h} \left(\Theta^i_x\right), \widehat{\widehat{ \mu_{x,t^\prime+h} \left(\Theta^i_x\right)}}\mid \calF_{t'}).
\]
For a longer discussion of the latter type of quadratic error, see, e.g., the discussion in \cite{lindholm20}.

\subsubsection{Connection to credibility models for non-life insurance claim frequencies}\label{sec: buhlmann comparison}

The model we present differs from the frequency model for non-life insurance pricing presented in Chapter 4 in \citet{buhlmann05}. In the non-life insurance setting, the main objective is typically to predict one-period-ahead claim frequencies for portfolios (super-populations) partitioned into smaller policyholder groups (sub-populations). In this framework, the underlying mean is usually assumed to be static, in the sense that it is not driven by a stochastic calendar-time trend. The main differences can be seen in \Cref{ass:CM1}, where we assume that there is an effect on the central mortality rate that depends on the calendar period and that the (future) mortality rates are stochastic. If we consider an age-only mortality model, i.e., $\mu_{x, t} := \mu_x =$  constant (deterministic), and let $E_x^i := \sum_{t \in \mathbb{T}} E_{x, t}^i$, which can be thought of as a Lee-Carter model where the calendar period effect has been dropped, \Cref{corollary:credibilityestimator} simplifies to the following:

\begin{corollary}
\label{cor:agemodel}
    Under the additional assumption that $\mu_{x,t}:=\mu_x$, the plug-in version of the linear credibility estimator for the central mortality rate from \Cref{corollary:credibilityestimator} reduces to
    $$
    	\widehat{\widehat{\mu_{x, t'+h} \left(\Theta^i_x\right)}} := (1 - \widehat z^i_x) \overline\mu_x + \widehat z^i_x \widehat{\mu}^i_x,
    $$
    where 

    $$
    \overline\mu_x := \mathbb{E}\left[\mu_{x,t} \mid \mathcal{F}_{t^\prime}\right] = \mu_x
    $$
    
    The credibility weights are
    $$
    	\widehat z^i_x := \frac{E^i_x}{\frac{1}{\overline\mu_x \widehat\Var(\Theta^i_x )}+ E^i_x},
    $$
    and where
    \[
    	\widehat{\mu}^i_x := \overline\mu_x \widehat \theta_x^i,
    \]
    with $\widehat \theta_x^i$ given by
    \[
    	\widehat \theta_x^i := \frac{1}{\overline\mu_x} \sum_v \frac{E^i_{x,v}}{E^i_x} F^i_{x,v}.
    \]
\end{corollary}
Thus, \Cref{cor:agemodel} retrieves the credibility estimator from Corollary~4.8 in \citet{buhlmann05}.
\vspace{3mm}

\section{Numerical illustrations}
\label{sec: numerical illustrations}

In this section, we illustrate the proposed credibility-based mortality predictor through a simulation study. The data-generating mechanism is described in Section~\ref{sec: simulating sub-population data}. It produces data for three sub-populations based on Italian mortality data from the Human Mortality Database \citep{wilmoth21}. The numerical study is divided into two parts. First, we investigate the behaviour of the credibility model on a single simulated data set. This includes a discussion of the estimation of the model components and an illustration of how, as more data are included in the training sample, the credibility model places increasing weight on the relative survival model. Second, we assess the predictive performance of the proposed approach systematically. We compare its expected quadratic forecast error with that of a separate mortality model fitted independently to each sub-population, and we evaluate its out-of-sample performance across 30 simulated data sets using mean squared error and Poisson deviance.

\subsection{Simulation of (sub-)population data}\label{sec: simulating sub-population data}

The synthetic (sub-)population data is based on crude Italian mortality data by ages and period from the Human Mortality Database \citep{hunt21}. The data consists of ages between $0$ and $110$ and the period $1872$ to $2021$. 

The observed one-year death probabilities for the Italian population are used as a basis for the simulation model. These probabilities are defined according to

$$
	q_{x,t}=\frac{d_{x,t}}{n_{x,t}},
$$
for $x=0, \ldots, 110$ and $t=1872, \ldots, 2021$, where $n_{x,t}=e_{x,t}+d_{x,t}/2$ denotes the observed initial exposure for age $x$ in calendar time $t$. The initial exposure for age and period for the data is estimated using the \texttt{central2initial} function from the \texttt{StMoMo} package.

{\color{black} The simulated data will be split into three sub-populations with sizes small, medium, and large. These sub-populations will be based on observed log-odds-ratios for the Italian population,  where the log-odds-ratio for age $x$ and calendar time $t$ is given by}%Based on the odds-ratio of the Italian population for age $x$ and calendar time $t$ given by
\[
	\log(q_{x,t}/(1-q_{x,t}))=\delta_{x,t},
\]
with $\delta_{x,t} \in \mathbb{R}$. Given these inital log-odds ratios, the log-odds-ratios for the sub-populations are defined according to
\[
	\log(q^i_{x,t}/(1-q^i_{x,t}))=\log(\Theta_x^i)+\delta_{x,t}, \quad i=1,2,3,
\]
where the $\Theta_x^i$s are non-negative random effects. 
That is, 
\begin{align}
\label{eq: q super-pop}
	q_{x,t} = \frac{\text{e}^{\delta_{x,t}}}{1 + \text{e}^{\delta_{x,t}}},
\end{align}
and
\begin{align}\label{eq: q sub-pop}
	q^i_{x,t} = \frac{\Theta_{x}^i\text{e}^{\delta_{x,t}}}{1 + \Theta_{x}^i\text{e}^{\delta_{x,t}}},~i = 1, 2, 3.
\end{align}

The above assumptions allow us to simulate the evolution of the different sub-populations using a binomial model based on an initial number of individuals given by $n_{0,t}^i$ together with the one-year death probabilities $q^i_{x,t}$, i.e.~raw deaths are simulated according to

\[
	D^i_{x,t} \mid  N_{x,t}, q_{x, t}^i \sim \text{Binomial}\left(N_{x,t} ,q^i_{x,t}\right).
\]

Further, concerning the $\Theta_{x}^i$s these are modelled as follows:
\begin{itemize}
    \item $\Theta_x^1 \sim \text{Uniform}(0.7,0.8)$,
    \item $\Theta_x^2 \sim \text{Uniform}(1.2,1.3)$,
    \item $\Theta_x^3 \equiv 1$.
\end{itemize}
Finally, the super-population, denoted ``0'', is constructed according to

\[
	D_{x, t}^0 := \sum_{i = 1}^3 D_{x, t}^i, \quad \text{and}~E_{x, t}^0 := \sum_{i = 1}^3 E_{x, t}^i,
\]
where the exposures, i.e.~the $E_{x,t}^i$s, are defined in \Cref{eq: pop size to exposure} in \Cref{app: life portfolio}. Note that by using the above described procedure for generating synthetic data, it is not assumed that the underlying data generating model complies with the model described in \Cref{sec:model}. However, for sufficiently small mortality rates the assumed Poisson likelihood equivalence will result in one-year death probabilities that are approximately equal to those used in the Binomial simulation model. For more details, see Appendix~\ref{app: life portfolio}. As described above, the underlying data that is being used comes from the Italian population, and an illustration of the $q_{x, t}^i$s is given in Figure~\ref{fig:lexis}. 
 
The above procedure is implemented by assuming an initial population size of $100~000$ individuals for {\color{black} each cohort of} the super-population 0, and that sub-population 1 {\color{black} cohorts} consist of $5~000$ individuals, and sub-population 2 {\color{black} cohorts} consist of $500$ individuals. That is, this corresponds to that population 3 {\color{black} cohorts} above consist of $94~500$ individuals. Further, the ``rule of thumb'' discussed in \citet{millosovich14}, which suggests that one should use an exposure of at least  $25~000$ lived years over $8$ calendar years of data in order to ascertain reliable estimates implies that the population size should not be smaller than about 3~000 individuals. In this respect sub-population 2 is of a size that could result in problematic estimates, whereas sub-population 1 is small but not obviously too small for reliable estimation.

\subsection{Selection of the global mortality model}
\label{ss:bic-comparison-models}

As  described above, the credibility  based mortality predictor does not rely on a specific underlying global mortality model for $\mu_{x, t}$ being used. In the numerical illustrations three different global mortality  models from the the generalised age-period-cohort family from \citet{hunt21} are considered: a Lee-Carter (LC) model, an age-period-cohort (APC) model and a Renshaw-Haberman (RH) model; see \Cref{tab:label_model_predictor_constraints}  for details. Furthermore, from \Cref{tab:label_model_predictor_constraints} it is seen that the LC model (on average) produces the best in-sample BIC scores when using the rolling window approach described in Section~\ref{sec:evaluation-metrics}. Due to this, and in order to keep the presentation of the results more precise, the illustrations below focus on the situation using an LC model to describe the global mortality.

\begin{table}[htb]
\centering
\begin{tabular}{|c|c|c|c|}
\hline
\textbf{Model} & \textbf{Predictor (log-scale)} & \textbf{Identification constraints}& \textbf{BIC  (average)} \\
\hline
 LC & $\alpha_x+\beta_x \kappa_t$ & $\sum_x \beta_x = 1$& \textbf{14.4} \\
\hline
 APC & $\alpha_x+ \kappa_t + \gamma_{t-x}.$ & $\sum_s \gamma_{s} = 0$, $ \sum_s s \gamma_{s} = 0$&15.6  \\
\hline
 RH & $\alpha_x+\beta_x \kappa_t + \gamma_{t-x}.$ & $\sum_x \beta_x = 1$&15.5 \\
\hline
\end{tabular}
\caption{\label{tab:label_model_predictor_constraints} The models (column one) and identification constraints (column three) for the different Generalised Age-Period-Cohort model configurations in column two; Lee-Carter (LC), Age-Period-Cohort (APC), Renshaw-Haberman (RH). The BIC score is displayed in column three divided by a thousand.}

\end{table}

\subsection{Illustration on a single simulated data set}
\label{ss:simulation-one-data-set}

In this subsection, we use a single simulated data set, generated as described in Section~\ref{sec: simulating sub-population data}, to mimic a real application. We first examine the estimation of the components $\widehat \theta_x^i$ and $\widehat \Var(\Theta_x^i)$ and illustrate how they influence the credibility predictor. We then study how the proposed approach changes as the sample size increases. Finally, we compare the expected quadratic forecast error from Theorem~\ref{thm: MSEP} with the bootstrap-based benchmark of \citet{koissi06}.

\subsubsection{Estimation of components in the credibility model}

\begin{figure}[ht]
  \centering
  \begin{subfigure}[t]{0.45\linewidth}
    \centering
    \includegraphics[width=\linewidth]{ 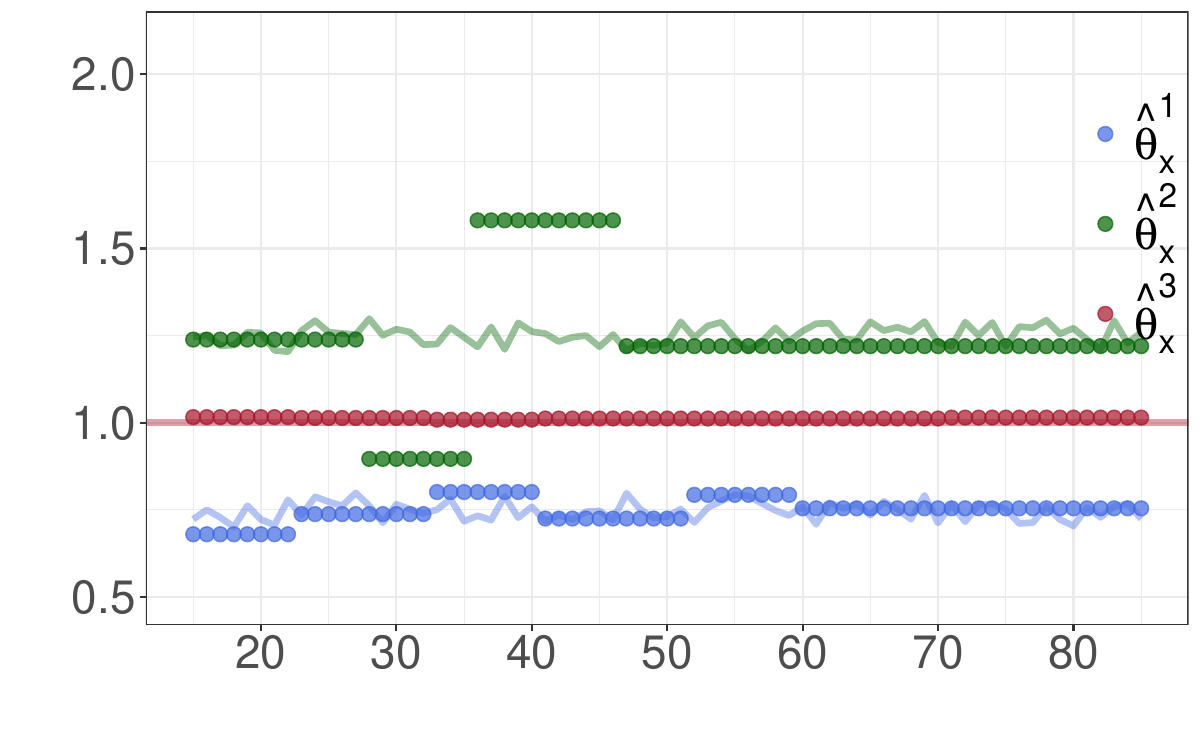}
    \caption{Estimated  sub-population effects $\widehat \theta^i_x$ for $i=1,2,3$ (dots) for different ages (x-axis) against the true values $ \Theta^i_x$ for $i=1,2,3$ (solid line).}
    \label{fig:allthetas}
  \end{subfigure}
  \hfill
  \begin{subfigure}[t]{0.45\linewidth}
    \centering
    \includegraphics[width=\linewidth]{ 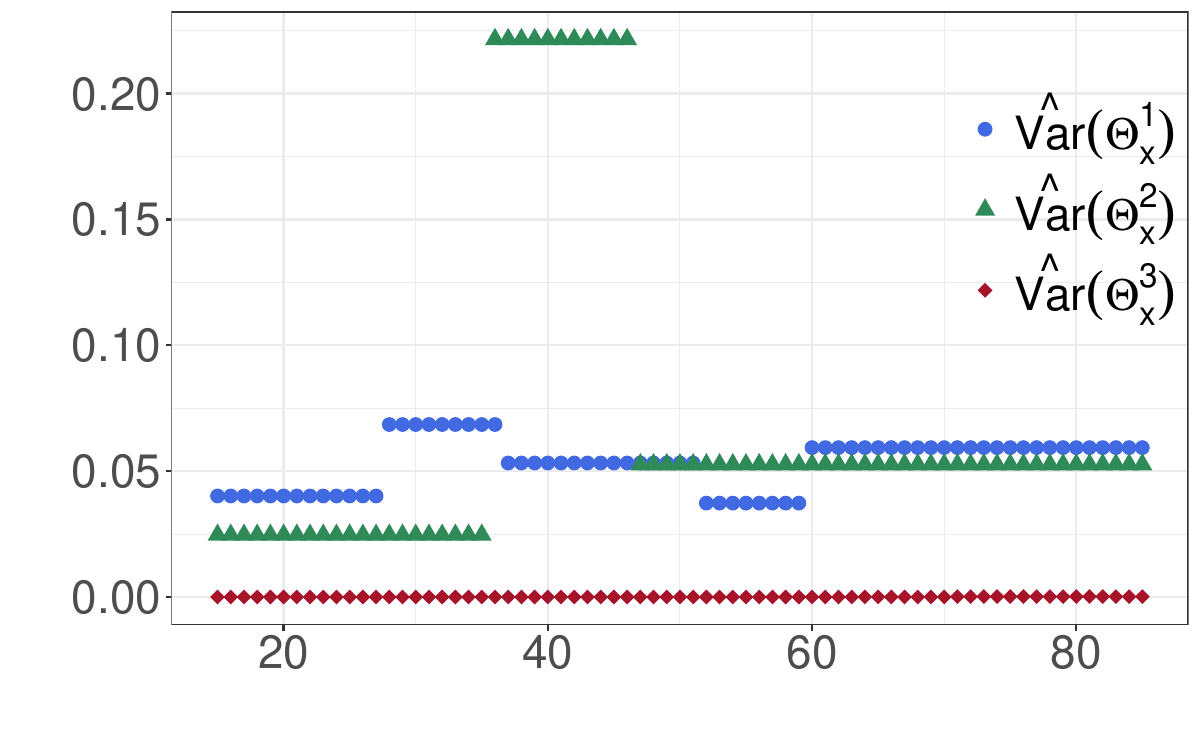}
    \caption{Estimated variance $\widehat \Var (\Theta_x^i)$ for $i=1,2,3$ by age (x-axis).}
    \label{fig:variances}
  \end{subfigure}

  \vspace{1em}

  \begin{subfigure}[t]{0.45\linewidth}
    \centering
    \includegraphics[width=\linewidth]{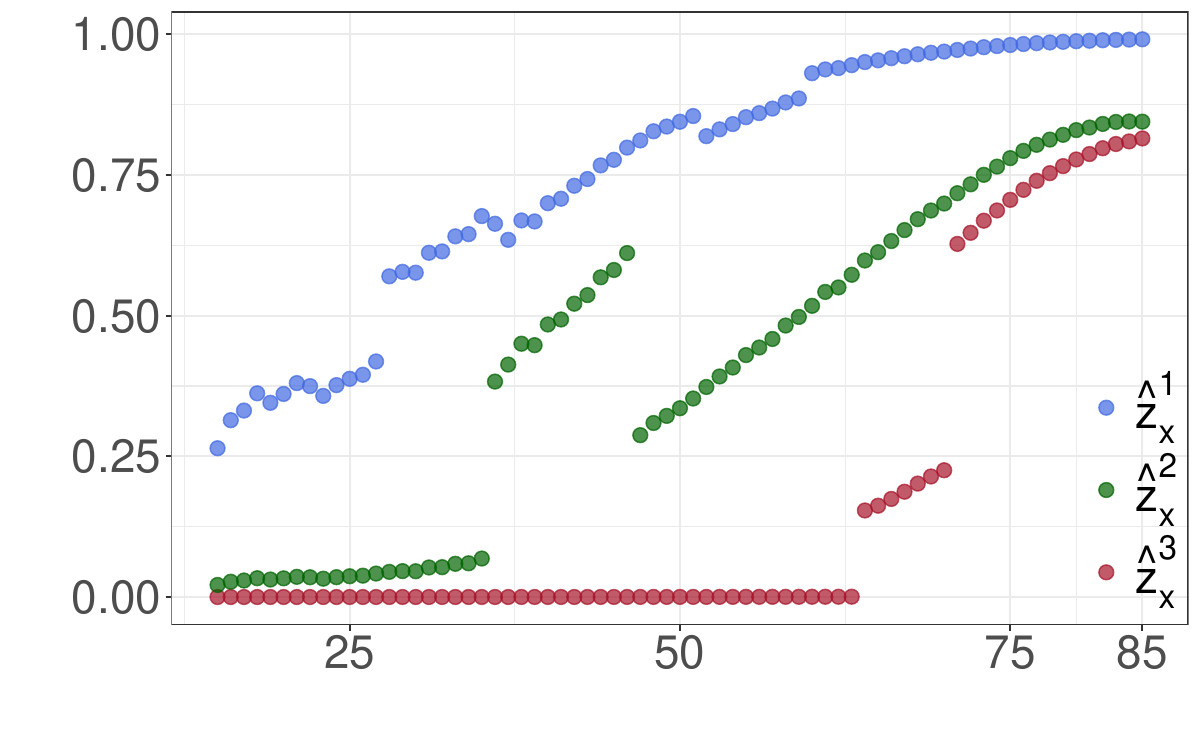}
    \caption{Credibility weights $\widehat z^i_x$ for $i=1,2,3$ by age (x-axis).}
    \label{fig:simzs}
  \end{subfigure}
  \hfill
  \begin{subfigure}[t]{0.45\linewidth}
    \centering
    \includegraphics[width=\linewidth]{ 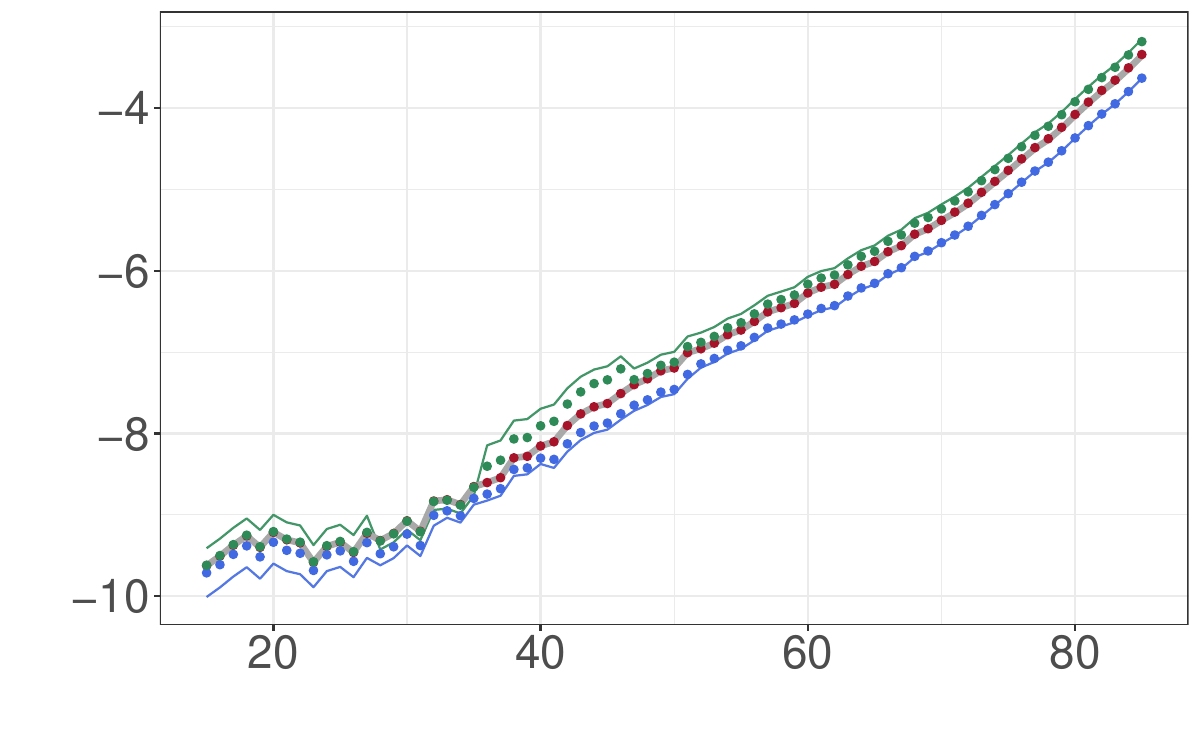}
    \caption{Logarithm of the predicted central mortality rate at calendar time (t' + 1) by age (x-axis). The credibility model is shown with dots, the global mortality model (Lee–Carter) is shown as a dark grey line, and the MLE model is shown with lines.}
    \label{fig:log-mortality-by-age}
  \end{subfigure}
  \caption{Credibility model results on a simulated data-set using the Lee-Carter as the predictor for the global mortality trend. In the sub-figures, sub-population 1 is shown in blue and was simulated with cohorts of 5000 individuals. Sub-population 2 is shown in green and was simulated with cohorts of 500 individuals. Sub-population 3 is shown in red and was simulated with cohorts of 94500 individuals. }
\end{figure}

This subsection illustrates how the estimated sub-population effects (\Cref{fig:allthetas}) and their estimated variances (\Cref{fig:variances}) determine the credibility weights, and how these weights in turn influence the one-year-ahead mortality predictions across ages. The corresponding age profile of the predicted mortality rates is shown on the logarithmic scale in \Cref{fig:log-mortality-by-age}.

In practice, the estimated $\widehat \theta_x^i$s and $\widehat \Var (\Theta_x^i)$s often tend to be noisy for small sub-populations, which suggests to use smoothing. When being familiar with data, this is usually done based on simple techniques and expert judgment. In the present paper we instead suggest to use a simple data driven binning strategy, where the estimated $\widehat \theta_x^i$s and $\widehat \Var (\Theta_x^i)$s are binned using an $L^2$ CART algorithm based on cross-validation. We believe that this data driven approach serves as a good benchmark technique, which does not assume prior knowledge of the data being used.

\Cref{fig:allthetas} shows the binned $\widehat \theta_x^i$s, together with the underlying age effects $\Theta_{x}^i$ used to generate the simulated data (thin solid line). As discussed above, and in Appendix~\ref{app: life portfolio}, for certain ages, the assumed Poisson approximation of the simulated binomial model closely agree.
An analogous plot for $\widehat \Var (\Theta_x^i)$ is given in \Cref{fig:variances}. From these figures it is clear that the signal is rather weak, since the binning strategy results in very few bins.

Continuing, recall that the values of $\widehat z_x^i$ depend on the joint effect of the exposure $E_{x, t}^i$, the  global population mortality rate $\mu_{x,t}$, and the estimated variances $\widehat \Var(\Theta^i_x)$, see \Cref{corollary:credibilityestimator}. When the variance grows, the values of $\widehat z_x^i$ also grow. Moreover, we have larger $\widehat z_x^i$ values when, given an age $x$ and a {\color{black} sub-population} $i$ (and fixed $\widehat \Var(\Theta^i_x )$), the total number of predicted deaths for age and group $\widehat D^i_x = \sum_{t \in \mathbb{T}} E^i_{x,t} \mu_{x,t}$ grows. 

In sub-population $1$, the estimated variance is between $5\%$ and $10\%$. Similarly, the estimated variance for sub-population $2$ fluctuates around the same levels, except for young age classes, where the variance is very small and for the age classes between $40$ and $60$, where the variance is close to $25\%$. The estimated variance in sub-population 3 is close to, but not identically, zero, which is consistent with the data simulation mechanism discussed in Appendix~\ref{app: life portfolio}. It is worth noting that the estimated values of $\widehat \theta_x^3$ are estimated to be very close to one, which suggests that the credibility model is not particularly relevant for this sub-population. 

Concerning sub-populations $1$ and $2$, as already noted, not having a sufficient amount of data tends to give credibility to the global mortality model. However, when age and variance grow, the credibility weights also grow.

Further, for sub-population $1$ there exists younger ages where the high estimated variances dominate the effect of the exposures, which makes the credibility weights increase to around $50\%$. 

In \Cref{fig:log-mortality-by-age}, the credibility model is shown by dots, the global mortality model (Lee--Carter) by a dark grey line, and the relative survival model by coloured lines. Sub-populations 1, 2, and 3 are shown in blue, green, and red, respectively. The behaviour of the credibility weights described above is reflected in the predicted central mortality rates. In particular, for sub-populations 1 and 2, the credibility predictor is closer to the global mortality model at younger ages and gradually moves towards the relative survival model at older ages. For sub-population 3, the estimated sub-population effect is close to one, and the credibility predictor therefore almost coincides with the global mortality model at all ages.

\subsubsection{Including more calendar years in the data}

\begin{figure}[ht]
  \centering

  \begin{subfigure}[t]{0.32\linewidth}
    \centering
    \includegraphics[width=\linewidth]{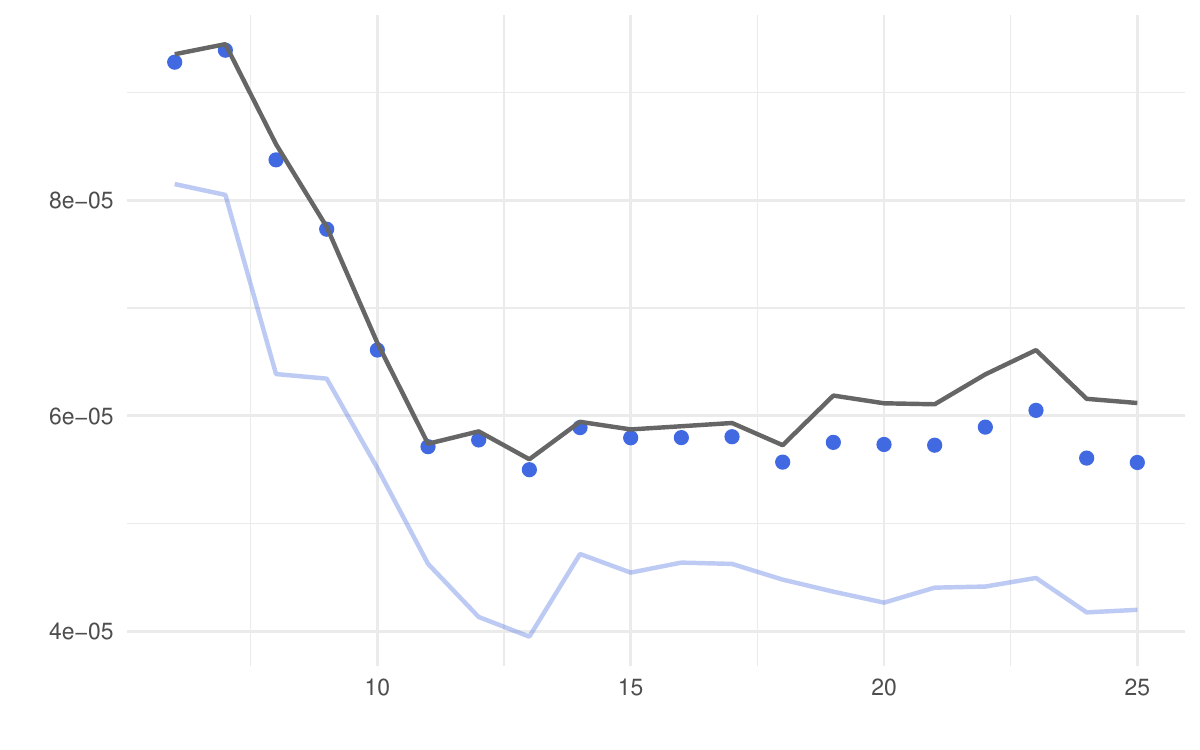}
    \caption{Sub-population 1, age 15.}
    \label{fig:an215}
  \end{subfigure}
  \hfill
  \begin{subfigure}[t]{0.32\linewidth}
    \centering
    \includegraphics[width=\linewidth]{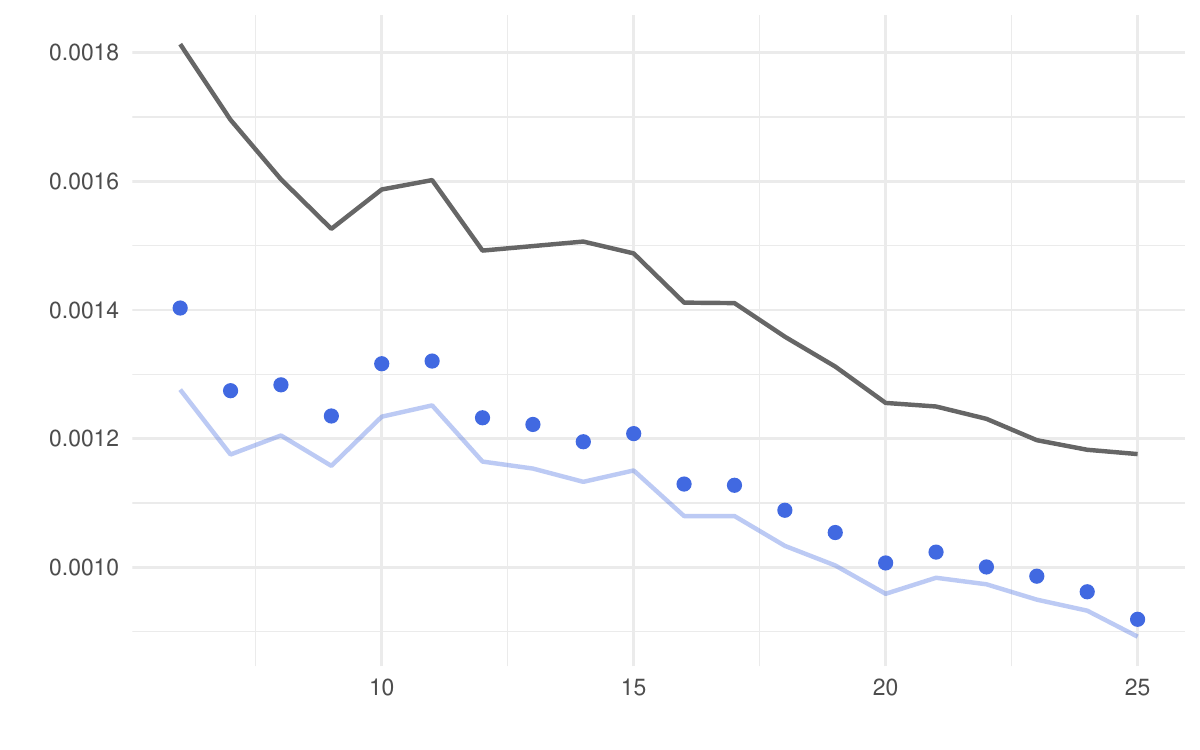}
    \caption{Sub-population 1, age 55.}
    \label{fig:an255}
  \end{subfigure}
  \hfill
  \begin{subfigure}[t]{0.32\linewidth}
    \centering
    \includegraphics[width=\linewidth]{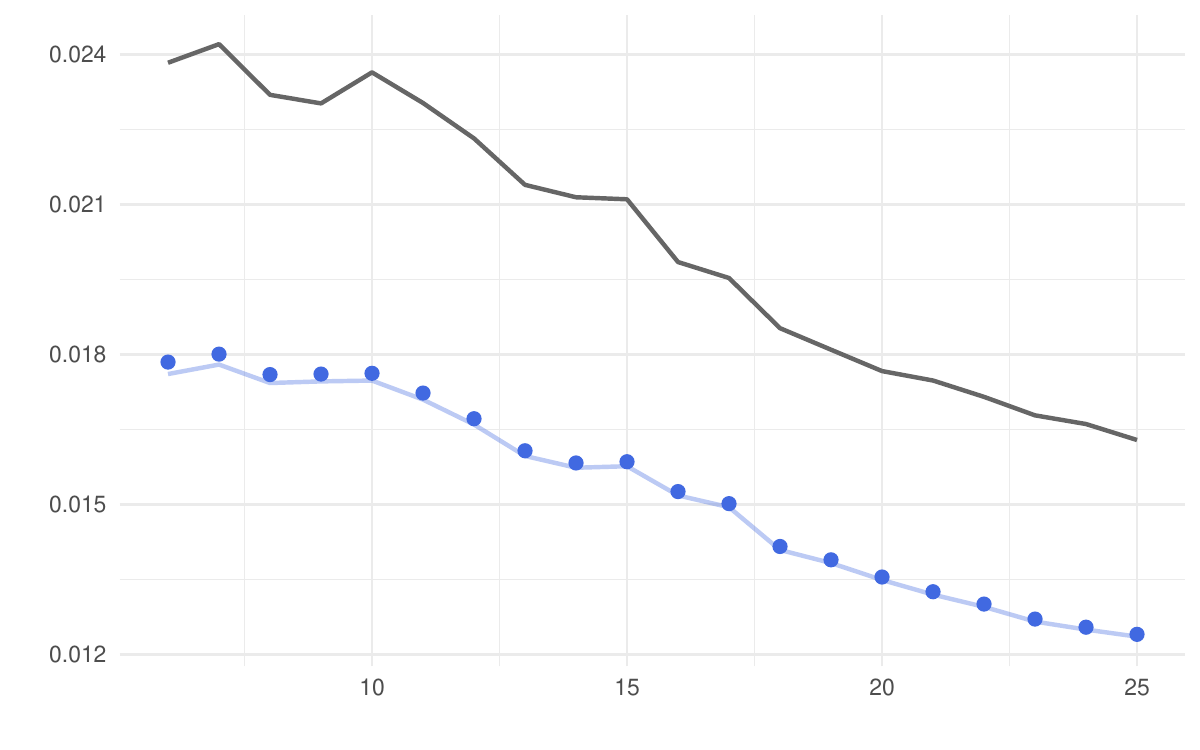}
    \caption{Sub-population 1, age 80.}
    \label{fig:an280}
  \end{subfigure}

  \vspace{1em}

  \begin{subfigure}[t]{0.32\linewidth}
    \centering
    \includegraphics[width=\linewidth]{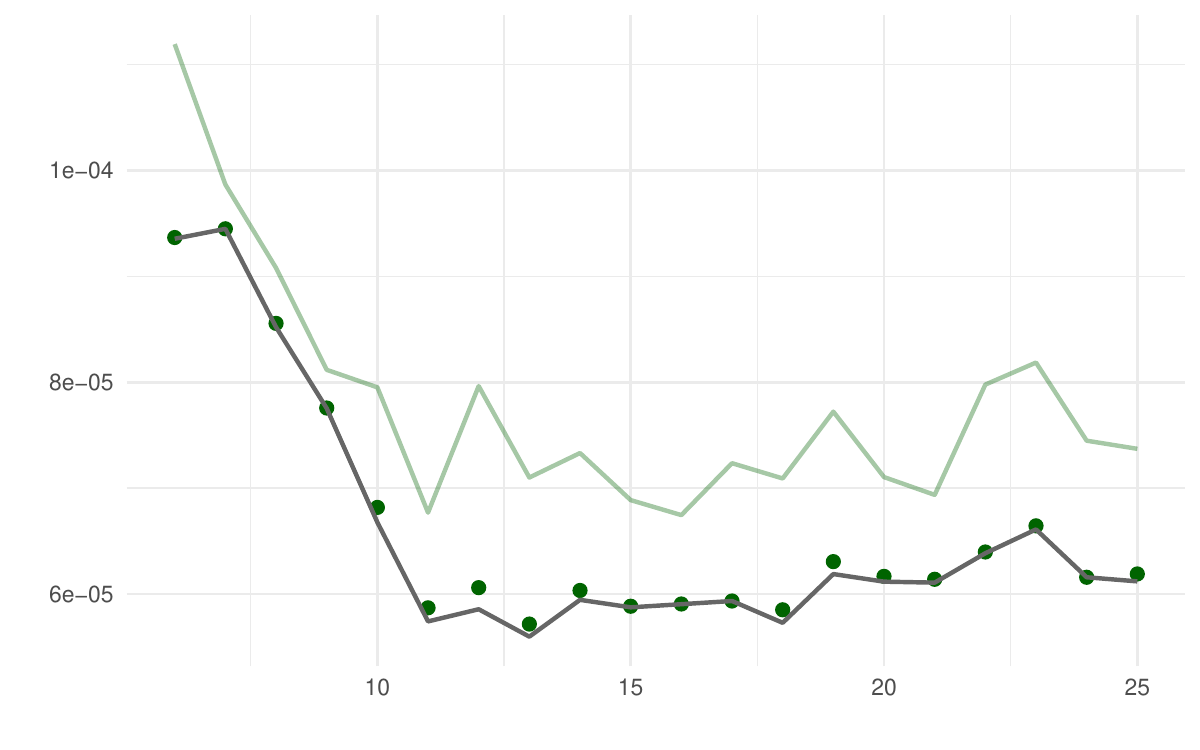}
    \caption{Sub-population 2, age 15.}
    \label{fig:an315}
  \end{subfigure}
  \hfill
  \begin{subfigure}[t]{0.32\linewidth}
    \centering
    \includegraphics[width=\linewidth]{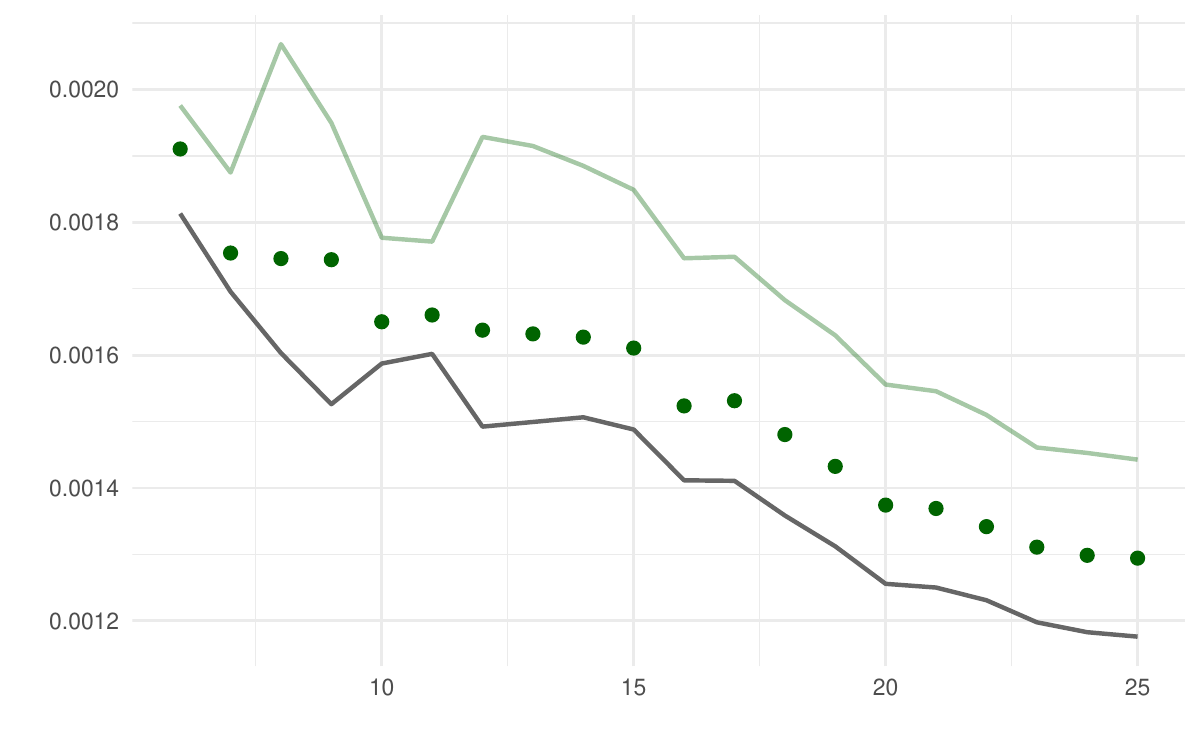}
    \caption{Sub-population 2, age 55.}
    \label{fig:an355}
  \end{subfigure}
  \hfill
  \begin{subfigure}[t]{0.32\linewidth}
    \centering
    \includegraphics[width=\linewidth]{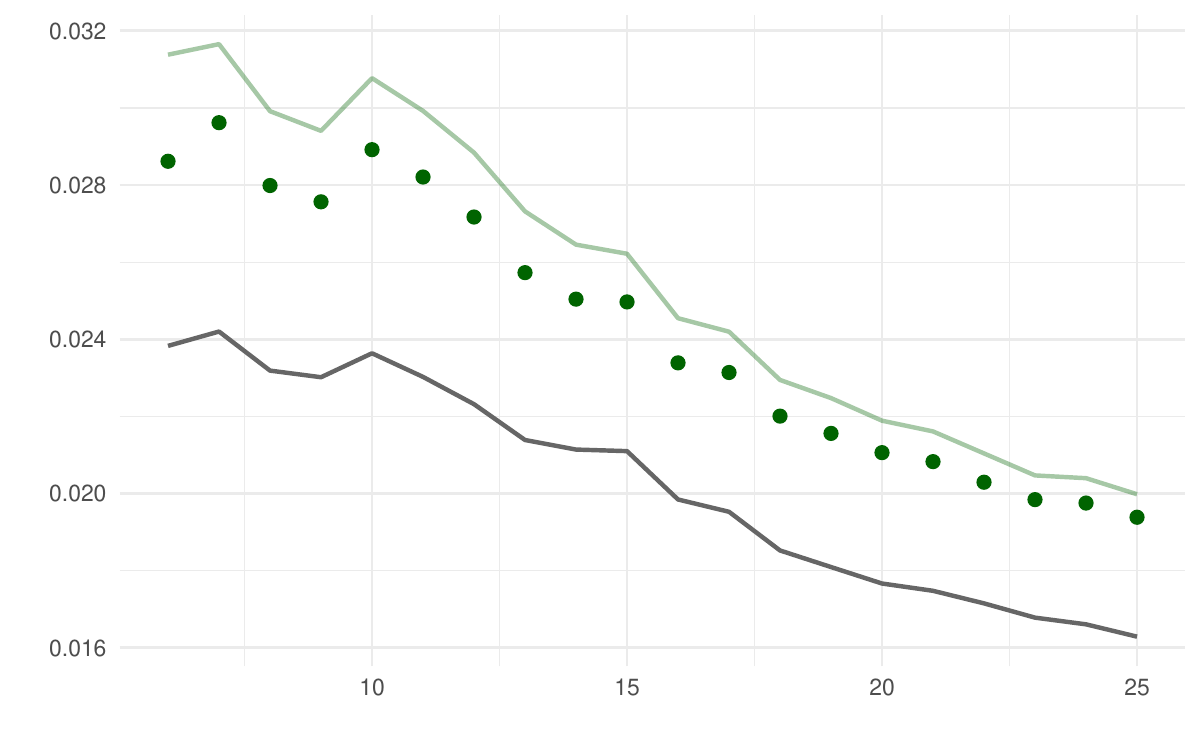}
    \caption{Sub-population 2, age 80.}
    \label{fig:an380}
  \end{subfigure}

  \vspace{1em}

  \begin{subfigure}[t]{0.32\linewidth}
    \centering
    \includegraphics[width=\linewidth]{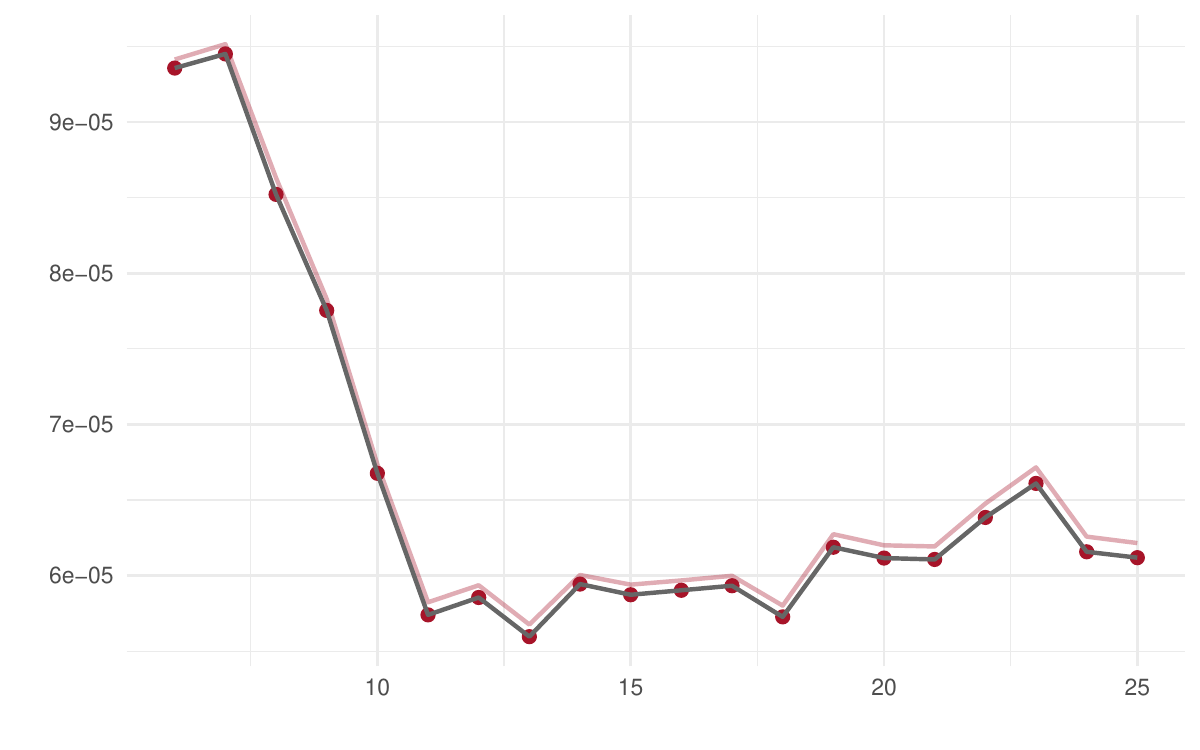}
    \caption{Sub-population 3, age 15.}
    \label{fig:an115}
  \end{subfigure}
  \hfill
  \begin{subfigure}[t]{0.32\linewidth}
    \centering
    \includegraphics[width=\linewidth]{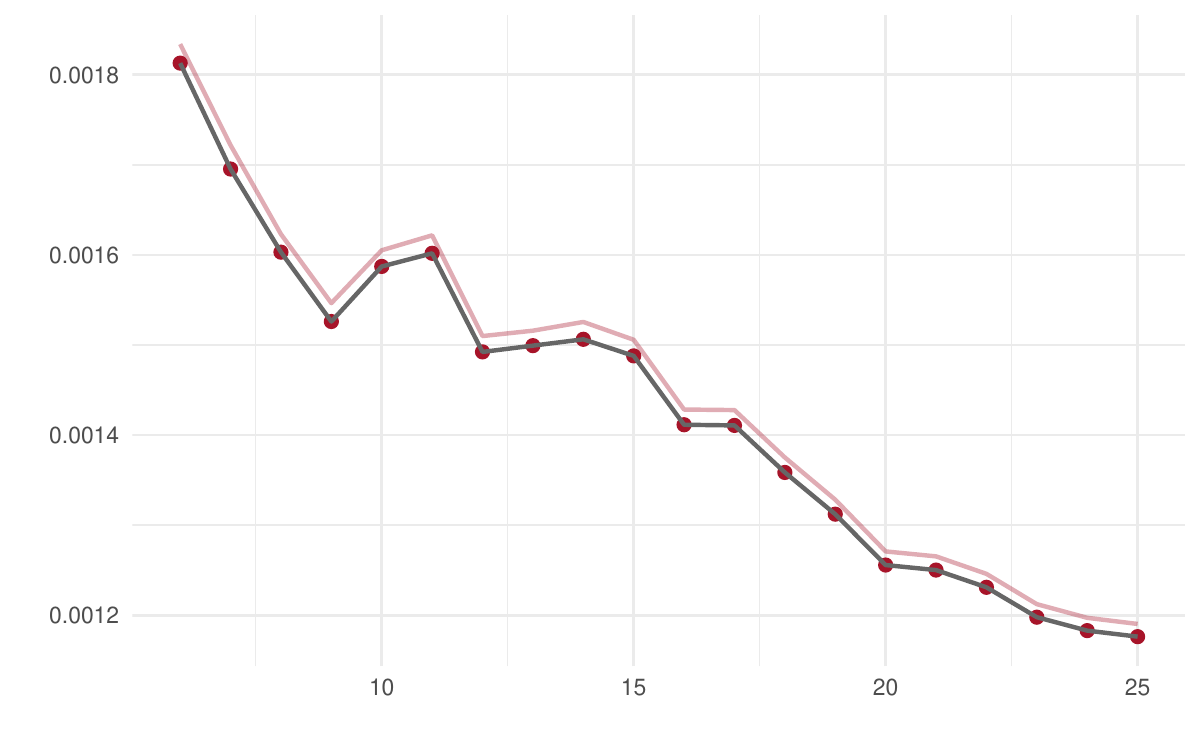}
    \caption{Sub-population 3, age 55.}
    \label{fig:an155}
  \end{subfigure}
  \hfill
  \begin{subfigure}[t]{0.32\linewidth}
    \centering
    \includegraphics[width=\linewidth]{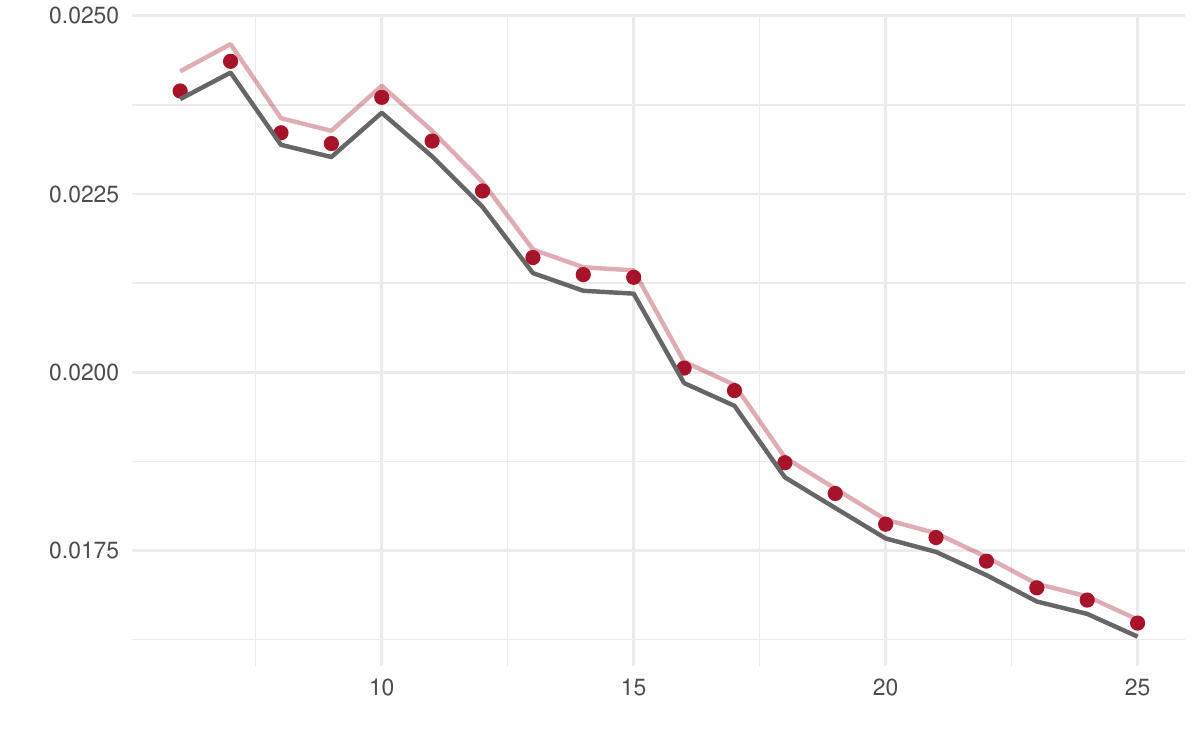}
    \caption{Sub-population 3, age 80.}
    \label{fig:an180}
  \end{subfigure}

\caption{Starting from an in-sample data set with 5 calendar periods and ages ranging from 15 to 85, we produce one-year-ahead mortality forecasts and then apply a rolling-window procedure in which one calendar year is added at each step before re-estimating the models and predicting the next one-year-ahead central mortality rate. We stop the illustration once the in-sample period reaches 25 years. The first row corresponds to Sub-population 1, the second row to Sub-population 2, and the third row to Sub-population 3. The columns display three representative ages, namely 15, 55, and 80. In each panel, the points represent the credibility estimate $\widehat{\widehat{\mu_{x, t+1} \left(\Theta^i_x\right)}}$, the transparent coloured line represents the relative survival model $\widehat{\mu}^i_{x, t' + h}$, and the solid grey line represents the global mortality trend $\overline \mu_{x, t' + h}$, all plotted against the number of sample years (x-axis).}
  \label{fig:sample_size_effect_grid}
\end{figure}

In this section, we study how the credibility predictor changes as more calendar years are added to the in-sample data. We begin with an in-sample data set consisting of 5 calendar periods and ages from 15 to 85, and then apply a rolling-window procedure in which one calendar year is added at each step, see Figure~\ref{fig:sample_size_effect_grid}. After each update, the models are re-estimated and used to produce a one-year-ahead forecast of the central mortality rate. The results are shown for the three sub-populations and for three representative ages, namely 15, 55, and 80. We stop the illustration once the number of in-sample periods reaches 25 years. In each panel, the points show the credibility estimate $\widehat{\widehat{\mu_{x,t+1}(\Theta_x^i)}}$, the transparent coloured line shows the relative survival model $\widehat{\mu}^i_{x,t'+h}$, and the solid grey line shows the global mortality trend $\overline{\mu}_{x,t'+h}$, all plotted against the number of sample years. The figure shows four main features. First, as more data are added, the credibility predictor moves smoothly towards the relative survival model. Second, the overall downward trend reflects the biological decline in mortality over calendar time. Third, the credibility predictor is generally closer to the global mortality model at younger ages and closer to the relative survival model at older ages. This age pattern is consistent with the expression for $\widehat z_x^i$ in \Cref{corollary:credibilityestimator}: as $\sum_{v \in \mathbb{T}} E^i_{x,v}\mu_{x,v}$ increases, the credibility weight $\widehat z_x^i$ increases, so that more weight is assigned to the relative survival model. Since the expected number of deaths is typically higher at older ages, the credibility predictor is pulled more strongly towards $\widehat{\mu}^i_{x,t'+h}$ in that part of the age range. Fourth, comparing sub-populations 1 and 2, the smaller sub-population places more weight on the global mortality model, whereas the larger sub-population places more weight on the relative survival model. As expected, the trajectories for sub-population 1 are relatively flat, and as commented on above, for sub-population 3 the difference between the global model and the sub-population is very small.

\subsubsection{Expected quadratic forecast error}

\begin{figure}[ht]
  \centering
  \begin{subfigure}[t]{0.3\textwidth}
    \centering
    \includegraphics[width=\linewidth]{ 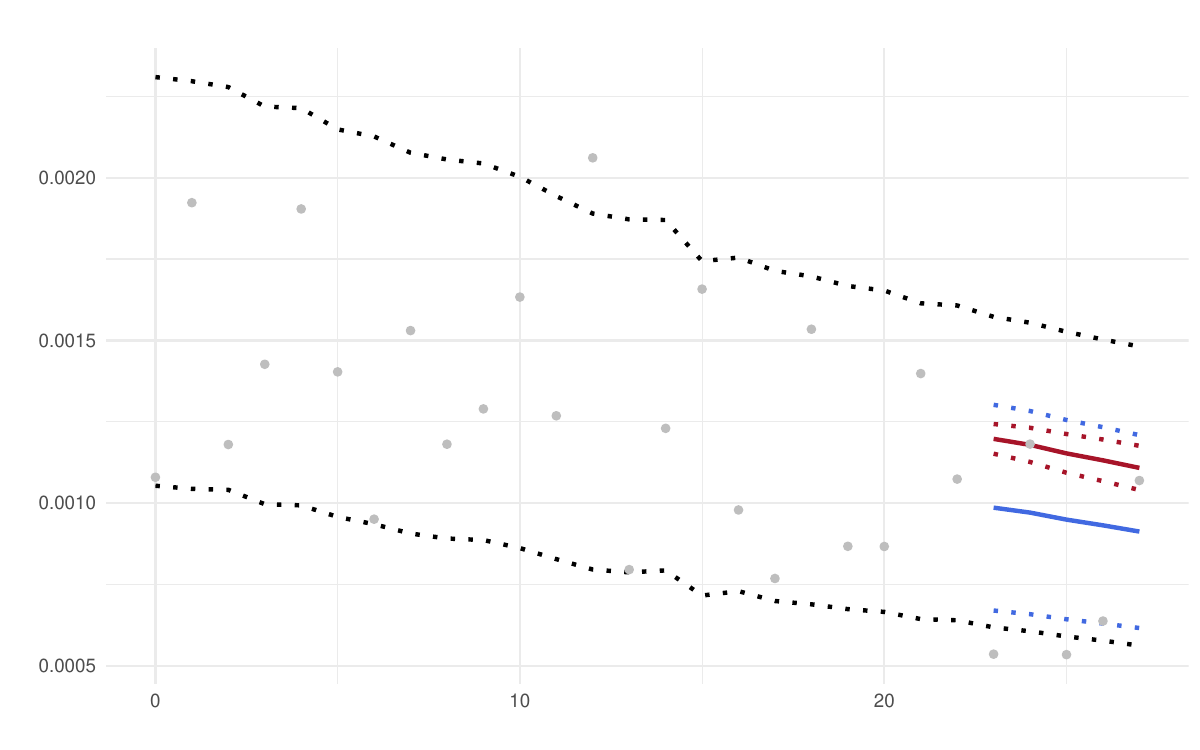}
    \caption{Sub-population 1.}
    \label{sf:group1_msep}
  \end{subfigure}
  \hfill
  \begin{subfigure}[t]{0.3\textwidth}
    \centering
    \includegraphics[width=\linewidth]{ 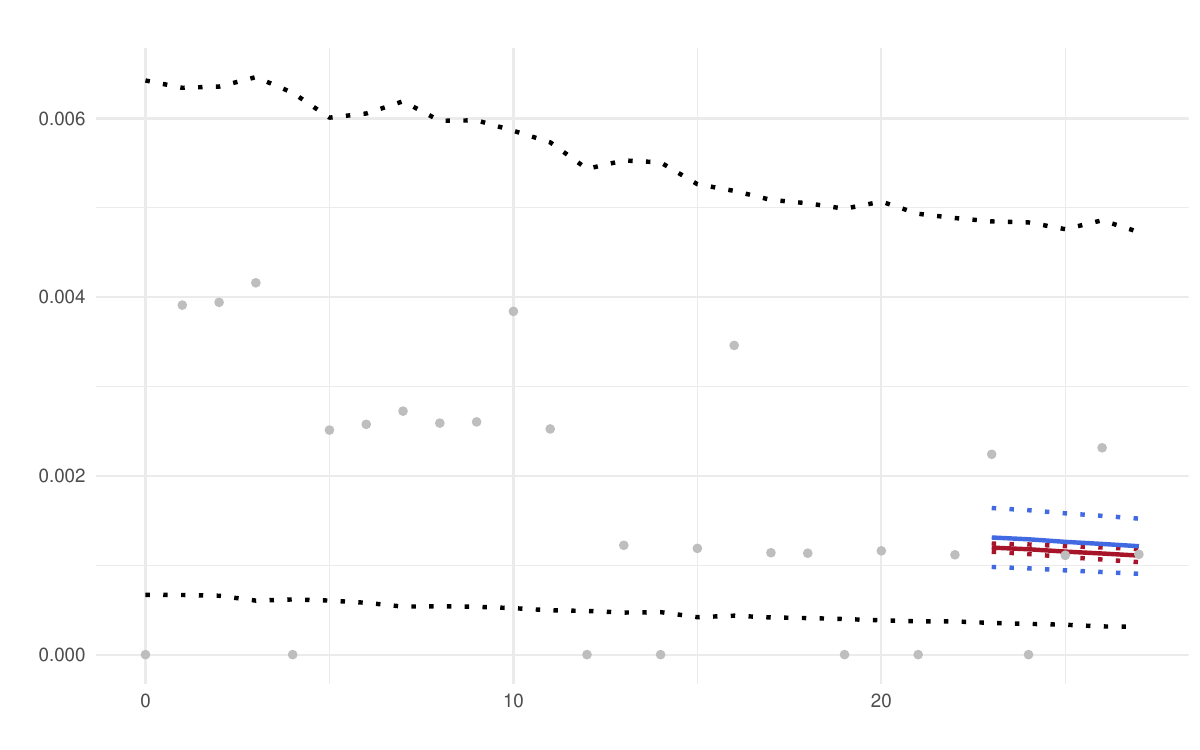}
    \caption{Sub-population 2.}
    \label{sf:group2_msep}
  \end{subfigure}
  \hfill
  \begin{subfigure}[t]{0.3\textwidth}
    \centering
    \includegraphics[width=\linewidth]{ 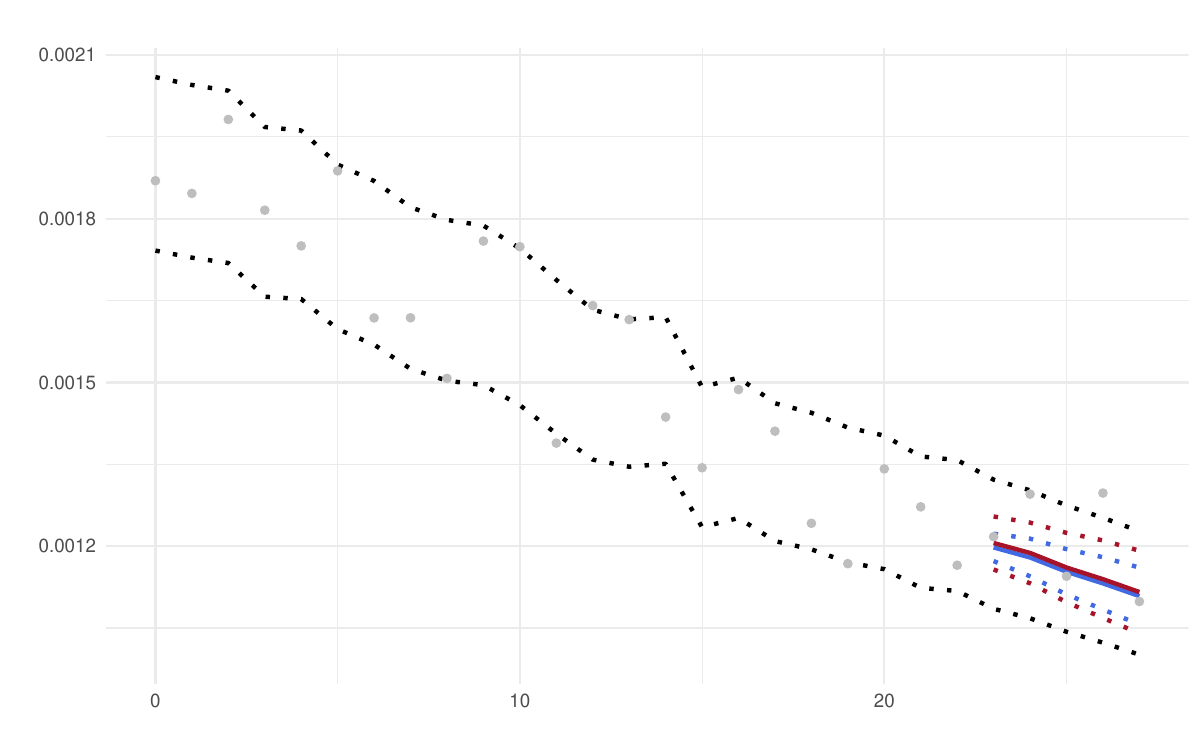}
    \caption{Sub-population 3.}
    \label{sf:group0_msep}
  \end{subfigure}
  \caption{\label{fig:mortality fan plots} The gray dots represent observed central mortality rates for the synthetic data for age $x = 55$. Solid lines correspond to  mortality rate predictions, dotted lines correspond to the root of the expected quadratic forecast error for mortality rates, where model A.~is in blue and model C.~is in red. 
  The black dashed lines corresponds to simulated standard error bounds for the credibility based Poisson model from \eqref{eq: std dev bounds Pois} using model A. }
\end{figure}

\Cref{fig:mortality fan plots} shows a numerical study on the expected quadratic forecast error for age $x=55$ for a forecasting horizon of $5$ years ($t \in \left\{t^\prime+1, \ldots, t^\prime+5\right\}$) for the different sub-populations. The application consists of a comparison of the prediction intervals calculated using the expected quadratic forecast error for the credibility approach as computed in \Cref{thm: MSEP} and the expected quadratic forecast error computed for the separate mortality models, for sub-populations 1 (\Cref{sf:group1_msep}) and 2 (\Cref{sf:group2_msep}) and on the sub-population 3, i.e.~the super-population excluding sub-populations 1 and 2 (\Cref{sf:group0_msep}). For the expected quadratic forecast error comparison, we focus on these two approaches only. This restriction is made partly to keep the graphical presentation clear, and partly because the separate sub-population model provides a benchmark with an established implementation. Moreover, this benchmark is particularly relevant in the present setting, since it constitutes the most direct alternative to the proposed credibility approach by fitting the mortality model separately to each sub-population.

Using a Lee-Carter model for the central mortality rates we compare the predictions of the credibility model~(blue solid line) with
the benchmark model~(red solid line) over a five-year prediction horizon in \Cref{fig:mortality fan plots}. Since we experienced issues in convergence for the separate mortality model for sub-population $2$, we use the predictor from global mortality model. The blue dotted lines represent the expected quadratic forecast error intervals for the credibility model,  computed as  $\widehat{\widehat{\mu_{x,t^\prime+h} \left(\Theta^i_x\right)}} \pm \sqrt{ \widehat{\operatorname{Q}}(\mu_{x,t^\prime+h} \left(\Theta^i_x\right), \widehat{ \mu_{x,t^\prime+h} \left(\Theta^i_x\right)}\mid \calF_{t'}) }$. The red dotted lines represent the expected quadratic forecast error intervals for the separate mortality model. The expected quadratic forecast error for the separate mortality models are computed on the sub-population data using the residual bootstrap procedure described in \citet{koissi06} and implemented in the \texttt{R} package \texttt{StMoMo}. 

Figure~\ref{fig:mortality fan plots} indicates that for age $55$ we have smaller expected quadratic forecast error bounds for sub-populations with larger exposure, compare e.g.~the bounds for sub-population $3$ in \Cref{sf:group0_msep} with those for sub-population $2$ in \Cref{sf:group2_msep}. The proposed credibility approach, has smaller or comparable expected quadratic forecast error intervals compared to the benchmark, for groups $1$ and $3$. The benchmark model shows smaller expected quadratic forecast error intervals for group $2$.

Further, the above discussion of expected quadratic forecast error concerns the predicted mortality rates in isolation. In order to capture the full variability in the raw mortality rates $F_{x, t}^i$, we also need to take the Poisson variation into account. Let us consider a Poisson model given by

\begin{align}\label{eq: Poisson sim}
    D^i_{x,t} \mid E^i_{x,t}, \tilde{\mu}_{x, t} \sim \text{Poisson}(E^i_{x,t} \tilde{\mu}_{x, t}).
\end{align}

By applying the functional delta method, see e.g.~\citep[Ch.~2]{andersen93} on the log-rate scale to the Poisson model in \Cref{eq: Poisson sim}, for some $t>0$ we obtain the log-transformed $\pm$ standard deviation interval

\begin{align}\label{eq: std dev bounds Pois}
  C_{x,t}^i
:=
\left(
\tilde{\mu}_{x,t}
\exp\!\left(
-\frac{1}{\sqrt{E_{x,t}^i\,\tilde{\mu}_{x,t}}}
\right),
\,
\tilde{\mu}_{x,t}
\exp\!\left(
\frac{1}{\sqrt{E_{x,t}^i\,\tilde{\mu}_{x,t}}}
\right)
\right).
\end{align}

The black-dotted lines represent the bounds from \eqref{eq: std dev bounds Pois} from the Poisson model in \Cref{eq: Poisson sim}, evaluated at $\tilde{\mu}_{x,t}=\widehat{\widehat{\mu_{x,t^\prime+h}\left(\Theta_x^i\right)}}$, with $t=0,\ldots, t^\prime+5$.
 
The gray dots represent the observed (``raw'') mortality rates obtained from the synthetic simulated data. By comparing the black dotted lines and the gray dots indicates that the observed rates seem to fall within the bounds obtained using $C_{x,t}^i$. In other words, the observed variation in the $F_{x, t}^i$s can be motivated by the suggested Poisson credibility model.

\subsection{Assessment of out-of-sample performance across repeated simulations}
\label{ss:assessment-oos}

In this section, we study the out-of-sample performance of the proposed credibility approach under the simulation setting described in Section~\ref{sec: simulating sub-population data}. More specifically, in light of the model selection step discussed in \Cref{ss:bic-comparison-models} we compare the predictive performance of the following four approaches to forecasting sub-population mortality:

\begin{enumerate}
    \item[A.] The proposed credibility approach from Proposition~\ref{proposition:credibilityestimator}, using the LC model for the global mortality trend.
    \item[B.] The relative survival model from equation \eqref{eq: relative survival} in combination with a LC predictor for the global mortality trend.
    \item[C.] A separate LC mortality model for each sub-population, with parameters estimated on the respective sub-population-specific data subsets.
    \item[D.] The LC mortality model fitted to the super-population 0, that is, a single global mortality model with parameters estimated from the aggregated super-population data.
\end{enumerate}

The comparison is carried using the proper scoring rules corresponding to Poisson deviance losses and the MSE, see e.g.~\cite{czado09}, using a rolling-window evaluation scheme. The results are averaged across 30 repetitions of the simulated data.

\subsubsection{Evaluation metrics}
\label{sec:evaluation-metrics}

\begin{figure}[htbp]
     \centering
     % First image
     \begin{subfigure}[b]{0.45\textwidth}
         \includegraphics[width=\textwidth]{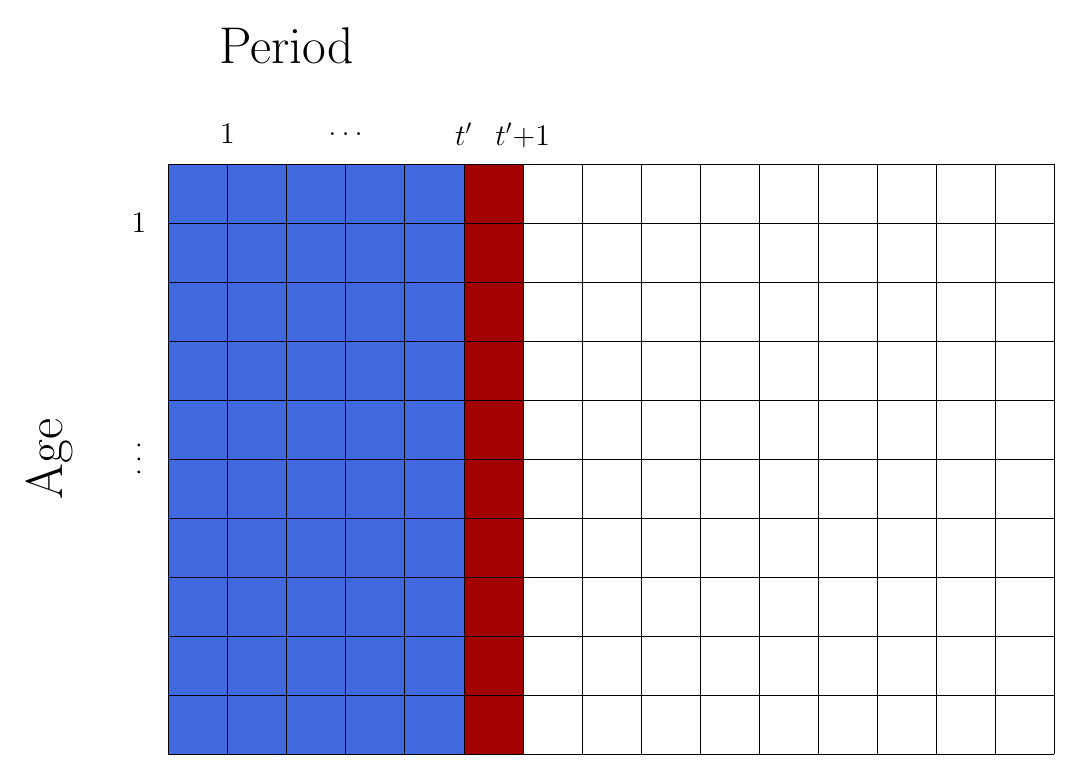}
         \caption{\label{fig:vs1} First step, 1 year ahead predictions.}
     \end{subfigure}
     \hfill
   % Second image
     \begin{subfigure}[b]{0.45\textwidth}
         \includegraphics[width=\textwidth]{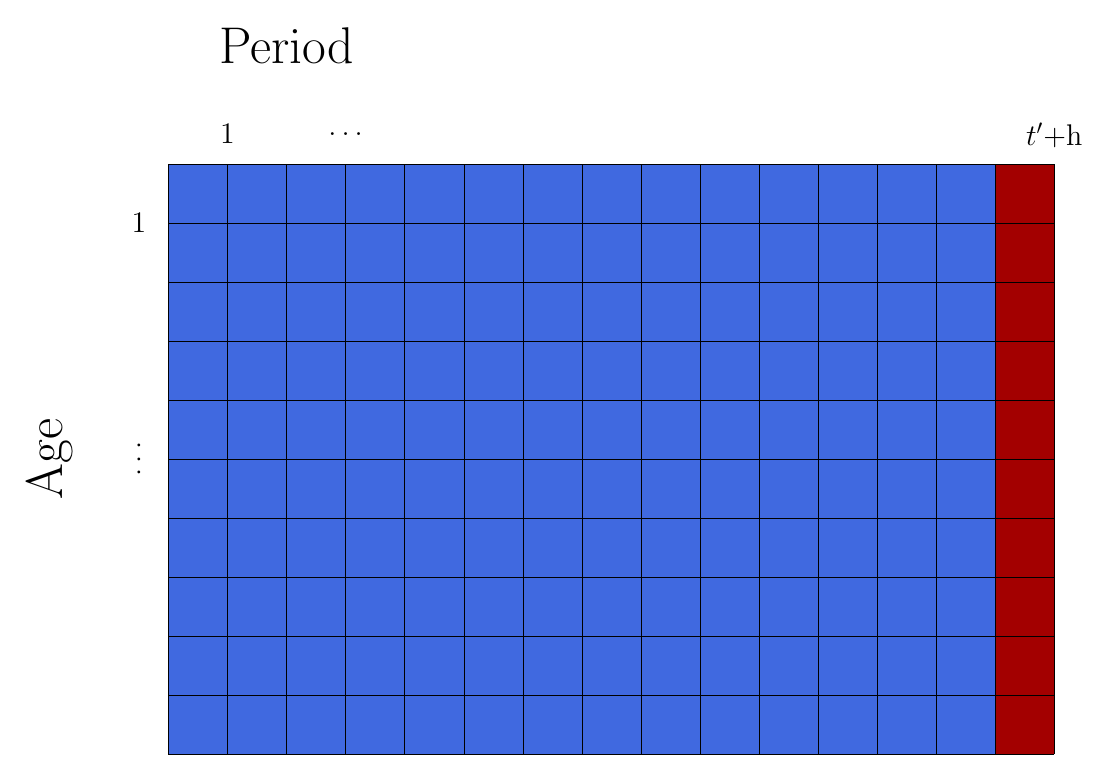}
         \caption{\label{fig:vs2} Last step, 1 year ahead predictions.}
     \end{subfigure}

     \caption{\label{fig:validation} Test (red) split of the out-of-sample data in the rolling window approach. The in-sample data are in blue. The data that are omitted from the computation are in white. In the example, we want to predict $1$ calendar period ahead. }
\end{figure}

The evaluation of models A. -- D., is based on the out-of-sample performance w.r.t.~mean Poisson deviance and mean squared error (MSE).  Although it would have been possible to consider a fixed time horizon, the time series structure of the data motivates using a rolling-window approach for model evaluation. The rolling window approach is schematically described in \Cref{fig:validation} for one sub-population representing data in an age-period tabulation. 
Row one depicts the rolling window approach for a forecasting horizon of one year. Starting from calendar period $t^\prime$, we forecast one period ahead (red column) and compute the performance metrics, and add period $t' + 1$ to the in-sample data to be used for predicting $t' + 2$. This procedure continues until we have forecasted the period $t' + h$, for a pre-specified window size $h$.

For $\ell \in \{A,B,C,D\}$, let $\widehat m^{i,\ell}_{x,t}$ denote the out-of-sample forecast of the central mortality rate for age $x$, period $t$, and sub-population $i$ obtained from method $\ell$. The mean squared error (MSE) over a subset of ages $\mathbb{X}' \subseteq \mathbb{X}$, group $i \in \mathbb{I}$, and a rolling window of size $h$ is defined as
\[
\mathrm{MSE}^{i,\ell}(\mathbb{X}',t^\prime,h)
=
\frac{1}{|\mathbb{X}'|}\frac{1}{h}
\sum_{x\in\mathbb{X}'}
\sum_{j=1}^{h}
\left(\widehat m^{i,\ell}_{x,t^\prime+j}-F^i_{x,t^\prime+j}\right)^2,
\]
where $F^i_{x,t}$ denotes the observed central mortality rate for age $x$, period $t$, and sub-population $i$.

The average out-of-sample Poisson deviance is defined analogously by
\[
\overline{\mathrm{dev}}^{i,\ell}(\mathbb{X}',t^\prime,h)
=
\frac{1}{|\mathbb{X}'|}\frac{1}{h}
\sum_{x\in\mathbb{X}'}
\sum_{j=1}^{h}
\mathrm{dev}^{i,\ell}_{x,t^\prime+j},
\]
where
\[
\mathrm{dev}^{i,\ell}_{x,t}
=
2E^i_{x,t}
\left(
\widehat m^{i,\ell}_{x,t}
-
F^i_{x,t}
+
F^i_{x,t}
\log\left(\frac{F^i_{x,t}}{\widehat m^{i,\ell}_{x,t}}\right)
\right),
\]
for $F^i_{x,t}>0$. When $F^i_{x,t}=0$, this reduces to
\[
\mathrm{dev}^{i,\ell}_{x,t}=2E^i_{x,t}\widehat m^{i,\ell}_{x,t}.
\]

\begin{remark}
\label{sub-population models}
As discussed above, for all approaches A.~-- D.~the mortality models were fitted using the \texttt{StMoMo} package in \texttt{R}, which relies on the \texttt{gnm} package for parameter estimation \citep{gnmpackage}. When applying model C.~to the simulated sub-population data, we encountered numerical issues mostly for the Renshaw-Haberman model, and particularly in sub-population $2$, where the number of observations is small. These issues arose either from lack of convergence or from numerical errors returned by the algorithm. When this occured, the corresponding super-population model (approach D.) was used as a proxy for model C., replacing the sub-population-specific predictions. Since the selected mortality model based on the BIC was the Lee-Carter, this proxy will not really impact on the results that we will show in the application Section of the paper. 
\end{remark}

\subsubsection{Results}
\label{ss:results-oos}

\begin{figure}[htp]
\centering
\captionsetup[subfigure]{font=small}

\begin{adjustbox}{max totalsize={\textwidth}{0.8\textheight},center}
\begin{minipage}{\textwidth}
\centering

\begin{subfigure}[b]{0.48\textwidth}
  \includegraphics[width=\linewidth]{ 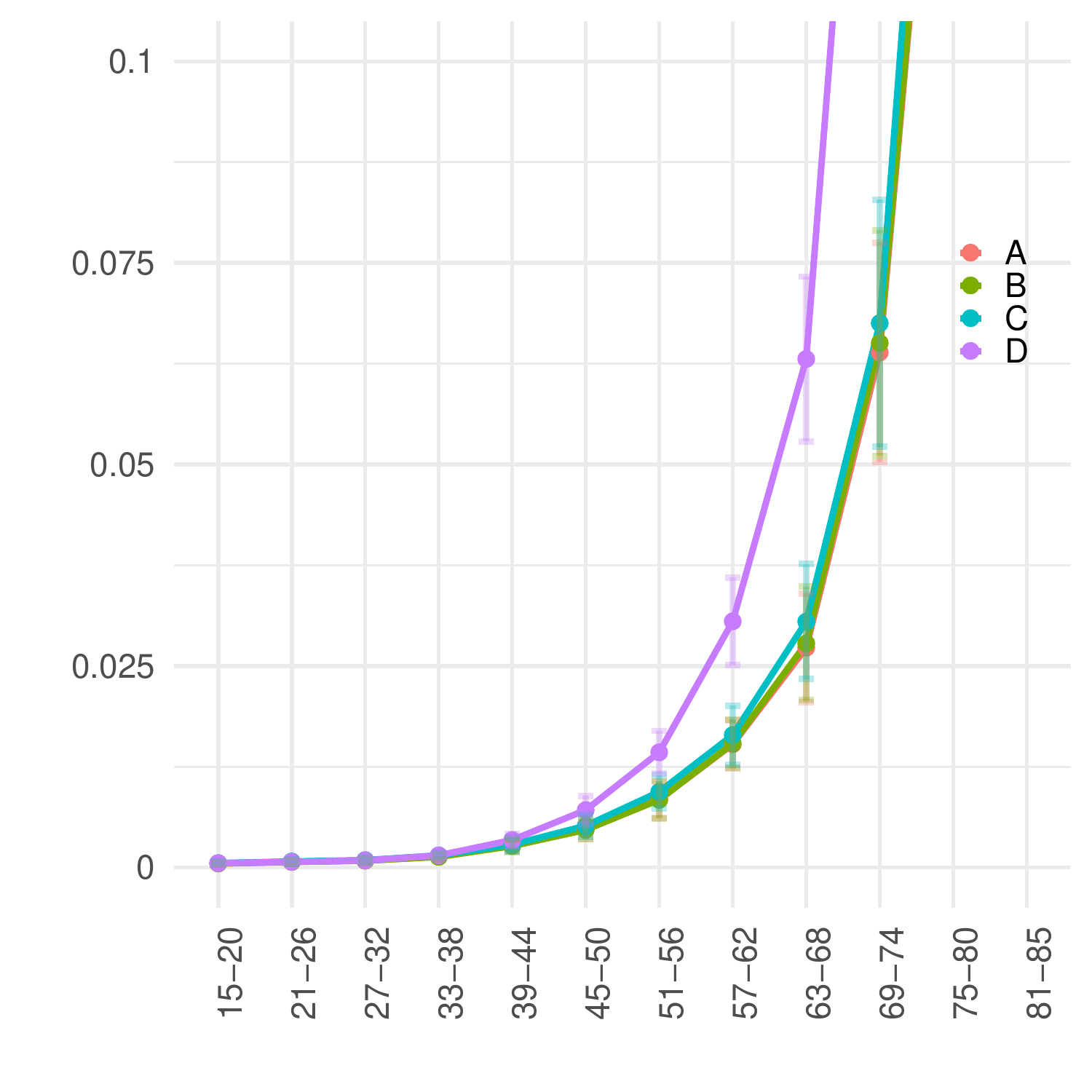}
  \caption{\label{subfig:subpop1-mse} Sub-population 1, MSE}
\end{subfigure}\hfill
\begin{subfigure}[b]{0.48\textwidth}
  \includegraphics[width=\linewidth]{ 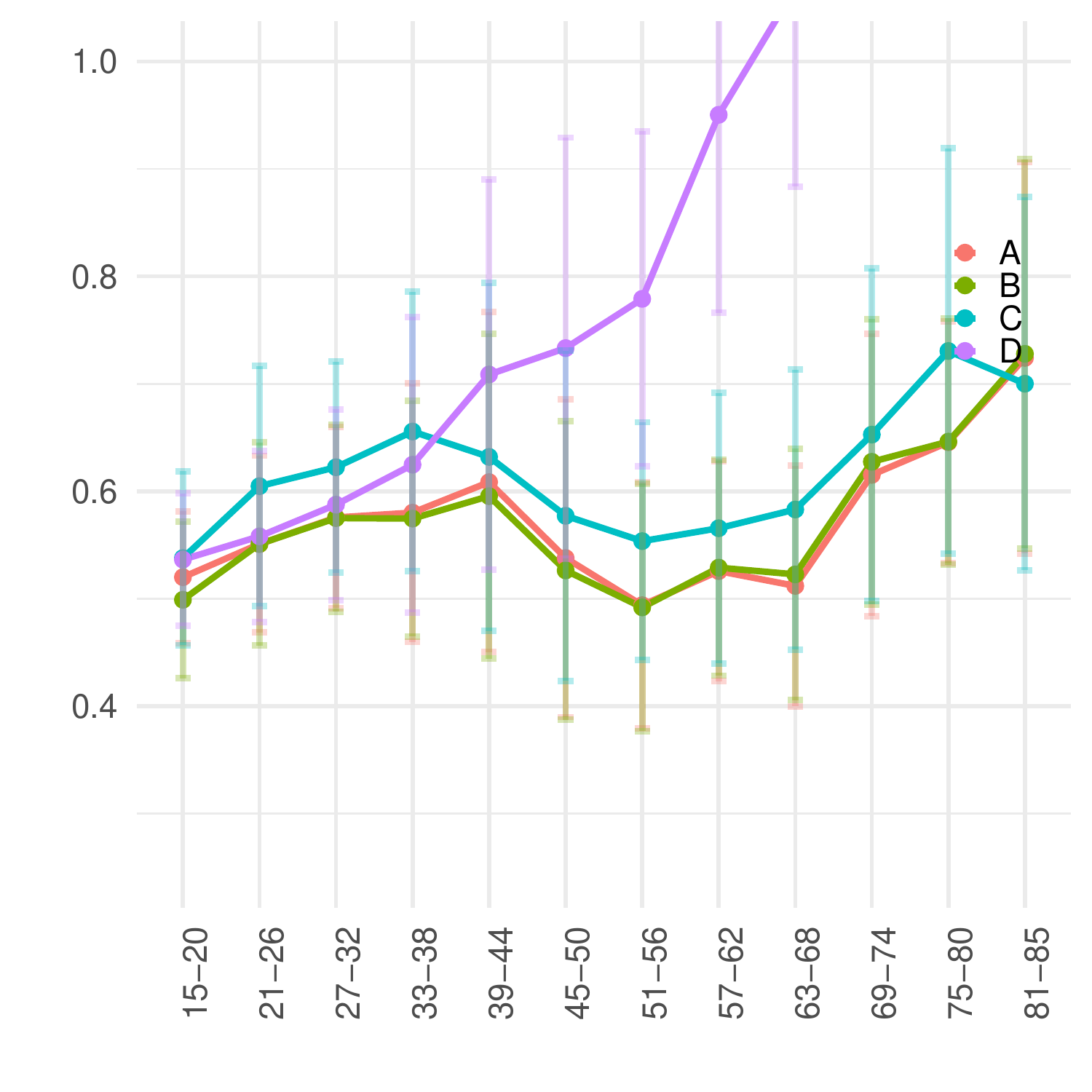}
  \caption{\label{subfig:subpop1-dev} Sub-population 1, Deviance}
\end{subfigure}

\vspace{0.5em}

\begin{subfigure}[b]{0.48\textwidth}
  \includegraphics[width=\linewidth]{ 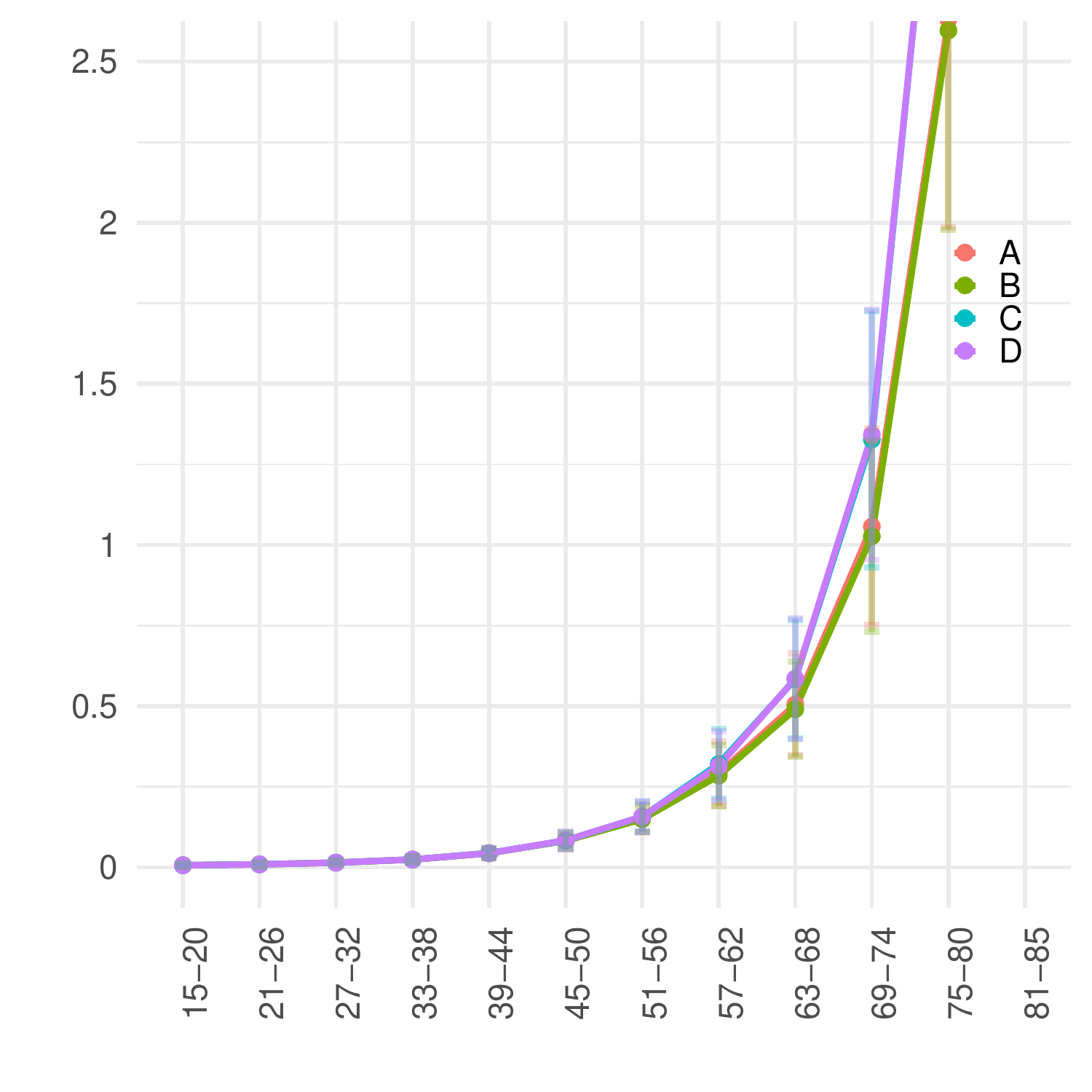}
  \caption{\label{subfig:subpop2-mse} Sub-population 2, MSE}
\end{subfigure}\hfill
\begin{subfigure}[b]{0.48\textwidth}
  \includegraphics[width=\linewidth]{ 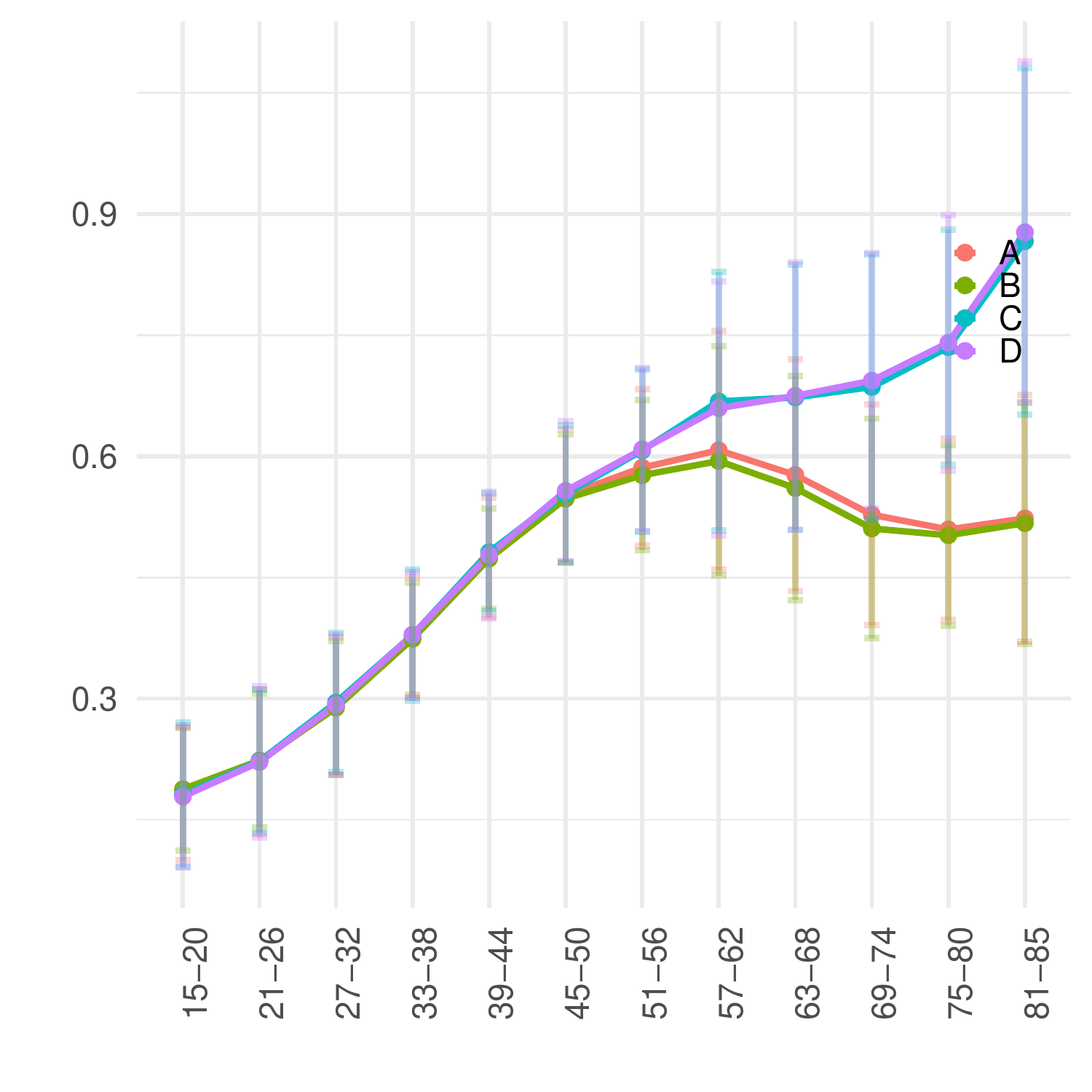}
  \caption{\label{subfig:subpop2-dev} Sub-population 2, Deviance}
\end{subfigure}

\vspace{0.5em}

\begin{subfigure}[b]{0.48\textwidth}
  \includegraphics[width=\linewidth]{ 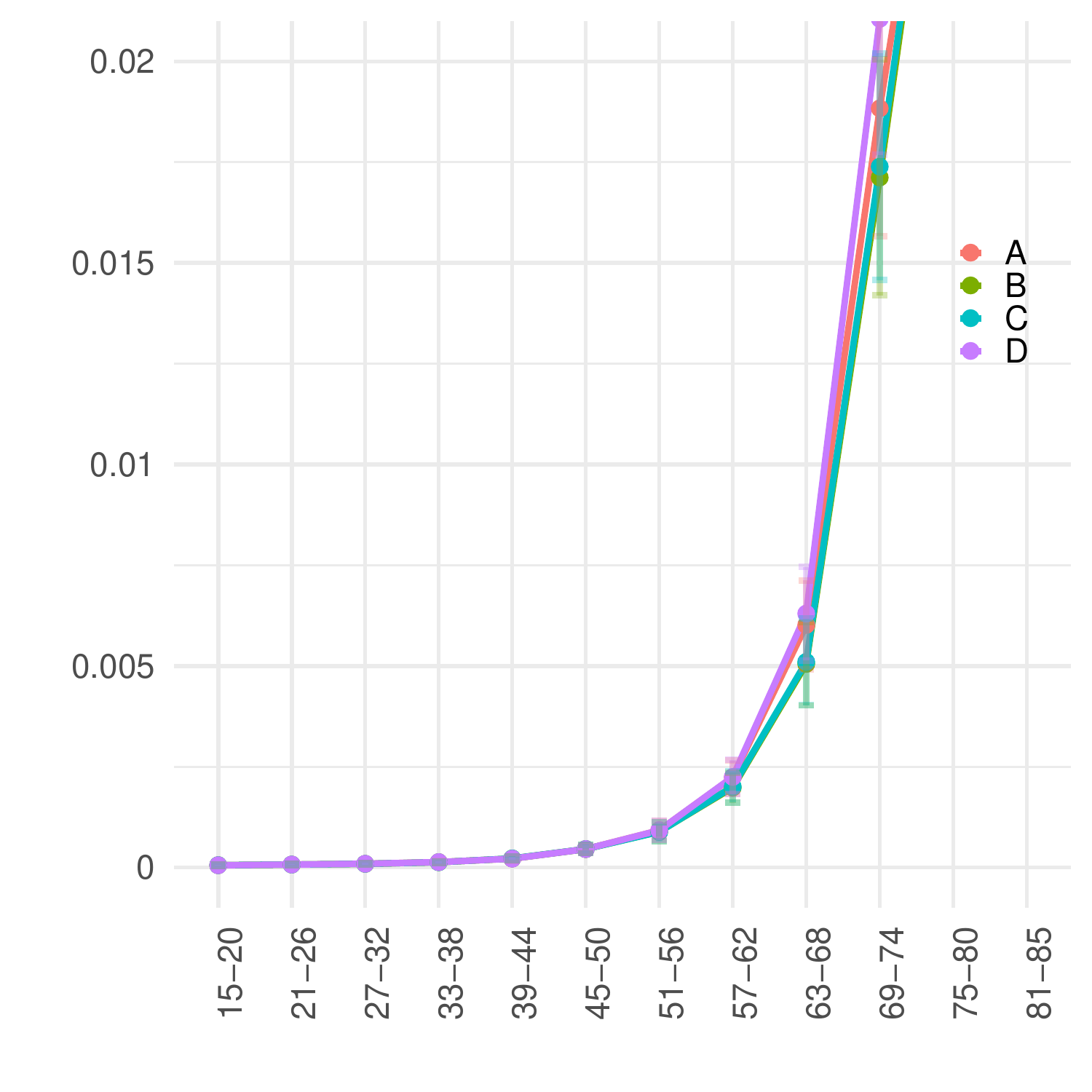}
  \caption{\label{subfig:subpop3-mse} Sub-population 3, MSE}
\end{subfigure}\hfill
\begin{subfigure}[b]{0.48\textwidth}
  \includegraphics[width=\linewidth]{ 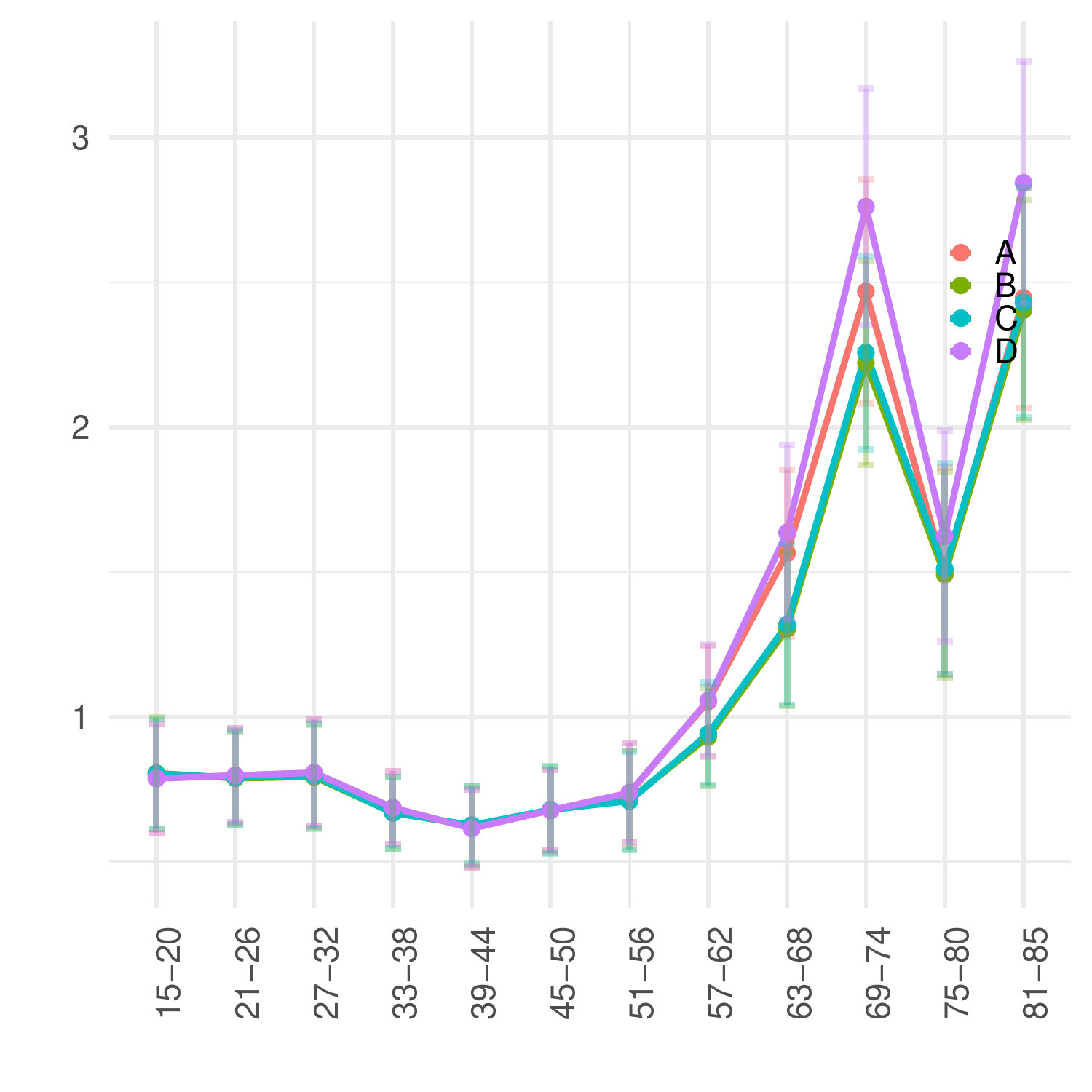}
  \caption{\label{subfig:subpop3-dev} Sub-population 3, Deviance}
\end{subfigure}

\end{minipage}
\end{adjustbox}
    \caption{\label{fig:oos-results} Out-of-sample mean squared error (left-hand side column) and out-of-sample average Poisson deviance (right-hand side column) under a rolling-window evaluation scheme using the Lee--Carter model as the global mortality model. The panels report results for sub-populations 1--3 by row. In each rolling-window step, the models are fitted on the available in-sample years and used to produce a one-year-ahead forecast for the next calendar year. This procedure is repeated over six successive rolling-window updates obtained by adding one calendar year at a time to the training sample. The x-axis shows the age brackets and the y-axis shows the corresponding average mean squared error. The four curves correspond to methods A.--D. Error bars indicate the empirical standard deviation across independently simulated data sets of the average performance over the six rolling-window updates.}
\end{figure}

When evaluating the performance metrics discussed in Section~\ref{sec:evaluation-metrics} we apply a rolling window prediction strategy for a total of six calendar periods subsequent to $t^\prime$, and the results are segmented in five-year age intervals for the ages between $16$ to $85$. In addition, thirty independent copies of the population data is generated, which is used to provide the whiskers in the MSE (\Cref{fig:oos-results}, left-hand side) and average Poisson deviance plots (\Cref{fig:oos-results}, right-hand side). To facilitate the discussion in this section, the figures are presented on sub-population and metric-specific scales. The corresponding results plotted on common scales are reported in \Cref{app:same-scale}.

Figure~\ref{fig:oos-results} shows that the credibility approach, model A., more or less coincides with the global mortality model D., for low to middle ages. For higher ages the credibility approach instead coincides with model B., which here outperforms the global model D. In terms of the credibility weights, this behaviour is intuitively reasonable: For low ages the death counts are low and the credibility weights will be close to 0, favouring the global mortality model D.; for higher ages the death counts are higher and the credibility weights are closer to 1 making model A.~closer to the relative survival model B. In terms of Poisson Deviance, in Figure~\ref{subfig:subpop1-dev} for sub-population 1, Model C.~performs worse than model D.~for younger ages and better than model D. but worse than models A.~and B.~for middle and old ages. It is also interesting to note that (on average) the relative survival model B.~for most ages outperform the separate sub-population model C. The intuition behind this is that, since the sub-population size is rather small, using sub-population specific models will introduce unreliable estimates. 

When considering the smallest sub-population, sub-population 2, Figures \ref{subfig:subpop2-mse} and \ref{subfig:subpop2-dev} show that the credibility approach provides a compromise between the global mortality model D.~and the relative survival model B. In this situation a relatively smooth transition from the global model to the relative survival model is seen with increasing age. It is also interesting to again see that the relative survival model B.~outperforms the sub-population specific models from approach C. Figures~\ref{subfig:subpop3-mse} and \ref{subfig:subpop3-dev} indicate only minor differences in performance among the models for the residual sub-population (sub-population 3). This is expected, since the sub-population effect is close to one, implying that the residual sub-population follows the mortality dynamics of the super-population closely.

\section{Conclusions}
\label{sec:conclusions}

In this paper, we introduce a credibility-based predictor for human mortality in a setting where a super-population consists of several small sub-populations. The predictor for each sub-population is defined as a weighted average of a global mortality model fitted to the super-population and a relative survival model.

Compared with classical credibility theory for non-life insurance data \citep{buhlmann05}, where exposures and average claim frequencies are assumed to be deterministic, our framework extends credibility theory to a stochastic intensity model and studies its estimation through plug-in techniques. The numerical illustrations show promising performance and indicate that the proposed method may serve as a useful data-driven approach for modelling mortality trends in small sub-populations.  

\section*{Data Availability Statement}

The code that we used to perform the analysis in this paper can be found in the GitHub repository \href{https://github.com/gpitt71/credMortality}{gpitt71/credMortality}. The GitHub folder is registered with the following Zenodo DOI: \href{https://doi.org/10.5281/zenodo.19615353}{https://doi.org/10.5281/zenodo.19615353}. The supplement is organized as a reproducible \texttt{targets} pipeline for regenerating the outputs of the paper \citep{targets}.

\section*{Conflicts of Interest}

No conflict of interest is declared.

\section*{Acknowledgments}

Gabriele Pittarello is funded by the Novo Nordisk Foundation grant NNF23OC0084961.

\printbibliography

\appendix

\section{Proofs}\label{app: proofs}

\subsection{Proof of the credibility estimator}

\label{appendix:proof_cm}

In this section we prove the credibility model estimator in \Cref{proposition:credibilityestimator} under the regularity conditions of \Cref{sec:model}. Given some sub-population given sub-populations $ i \in \mathbb{I}$ and ages $x \in \mathbb{X}$, is expressed in terms of parameters $\boldsymbol\omega_x^i :=  (\omega^i_{x,0},\omega^i_{x,1}, \ldots, \omega^i_{x, t'})^\Tr$ the credibility estimator of the central mortality rate in \Cref{definition: linear cred} minimises, conditionally on the filtration  $\mathcal{F}_{t^{\prime}}:=\sigma\left\{\mu_{x, s}: s \leq t^{\prime}, x \in \mathcal{X}\right\}$, the following expectation

\begin{align*}
    \min_{\boldsymbol{\omega}^i_x \in \mathbb{R}^{|\mathbb{T}|+1}} \; \mathbb{E}\left[\left(\mu_{x, t'+h} \left(\Theta^i_x\right) - \omega^i_{x,0} - \sum_{v \in \mathbb{T}} \omega^i_{x,v} F^i_{x,v} \right)^2 ~ \Bigg|~ \mathcal{F}_{t^{\prime}}  \right],
\end{align*}

Here, $\mu_{x,t} \left(\Theta^i_x\right)$  is the central mortality rate in calendar period $t$, age $x$ and sub-population $i$ that we want to estimate as a function of the observed population specific mortality rates. 

For ease of notation, let $\mathcal{E}(\boldsymbol{\omega}^i_x) := \mathbb{E}\left[\left(\mu_{x, t'+h} \left(\Theta^i_x\right) - \omega^i_{x,0} - \sum_{v \in \mathbb{T}} \omega^i_{x,v} F^i_{x,v} \right)^2 ~ \Bigg|~ \mathcal{F}_{t^{\prime}}  \right]$.

Let us consider the first order derivative with respect to  $\omega^i_{x,0}$ 

\begin{equation}
\label{eq:der1}
    %\frac{\partial \mathcal{E}(\boldsymbol{\omega}^i_x;  \boldsymbol{F}_x^i)}{\partial \omega^i_{x,0}}
    \frac{\partial \mathcal{E}(\boldsymbol{\omega}^i_x)}{\partial \omega^i_{x,0}}
    = \mathbb{E}\left[-2\left(\mu_{x,t^\prime+h} \left(\Theta^i_x\right) - \omega^i_{x,0} - \sum_{v \in \mathbb{T}} \omega^i_{x,v} F^i_{x,v} \right)  ~ \Bigg|~ \mathcal{F}_{t^{\prime}}\right],
\end{equation}
and with respect to $ \omega^i_{x,t}$ with $t \in \mathbb{T}$ 
\begin{equation}
\label{eq:der2}
    %\frac{\partial \mathcal{E}(\boldsymbol{\omega}^i_x;  \boldsymbol{F}_x^i)}{\partial \omega^i_{x,t}}
    \frac{\partial \mathcal{E}(\boldsymbol{\omega}^i_x)}{\partial \omega^i_{x,t}}
    = \mathbb{E}\left[-2F^i_{x,t}\left(\mu_{x,t^\prime+h} \left(\Theta^i_x\right) - \omega^i_{x,0} - \sum_{v \in \mathbb{T}} \omega^i_{x,v} F^i_{x,v} \right) ~ \Bigg|~ \mathcal{F}_{t^{\prime}}\right].
\end{equation}
By setting \Cref{eq:der1} equal to zero yields

\begin{equation}
\label{eq:neq1}
	\widehat\omega^i_{x,0} = \bar \mu_{x,t^\prime+h}-\sum_{v \in \mathbb{T}} \widehat  \omega^i_{x,v} \mu_{x,v}.
\end{equation}

Analogously, setting \Cref{eq:der2} equal to zero and subtracting the product of \Cref{eq:der1} and $\mathbb{E}\left[F^i_{x,t}~|~ \mathcal{F}_{t^{\prime}}\right]$ provides us with
\begin{equation}
\label{eq:omegaxtisolution}
	\text{Cov}(F^i_{x,t}, \mu_{x,t^\prime+h}\left(\Theta^i_x\right) ~|~ \mathcal{F}_{t^{\prime}}) = \sum_{v\neq t}  \widehat\omega^i_{x,v}\text{Cov}(F^i_{x,t}, F^i_{x,v} ~|~ \mathcal{F}_{t^{\prime}})+ \widehat\omega^i_{x,t} \Var\left(F^i_{x,t} ~|~ \mathcal{F}_{t^{\prime}}\right).
\end{equation}

Under \Cref{ass:CM1}, recall \eqref{eq: covariance raw mortality rates}, i.e.
$$
	\text{Cov}(F^i_{x,t}, F^i_{x,v}~|~ \mathcal{F}_{t^{\prime}})= \mu_{x,t} \; \mu_{x,v} \Var(\Theta^i_x ),
$$
and, analogously, it holds that

$$
	\text{Cov}(F^i_{x,t}, \mu_{x,t^\prime+h}\left(\Theta^i_x\right)~|~ \mathcal{F}_{t^{\prime}})= \mu_{x,t} \mathbb{E} \; \left[ \mu_{x,t^\prime + h} \Var(\Theta_x^i) \mid \mathcal{F}_{t^\prime}\right]$$
According to \Cref{ass:CM2}, we write

$$
\Var\left(F^i_{x,t} ~|~ \mathcal{F}_{t^{\prime}}\right) = \mu_{x,t}\; \left( \mathbb{E}\left[ \frac{\Theta^i_x}{E^i_{x,t}} \mid  \mathcal{F}_{t^{\prime}}\right] + \mu_{x,t} \Var(\Theta^i_x )) \right).
$$

Using this, we rewrite \Cref{eq:omegaxtisolution} as 

\begin{equation}
\label{eq:neq2}
    \left(\mathbb{E}\left[\mu_{x,t^\prime+h} \Var(\Theta^i_x )\mid \mathcal{F}_{t^\prime} \right] - \sum_{v \in \mathbb{T}} \widehat  \omega^i_{x,v} \mu_{x,v} \Var(\Theta^i_x ) \right) =  \widehat  \omega^i_{x,t} \mathbb{E}\left[ \frac{\Theta^i_x}{E^i_{x,t}} \mid  \mathcal{F}_{t^{\prime}}\right],
\end{equation}

and observe that for any $t,s \in \mathbb{T}$ with $t\neq s$

$$
 \widehat  \omega^i_{x,t} = \left( \mathbb{E}\left[ \frac{\Theta^i_x}{E^i_{x,s}} \mid  \mathcal{F}_{t^{\prime}}\right] / \mathbb{E}\left[ \frac{\Theta^i_x}{E^i_{x,t}} \mid  \mathcal{F}_{t^{\prime}}\right]\right) \widehat  \omega^i_{x,s}.
$$

By plugging $\widehat  \omega^i_{x,t}$ into \Cref{eq:neq2}, we obtain for some $t \in \mathbb{T}$

$$
 \widehat  \omega^i_{x,t} = \frac{\bar\mu_{x,t^\prime+h}/ \mathbb{E}\left[ \frac{\Theta^i_x}{E^i_{x,t}} \mid  \mathcal{F}_{t^{\prime}}\right]}{ \frac{1}{\Var(\Theta^i_x)} +  \sum_{v \in \mathbb{T}}  \mu_{x,v} / \mathbb{E}\left[ \frac{\Theta^i_x}{E^i_{x,v}} \mid  \mathcal{F}_{t^{\prime}}\right] }.
$$

Under Assumptions \ref{ass:CM1}-- \ref{ass:CM2}, the estimated weights for the credibility model provide the solution

$$
\widehat{\mu_{x, t'+h} \left(\Theta^i_x\right)} = \frac{\bar\mu_{x,t^\prime+h}}{\frac{1}{\Var(\Theta^i_x  )} +  \sum_{v \in \mathbb{T}}  \mu_{x,v} / \mathbb{E}\left[ \frac{\Theta^i_x}{E^i_{x,v}} \mid  \mathcal{F}_{t^{\prime}}\right]} \left( \frac{1}{\Var(\Theta^i_x)} + \sum_{v \in \mathbb{T}}  F^i_{x,v} / \mathbb{E}\left[ \frac{\Theta^i_x}{E^i_{x,v}} \mid  \mathcal{F}_{t^{\prime}}\right]\right).
$$

Few algebraic steps lead to 

$$
 \widehat{\mu_{x, t'+h} \left(\Theta^i_x\right)} := (1 - z_x^i) \overline \mu_{x, t' + h} + z_x^i \widehat \mu_{x, t' + h}^i,
$$

with 

$$
z_x^i := \frac{\sum_{v \in \mathbb{T}}  \mu_{x,v} / \mathbb{E}\left[ \frac{\Theta^i_x}{E^i_{x,v}} \mid  \mathcal{F}_{t^{\prime}}\right]}{ \frac{1}{\Var(\Theta^i_x)} +  \sum_{v \in \mathbb{T}}  \mu_{x,v} / \mathbb{E}\left[ \frac{\Theta^i_x}{E^i_{x,v}} \mid  \mathcal{F}_{t^{\prime}}\right] },
$$

and $\widehat \mu_{x, t' + h}^i := \overline \mu_{x, t' + h} R_x^i,$ where $ R_x^i = \frac{\sum_{v \in \mathbb{T}}  F_{x,v}^i / \mathbb{E}\left[ \frac{\Theta^i_x}{E^i_{x,v}} \mid  \mathcal{F}_{t^{\prime}}\right]}{ \sum_{v \in \mathbb{T}}  \mu_{x,v} / \mathbb{E}\left[ \frac{\Theta^i_x}{E^i_{x,v}} \mid  \mathcal{F}_{t^{\prime}}\right] }.$

\subsection{Proof of the expected quadratic forecast error }\label{app: proof MSEP}

In this section we prove the expression for the quadratic forecast error from \Cref{thm: MSEP}. Let us consider, for some sub-populations $i, i \in \mathbb{I}$, ages $x \in \mathbb{X}$, given the filtration $\mathcal{F}_{t^{\prime}}$, the following expectation 
\begin{align*}
 \operatorname{Q}(\mu_{x,t^\prime+h} \left(\Theta^i_x\right), \widehat{ \mu_{x,t^\prime+h} \left(\Theta^i_x\right)} \mid \calF_{t'}) &= \mathbb{E}\left[(\mu_{x,t^\prime+h} \left(\Theta^i_x\right)-\widehat{ \mu_{x,t^\prime+h} \left(\Theta^i_x\right)})^2 \mid \mathcal{F}_{t^{\prime}}\right] \\
&=\mathbb{E}\left[(\mu_{x,t^\prime+h} \left(\Theta^i_x\right)-\left((1- z^i_{x}) \overline \mu_{x, t'+h} + z^i_{x} \widehat\mu_{x, t'+h}^i \right)^2 \mid \mathcal{F}_{t^{\prime}}\right].
\end{align*}
With a few algebraic steps it follows that 
\begin{align*}
&\mathbb{E}\left[(\mu_{x,t^\prime+h} \left(\Theta^i_x\right)-\left((1- z^i_{x}) \overline \mu_{x, t'+h} + z^i_{x} \widehat\mu_{x, t'+h}^i \right)^2 \mid \mathcal{F}_{t^{\prime}}\right] \\
&= \Var\left(\mu_{x,t^\prime+h} \left(\Theta^i_x\right) \mid \mathcal{F}_{t^{\prime}} \right) \\
&\quad + \mathbb{E}\left[ \left( z_x^i \; (\overline\mu_{x,t^\prime+h}  - \widehat\mu^i_{x, t' + h}) \right)^2 \mid \mathcal{F}_{t^{\prime}} \right] \\
&\quad + 2 \; \mathbb{E}\left[ z_x^i \left( \mu_{x,t^\prime+h} \left(\Theta^i_x\right) - \overline\mu_{x, t' + h} \right) \left(\overline\mu_{x, t' + h} - \widehat\mu^i_{x, t' + h} \right)  \mid \mathcal{F}_{t^{\prime}} \right],
\end{align*}

with $\widehat\mu^i_{x, t' + h}:= R_x^i \; \overline\mu^i_{x, t' + h}$.
Next, note that one can rewrite according to

\begin{align*}
\mathbb{E}\left[ \left( z_x^i \; (\overline\mu_{x,t^\prime+h}  - \widehat\mu^i_{x, t' + h}) \right)^2 \mid \mathcal{F}_{t^{\prime}} \right] = (z_x^i \; \overline\mu_{x,t^\prime+h})^2 \;  \Var\left( R_x^i \mid \mathcal{F}_{t^{\prime}} \right).
\end{align*}

Lastly, 

\begin{align*}
    \mathbb{E}&\left[  z_x^i \left( \mu_{x,t^\prime+h} \left(\Theta^i_x\right) - \overline\mu_{x, t' + h} \right) \left(\overline\mu_{x, t' + h} - \widehat\mu^i_{x, t' + h} \right)  \mid \mathcal{F}_{t^{\prime}} \right]\\
    &= z_x^i \; \overline\mu_{x, t' + h} \; \mathbb{E}\left[\left( \mu_{x,t^\prime+h} \left(\Theta^i_x\right) - \overline\mu_{x, t' + h} \right) \left(1- R^i_x \right)  \mid \mathcal{F}_{t^{\prime}} \right]\\
    &\quad -z_x^i \; \overline\mu_{x, t' + h} \mathbb{E}\left[\left( \mu_{x,t^\prime+h} \; \Theta^i_x - \overline\mu_{x, t' + h} \right) R^i_x  \mid \mathcal{F}_{t^{\prime}} \right]\\
    &\quad -z_x^i \; \overline\mu_{x, t' + h} \mathbb{E}\left[\mathbb{E}\left[\left( \mu_{x,t^\prime+h} \; \Theta^i_x - \overline\mu_{x, t' + h} \right) R^i_x \mid \Theta_x^i, \mu_{x,t^\prime+h}, \mathcal{F}_{t^\prime}, \left(E_{x,v}\right)^{t^\prime}_{v=1} \right] \mid \mathcal{F}_{t^{\prime}} \right] \\
    & = 0
\end{align*}

\begin{align*}
    & \mathbb{E}\left[ \left( \mu_{x,t^\prime+h} \left(\Theta^i_x\right) - \widehat \theta_x^i \overline\mu_{x, t' + h} \right)^2 \mid \mathcal{F}_{t^{\prime}} \right] = \Var\left(\mu_{x,t^\prime+h} \left(\Theta^i_x\right) \mid \mathcal{F}_{t^{\prime}}  \right) + \overline\mu_{x, t' + h}  \Var\left(\widehat \theta_x^i  \mid \mathcal{F}_{t^{\prime}}  \right) \\
    &\quad = \overline{\sigma}^2_{x,t'  + h}(\Var\left(\Theta^i_x\right)+1)+(\overline{\mu}_{x,t^\prime+h})^2 \Var\left(\Theta^i_x\right)\\
    &\quad\quad + \overline\mu_{x, t' + h} \left( \Var\left(\Theta^i_x\right)\frac{\sum_v (E^i_{x,v}\mu_{x,v})^2}{(\sum_v E^i_{x,v} \mu_{x,v})^2}+\frac{1}{\sum_v E^i_{x,v} \mu_{x,v}} \right)\\
    &\quad = \Var\left(\Theta^i_x\right) \; \left(\frac{\sum_v (E^i_{x,v}\mu_{x,v})^2}{(\sum_v E^i_{x,v} \mu_{x,v})^2}+\overline\sigma_{x, t' + h}^2+\overline\mu_{x, t' + h}\right) + \overline\sigma_{x, t' + h}^2 + \frac{1}{\sum_v E^i_{x,v} \mu_{x,v}},
\end{align*}
with $\overline\sigma_{x, t' + h}^2 := \Var(\mu_{x, t' + h} \mid \mathcal{F}_{t^{\prime}}) = \Var(\mu_{x, t' + h} \mid  \mathcal{F}_{t^{\prime}}, t \in \mathbb{T})$. Further, we have that
\begin{align*}
    &\mathbb{E}\left[ \left( \mu_{x,t^\prime+h} \left(\Theta^i_x\right) - \widehat \theta_x^i \overline\mu_{x, t' + h} \right) \left( \mu_{x,t^\prime+h} \left(\Theta^i_x\right) - \overline\mu_{x, t' + h} \right)  \mid \mathcal{F}_{t^{\prime}} \right] \\
    &\quad = (\overline\sigma_{x, t' + h}^2 + \overline\mu_{x, t' + h}^2) \Var\left(\Theta^i_x\right) + \overline\sigma_{x, t' + h}^2 \\
    &\quad = \Var\left(\mu_{x,t^\prime+h} \left(\Theta^i_x\right) \mid \mathcal{F}_{t^{\prime}}  \right).
\end{align*}
Finally, we can write that 
\begin{align*}
        \operatorname{Q}(\mu_{x,t^\prime+h} \left(\Theta^i_x\right), \widehat{ \mu_{x,t^\prime+h} \left(\Theta^i_x\right)}\mid \calF_{t'}) = \Var\left(\mu_{x,t^\prime+h} \left(\Theta^i_x\right) \mid \mathcal{F}_{t^{\prime}}\right)+(z^i_x)^2(\overline{\mu}_{x,t^\prime+h})^2 \Var\left(R^i_x \mid \mathcal{F}_{t^{\prime}}\right),
    \end{align*}

with $\Var\left(R^i_x \mid \mathcal{F}_{t^{\prime}}\right) = \Var\left(\Theta^i_x\right)+\frac{1}{\sum_v \mu_{x,v} / \mathbb{E}\left[ \frac{\Theta^i_x}{E^i_{x,v}} \mid  \mathcal{F}_{t^{\prime}}\right]}$ and

\begin{align*}
\Var\left(\mu_{x,t^\prime+h} \left(\Theta^i_x\right) \mid \mathcal{F}_{t^{\prime}} \right) = \overline{\sigma}^2_{x,t'  + h}(\Var\left(\Theta^i_x\right)+1)+(\overline{\mu}_{x,t^\prime+h})^2 \Var\left(\Theta^i_x\right),
\end{align*}

with $\overline{\sigma}^2_{x,t'  + h} = \Var \left( \mu_{x,t^\prime+h} \mid \mathcal{F}_{t^{\prime}} \right)$.

\section{Auxiliary results and information}

{
\color{black}
\subsection{Generalised age-period-cohort models}
\label{appendix:gapc}

We refer to the generalised age-period-cohort (GAPC) class of models as a class of generalised non-linear models for the central mortality rate that assume the following structure on the predictor $\mu_{x,t},$ for some $x \in \mathbb{X}$ and $t \in \mathbb{T}$:

$$
\mu_{x,t} = \alpha_x + \sum^K_{k=1}\beta^{(k)}_x \kappa^{(k)}_t + \beta^{(0)}_x\gamma_{t-x}.
$$

The GAPC models were first discussed in \citet{hunt21} and allow to model a static age effect $\alpha_x$, $K$ interaction components between age and time ($\beta^{(k)}_x \kappa^{(k)}_t $) and interactions between age and cohorts ($\beta^{(0)}_x\gamma_{t-x}$). The models that we included in our application and that we reported in \Cref{tab:label_model_predictor_constraints} (the Lee-Carter, the Age-Period-Cohort, and the Renshaw-Haberman models) belong to the GAPC models class.

}

\subsection{Sub-population simulation}\label{app: life portfolio}

In order to ease the exposition, all sub- and superscripts are dropped. To start off, when using the Poisson assumption, it holds that
\[
	\operatorname{P}(D = 0 \mid E) = \text{e}^{-E \mu},
\]
whereas the corresponding probability under the binomial assumption provides us
\[
	\operatorname{P}(D = 0 \mid N) = (1 - q)^N.
\]
Further, from the formulation of the binomial model used for simulating the data, it was in addition assumed that
\[
	\log\frac{q}{1 - q} = \delta,~ \text{or equivalently, that}~ q = \frac{\text{e}^\delta}{1 + \text{e}^\delta}.
\]
Hence, by assuming that $\delta \ll 0$, i.e.~that $q \approx 0$, it follows that
\begin{align*}
	(1 - q)^N &= \exp\left\{N \log\left(1 -  \frac{\text{e}^\delta}{1 + \text{e}^\delta}\right)\right\} \approx \text{e}^{- N \text{e}^\delta}.
\end{align*}
This in turn means that if we replace $\delta$ with $\log(\Theta) + \delta$, we arrive at
\[
	(1 - q)^N \approx  \text{e}^{- N \Theta \text{e}^\delta}.
\]
That is, $\Theta \text{e}^\delta$ is an approximation of the relative survival model from \eqref{eq: relative survival} for sufficiently small $q$s.

\subsection{Age, period and cohort data classification}

While the data from epidemiological studies are tabulated by cohorts, insurance mortality data, human mortality data and registry data are tabulated by calendar period (the year of event). In this Section, we discuss the potential issues with tabulated data and show how to adjust the raw data for modelling central mortality rates. 

If the individual mortality data were available, the exact value of the time at risk for each individual in the population would be known. However, as mentioned above, our data consisted of an aggregation of the individual lives by age and period. For each age and period, we only observe the the raw lives (and deaths). Modelling mortality rates directly on the raw lives might lead to biased estimates of the underlying true mortality rates \citep{hunt21}.  %We recall that for a population $i$ at age $x$ and calendar period $t$, the total raw lives were denoted as $l^i_{x,t}$.

\begin{figure}[ht]
    \begin{subfigure}{0.45\textwidth}
        \includegraphics[width=\linewidth]{ 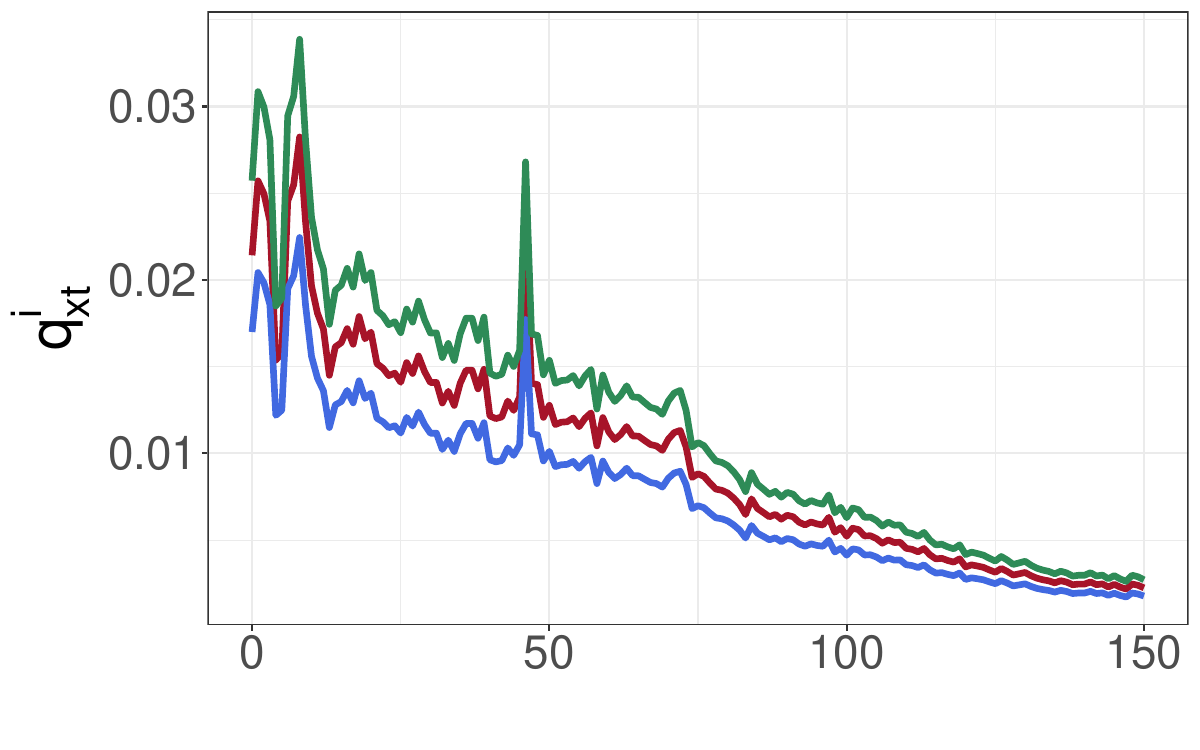}
        \caption{}
    \end{subfigure}   
    \hfill
    \begin{subfigure}{0.45\textwidth}
        \begin{tikzpicture}
        \draw[->] (0,0) to (0,4);
        \draw[->] (0,0) to (4,0);
        %cohort arrow
        \draw[->] (2,.5) to (3,1.5);

        %text
        \node[text width=4cm,rotate= 90] at (-1.2,3.5) 
            {Age };
        \node[text width=4cm] at (3.5,-.8) 
            {Period};
        \node[text width=4cm] at (5,.8) 
            {Cohort};
        \node[text width=.1cm] at (-.3,1) 
            {x};
        \node[text width=1cm] at (-.3,2) 
            {x+1};
        \node[text width=1cm] at (-.3,3) 
            {x+2};
        \node[text width=.1cm] at (1,-.3) 
            {t};
        \node[text width=.5cm] at (2,-.3) 
            {t+1};
                    
        %parallelogram
        \filldraw[draw=black, fill=gray!20] (1,1) -- (2,2) -- (2,3) -- (1,2)  -- cycle;
        
        \node[text width=.5cm] at (.9,.9) 
            {A};
        \node[text width=.5cm] at (2.3,.9) 
            {B};
        \node[text width=.5cm] at (2.3,2.1) 
            {C};
        \node[text width=.5cm] at (2.3,3.1) 
            {D};
        \node[text width=.5cm] at (.9,2.1) 
            {E};
            
        % block
        \draw[dashed] (1,0) -- (1,4);
        \draw[dashed] (2,0) -- (2,4);
        \draw[dashed] (0,1) -- (4,1);
        \draw[dashed] (0,2) -- (4,2);
        \draw[dashed] (0,3) -- (4,3);
    \end{tikzpicture}
    \caption{}
    \end{subfigure}
    \caption{\label{fig:lexis} On the left-hand side, assumptions on the mortality trend for $x=55$ across the different periods, for the super-population (red), sub-population 1 (blue) and sub-population 2 (green). On the right-hand side, sketch of a Lexis diagram, an age-period representation of mortality data. Periods are represented on the x-axis and ages are on the y-axis. The diagonals on the diagram represent the different cohorts. Mortality data are tabulated by squares. For example, for age $x$ and calendar period $t$ we only observe the total deaths represented in the square with vertices A,B,C,E. However the total individuals who have been exposed to risk in period $t$ are those represented with the parallelogram A,C,D,E.}
\end{figure}

Let us represent our data in the Lexis diagram in \Cref{fig:lexis}. The figure is an age-period representation of the tabular data for a period $t$ and age $x$. In the picture they are represented represented in the square with vertices A,B,C,E. However, the total individuals who have been exposed to the risk of death in period $t$ are those represented with the parallelogram A,C,D,E.  

Assuming a uniform distribution of the deaths on yearly parallelograms (A,C,D,E in the picture) we compute the average exposure as follows. For $i \in \left\{0,1,2\right\}$, $x \in \mathbb{X}$ and $t \in \mathbb{T}$ 
\begin{align}\label{eq: pop size to exposure}
	E^i_{x,t}= N^i_{x,t}/2+N^i_{x+1,t+1}/2.
\end{align}
The average deaths are computed as $D^i_{x,t}=(N^i_{x,t}-N^i_{x+1,t+1})/2$. For details on the calculation, we refer to \citet{carstensen05}. Further adjustments are required for computing the exposure at age $0$ and in elder classes $89+$, see, e.g., \citet{hunt21, carstensen05} for details, as modelling youngest and elderly ages goes beyond the scope of this manuscript. 

The pre-processing step described in this section enables us to model the central mortality rates that we discussed in this paper.

\subsection{Figures on comparable scales}
\label{app:same-scale}

For ease of comparison, this section reports \Cref{fig:mortality fan plots} and \Cref{fig:oos-results} using common scales.

\begin{figure}[ht]
  \centering
  \begin{subfigure}[t]{0.3\textwidth}
    \centering
    \includegraphics[width=\linewidth]{ 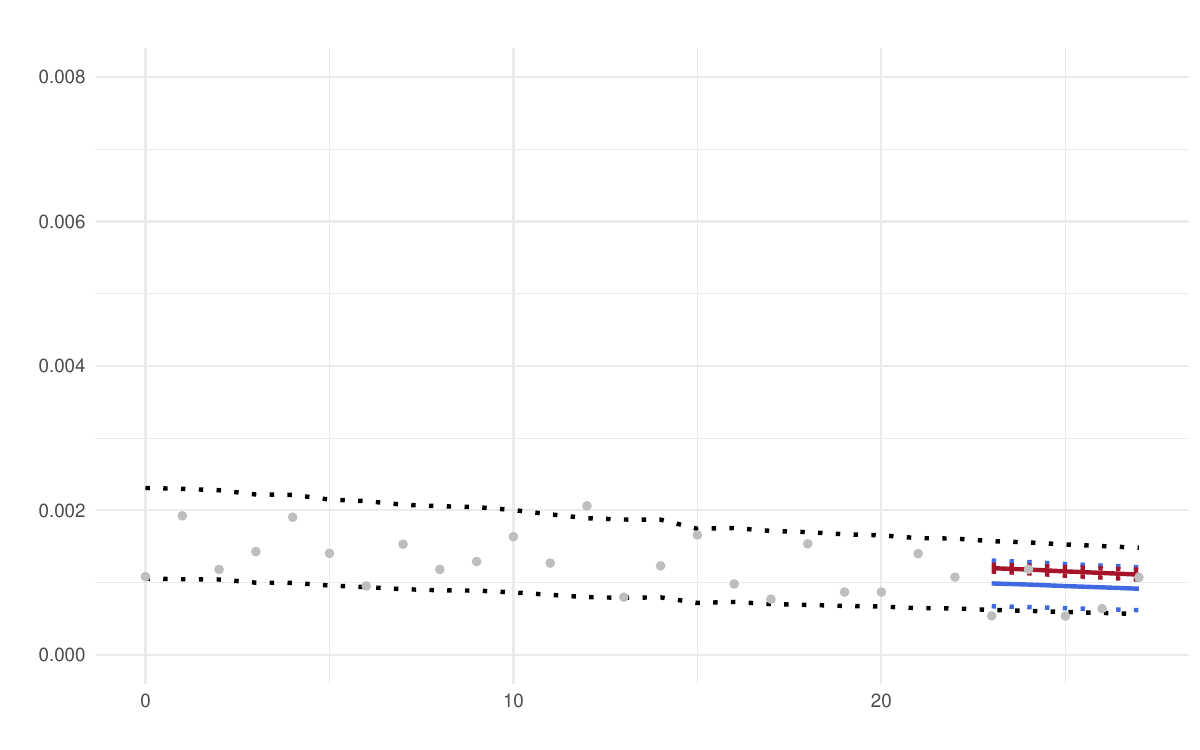}
    \caption{Sub-population 1.}
    \label{sf:group1_msep_app}
  \end{subfigure}
  \hfill
  \begin{subfigure}[t]{0.3\textwidth}
    \centering
    \includegraphics[width=\linewidth]{ 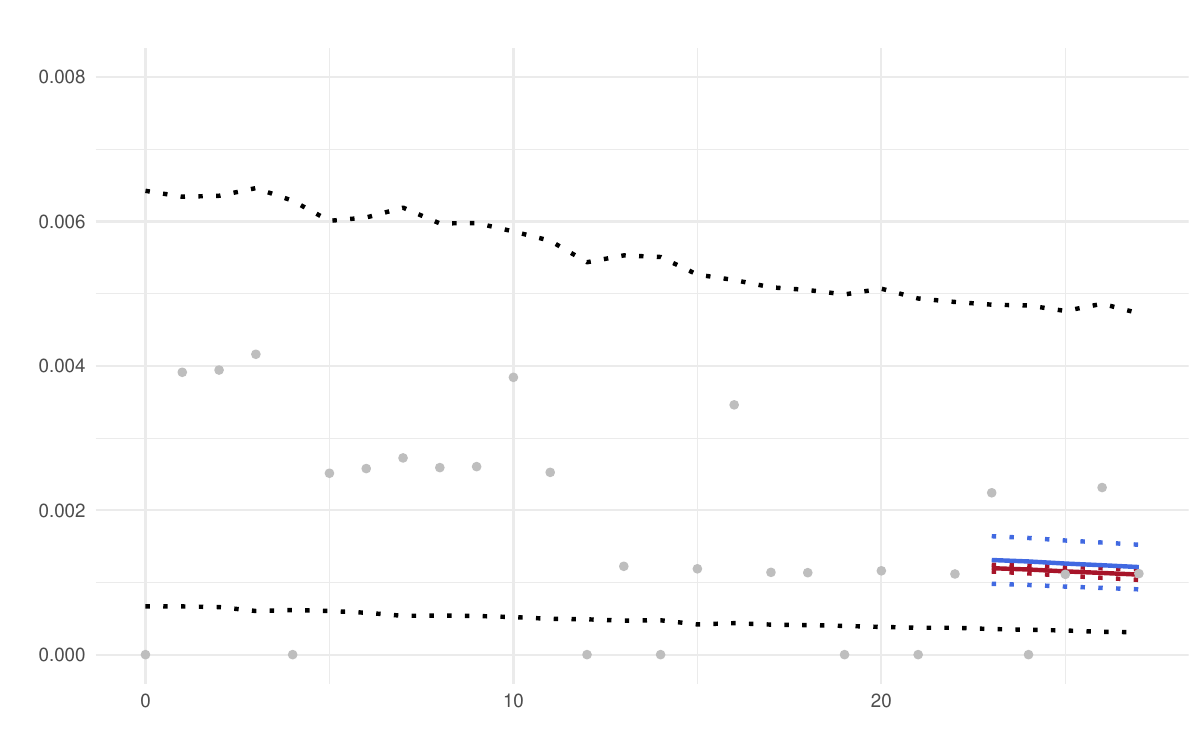}
    \caption{Sub-population 2.}
    \label{sf:group2_msep_app}
  \end{subfigure}
  \hfill
  \begin{subfigure}[t]{0.3\textwidth}
    \centering
    \includegraphics[width=\linewidth]{ 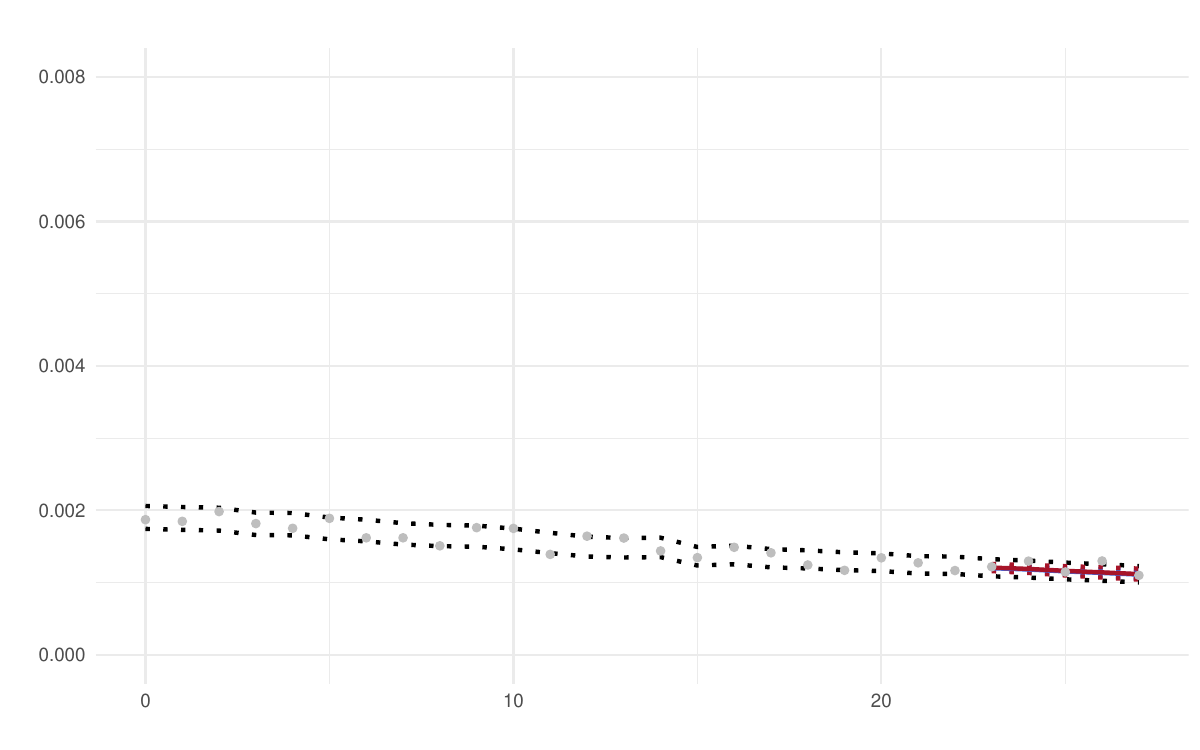}
    \caption{Sub-population 3.}
    \label{sf:group0_msep_app}
  \end{subfigure}
  \caption{\label{fig:mortality fan plots appendix} The gray dots represent observed central mortality rates for the synthetic data for age $x = 55$. Solid lines correspond to  mortality rate predictions, dotted lines correspond to the root of the expected quadratic forecast error for mortality rates, where model A.~is in blue and model C.~is in red. 
  The black dashed lines corresponds to simulated standard error bounds for the credibility based Poisson model from \eqref{eq: Poisson sim} using model A. }
\end{figure}

\begin{figure}[htp]
\centering
\captionsetup[subfigure]{font=small}

\begin{adjustbox}{max totalsize={\textwidth}{0.8\textheight},center}
\begin{minipage}{\textwidth}
\centering

\begin{subfigure}[b]{0.48\textwidth}
  \includegraphics[width=\linewidth]{ 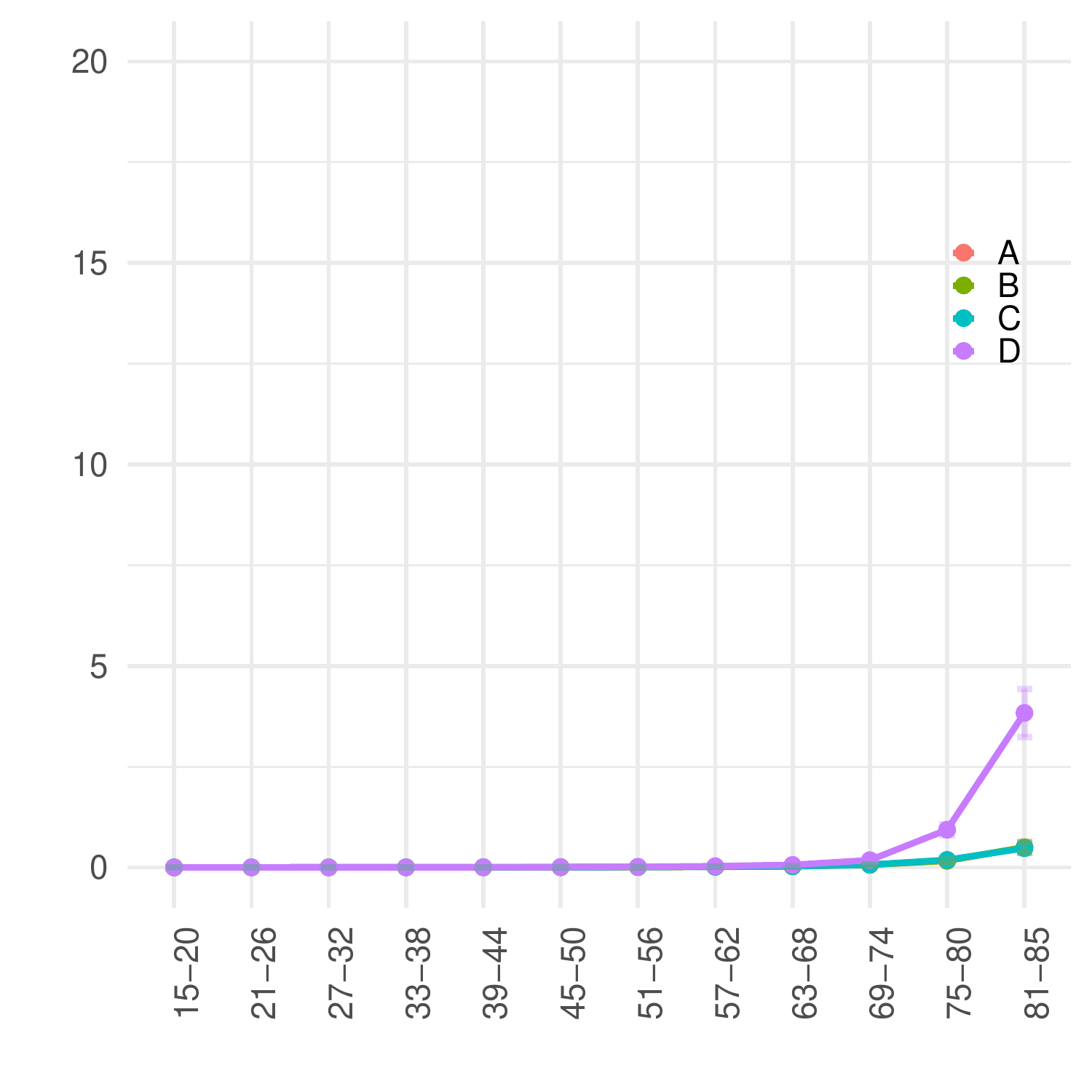}
  \caption{\label{subfig:subpop1-mse-app} Sub-population 1, MSE}
\end{subfigure}\hfill
\begin{subfigure}[b]{0.48\textwidth}
  \includegraphics[width=\linewidth]{ 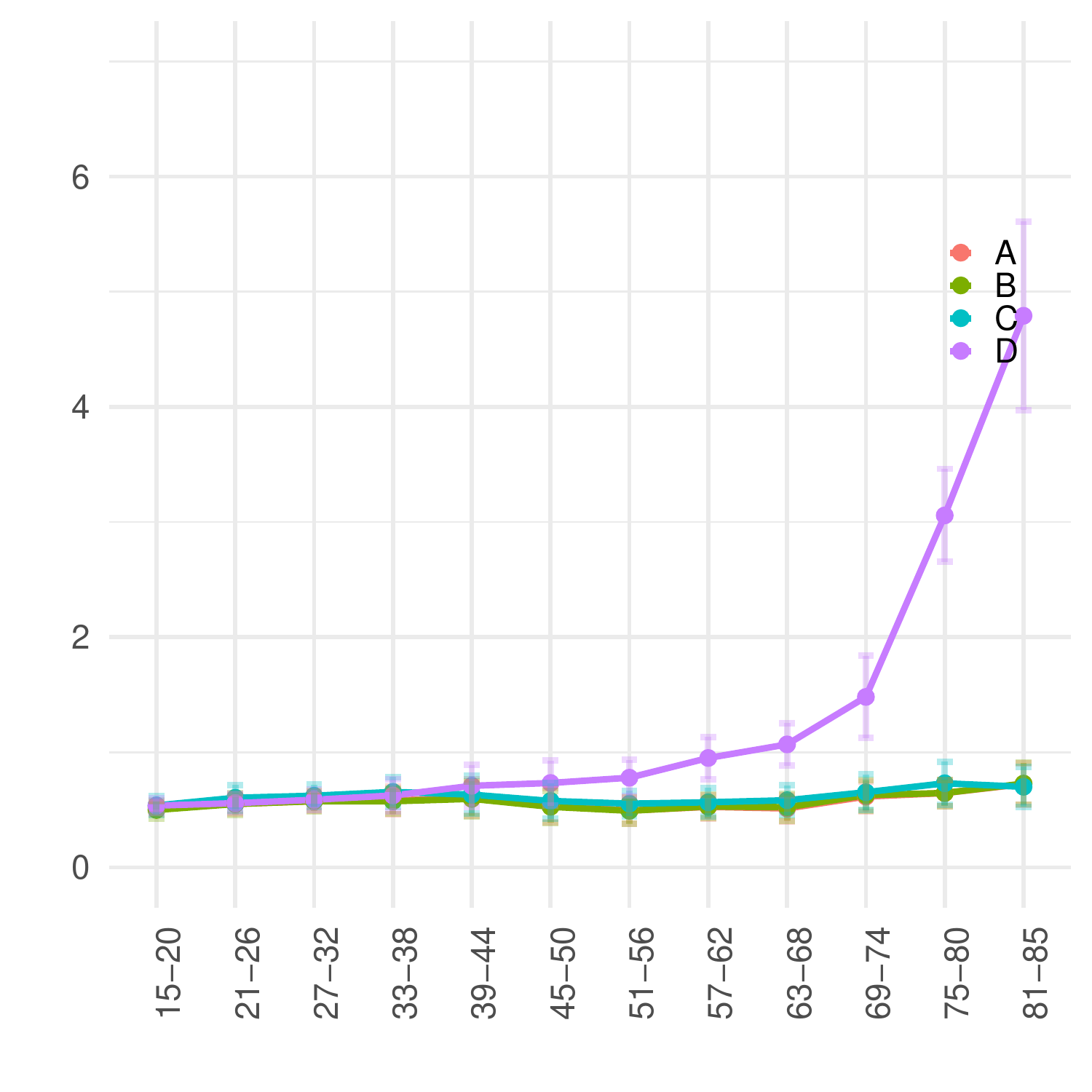}
  \caption{\label{subfig:subpop1-dev-app} Sub-population 1, Deviance}
\end{subfigure}

\vspace{0.5em}

\begin{subfigure}[b]{0.48\textwidth}
  \includegraphics[width=\linewidth]{ 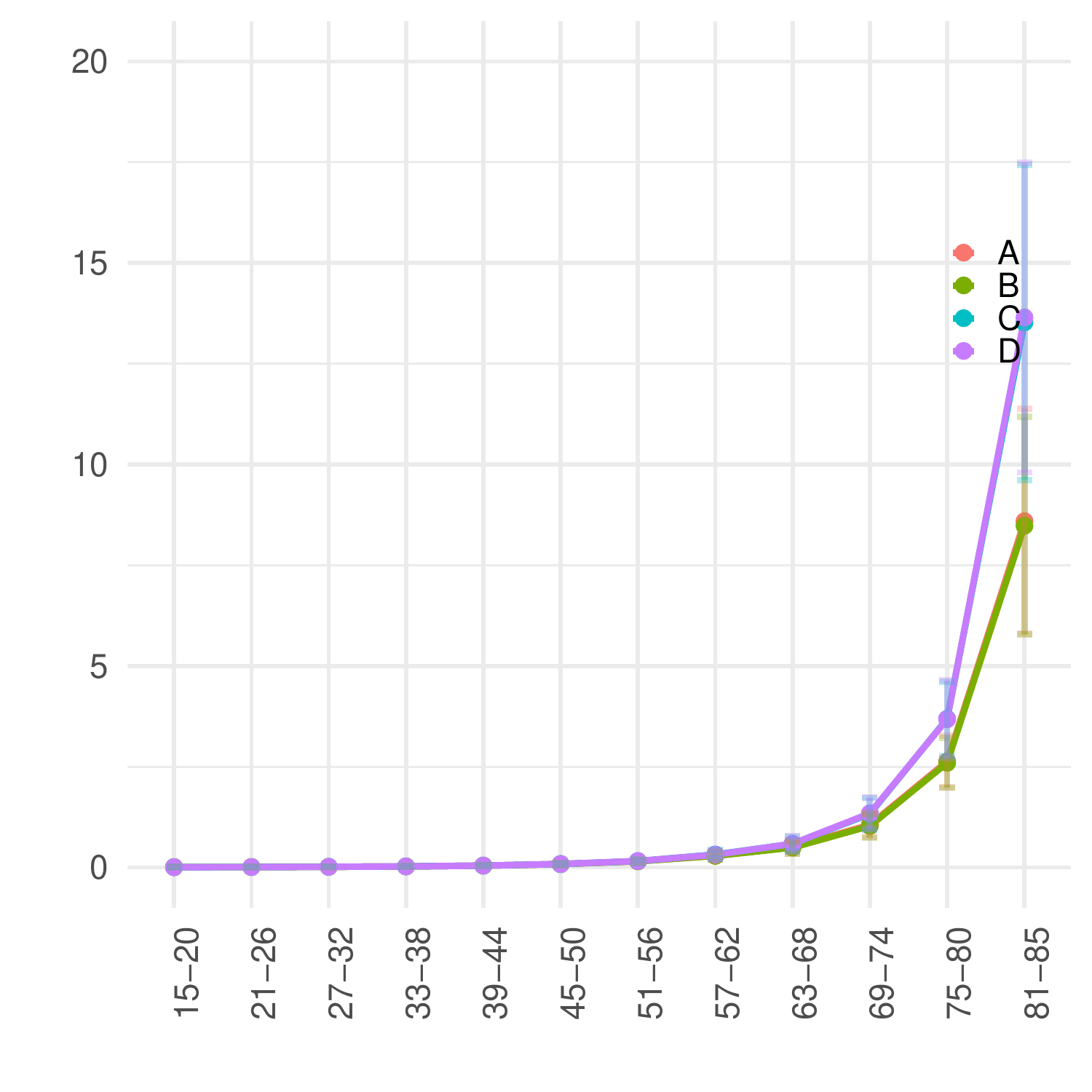}
  \caption{\label{subfig:subpop2-mse-app} Sub-population 2, MSE}
\end{subfigure}\hfill
\begin{subfigure}[b]{0.48\textwidth}
  \includegraphics[width=\linewidth]{ 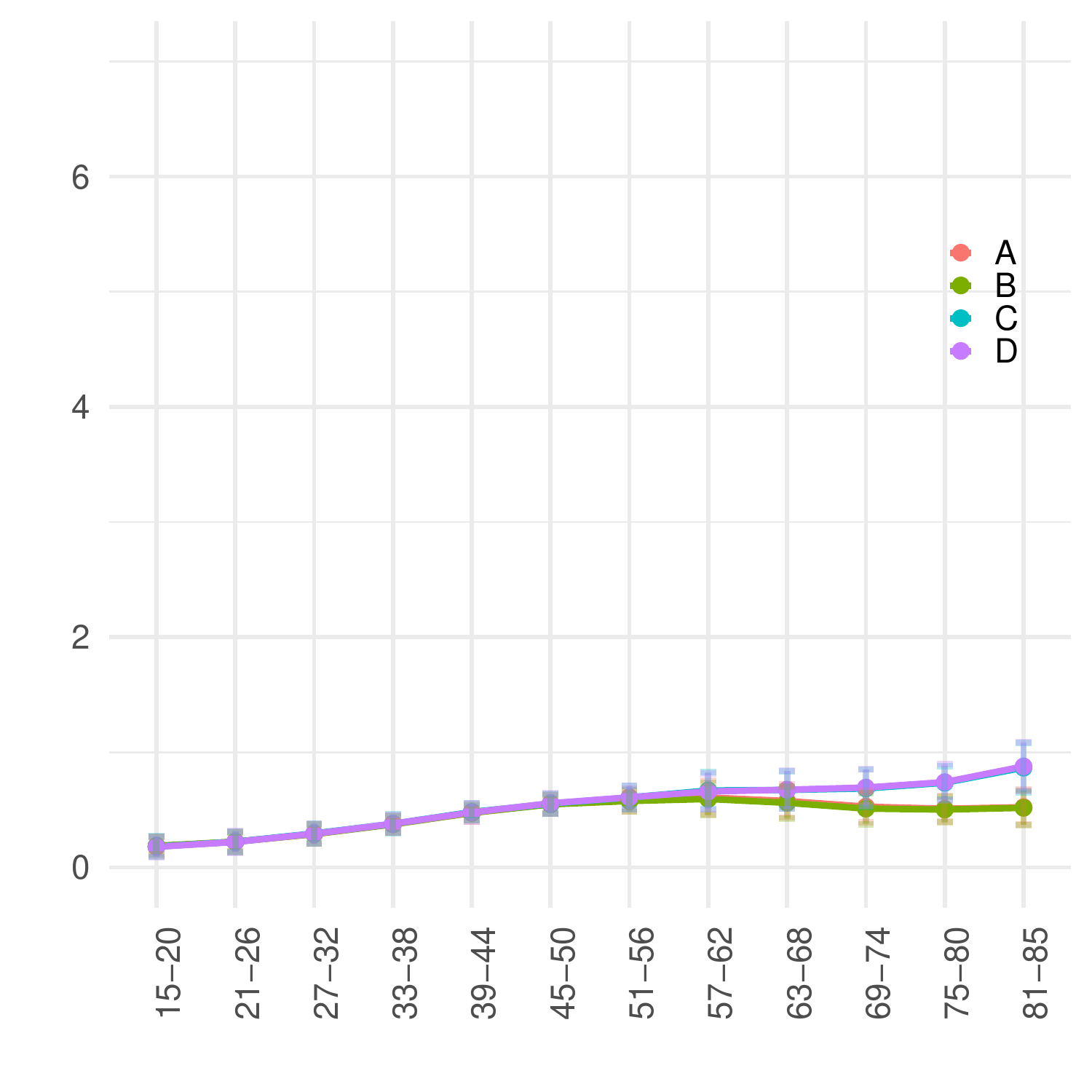}
  \caption{\label{subfig:subpop2-dev-app} Sub-population 2, Deviance}
\end{subfigure}

\vspace{0.5em}

\begin{subfigure}[b]{0.48\textwidth}
  \includegraphics[width=\linewidth]{ 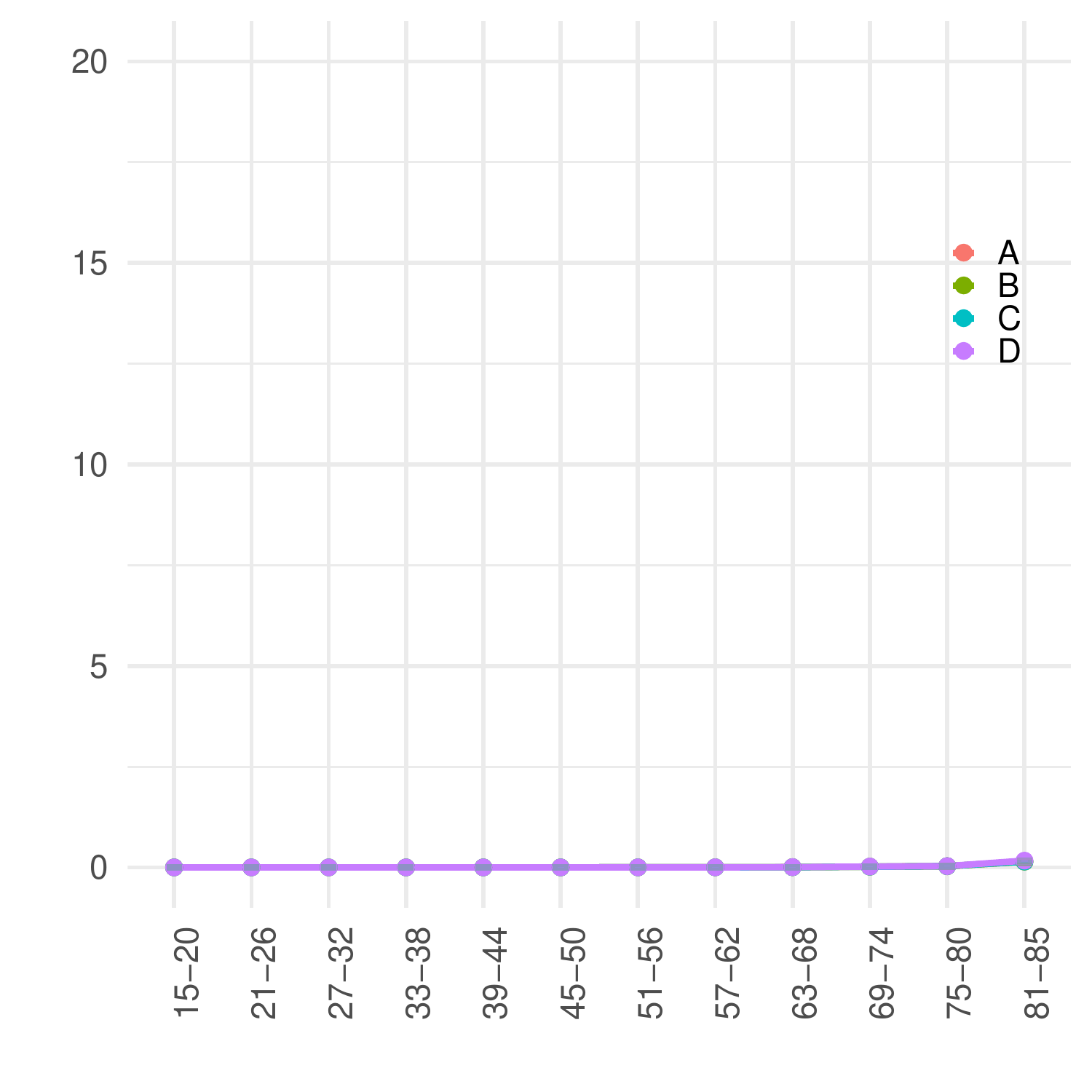}
  \caption{\label{subfig:subpop3-mse-app} Sub-population 3, MSE}
\end{subfigure}\hfill
\begin{subfigure}[b]{0.48\textwidth}
  \includegraphics[width=\linewidth]{ 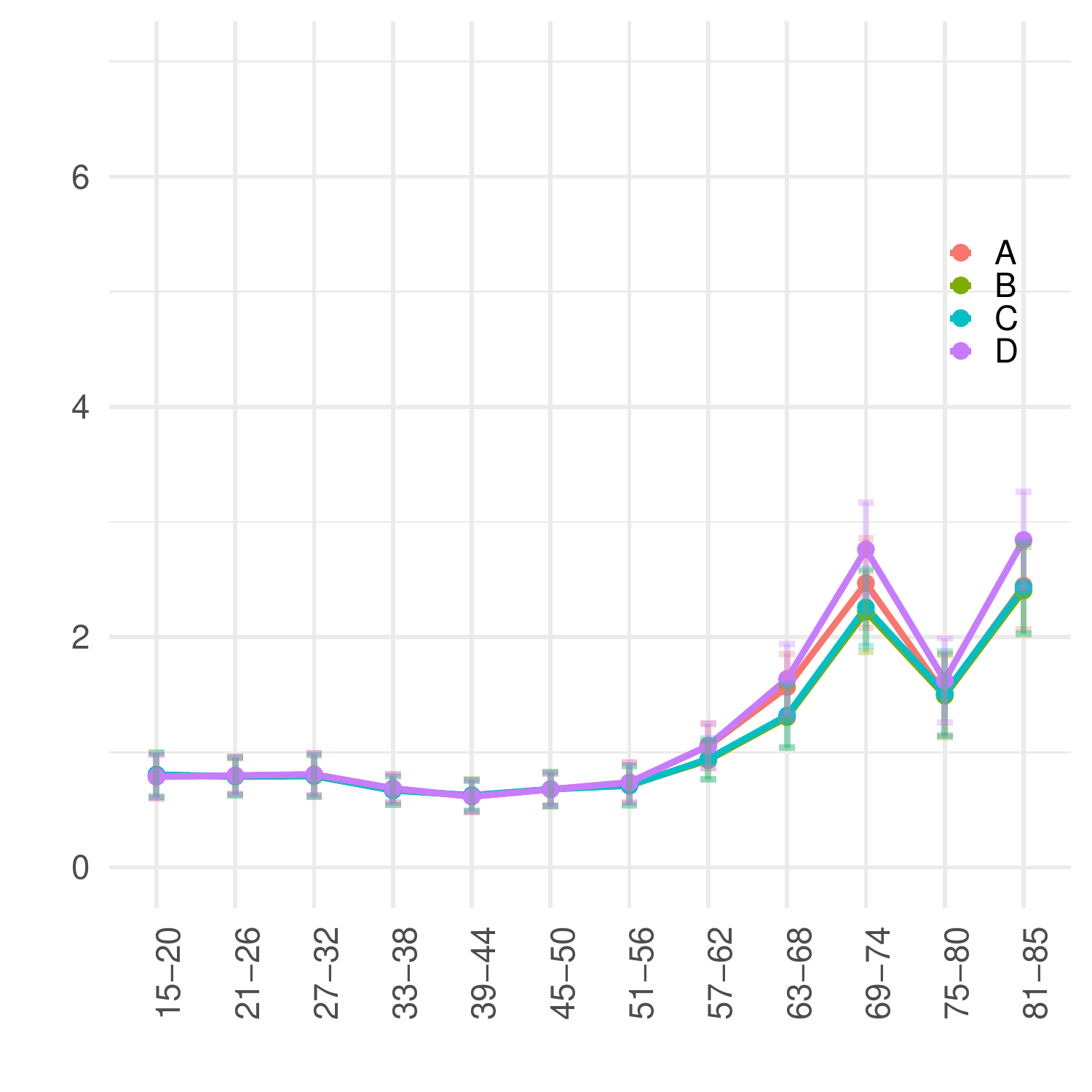}
  \caption{\label{subfig:subpop3-dev-app} Sub-population 3, Deviance}
\end{subfigure}

\end{minipage}
\end{adjustbox}
    \caption{\label{fig:oos-results-app} Out-of-sample mean squared error (left-hand side column) and out-of-sample average Poisson deviance (right-hand side column) under a rolling-window evaluation scheme using the Lee--Carter model as the global mortality model on a homogeneous scale. The panels report results for sub-populations 1--3 by row. In each rolling-window step, the models are fitted on the available in-sample years and used to produce a one-year-ahead forecast for the next calendar year. This procedure is repeated over six successive rolling-window updates obtained by adding one calendar year at a time to the training sample. The x-axis shows the age brackets and the y-axis shows the corresponding average mean squared error. The four curves correspond to methods A.--D. Error bars indicate the empirical standard deviation across independently simulated data sets of the average performance over the six rolling-window updates.}
\end{figure}

\end{document}